\RequirePackage{fix-cm}
\documentclass{svjour3}                     
\smartqed  
\usepackage{graphicx}
\usepackage{color}
\usepackage{cite}
\usepackage[sort&compress, square]{natbib}
\usepackage{har2nat}
\usepackage{amssymb}
\setcitestyle{square}
 \journalname{Space Science Review}

\newcommand{\grl}{    {Geophys. Res. Lett.}}
\newcommand{\jgr}{    {J. Geophys. Res.}}
\newcommand{\ssr}{    {Space Sci. Rev.}}
\newcommand{\planss}{    {Plan. Sp. Sci.}}

\newcommand{\nat}{ {Nature}}
\newcommand{\pre}{ {Phys. Rev. E}}

\begin{document}

\title{Energetic electron precipitation driven by electromagnetic ion cyclotron waves from ELFIN’s low altitude perspective
}
\subtitle{}

\titlerunning{Energetic electron precipitation driven by EMIC waves: ELFIN's perspective}        






\author{V. Angelopoulos\textsuperscript{1} \and X.-J. Zhang\textsuperscript{1,2} \and A. V. Artemyev\textsuperscript{1} \and D. Mourenas\textsuperscript{3} \and E. Tsai\textsuperscript{1} \and C. Wilkins\textsuperscript{1} \and A. Runov\textsuperscript{1} \and J. Liu\textsuperscript{1,4} \and D. L. Turner\textsuperscript{1,5} \and W. Li\textsuperscript{4} \and K. Khurana\textsuperscript{1} \and R. E. Wirz\textsuperscript{7,8} \and V. A. Sergeev\textsuperscript{9} \and X. Meng\textsuperscript{10} \and J. Wu\textsuperscript{1} \and M. D. Hartinger\textsuperscript{1,11} \and T. Raita\textsuperscript{12} \and Y. Shen\textsuperscript{1} \and X. An\textsuperscript{1} \and X. Shi\textsuperscript{1} \and M. F. Bashir\textsuperscript{1} \and X. Shen\textsuperscript{6} \and L. Gan\textsuperscript{6} \and M. Qin\textsuperscript{6} \and L. Capannolo\textsuperscript{6} \and Q. Ma\textsuperscript{6} \and C. L. Russell\textsuperscript{1} \and E. V. Masongsong\textsuperscript{1} \and R. Caron\textsuperscript{1} \and I. He\textsuperscript{1,13} \and L. Iglesias\textsuperscript{1,14} \and S. Jha\textsuperscript{1,15,16} \and J. King\textsuperscript{1,15} \and S. Kumar \textsuperscript{1,17} \and 18 \and K. Le\textsuperscript{1,13} \and J. Mao\textsuperscript{1,15,19} \and A. McDermott\textsuperscript{1,7} \and K. Nguyen\textsuperscript{1,7,20} \and A. Norris\textsuperscript{1} \and A. Palla\textsuperscript{1,15,21} \and Roosnovo\textsuperscript{1,17,22} \and J. Tam\textsuperscript{1,7} \and E. Xie\textsuperscript{1,14,15} \and R. C. Yap\textsuperscript{1,23,24} \and S. Ye\textsuperscript{1,7} \and C. Young\textsuperscript{1,15,16} \and L. A. Adair\textsuperscript{1,17,25} \and C. Shaffer\textsuperscript{1,7,26} \and M. Chung\textsuperscript{1,27} \and P. Cruce\textsuperscript{1,28} \and M. Lawson\textsuperscript{1} \and D. Leneman\textsuperscript{1} \and M. Allen\textsuperscript{1,7,29} \and M. Anderson\textsuperscript{1,23,30} \and M. Arreola-Zamora\textsuperscript{1,27} \and J. Artinger\textsuperscript{1,17,31} \and J. Asher\textsuperscript{1,7,32} \and D. Branchevsky\textsuperscript{1,32,33} \and M. Cliffe\textsuperscript{1,20,33} \and K. Colton\textsuperscript{1,23,24} \and C. Costello\textsuperscript{1,15,34} \and D. Depe\textsuperscript{1,33,35} \and B. W. Domae\textsuperscript{1,33} \and S. Eldin\textsuperscript{1,16,33} \and L. Fitzgibbon\textsuperscript{1,17,36} \and A. Flemming\textsuperscript{1,7,27} \and D. M. Frederick\textsuperscript{1,7,37} \and A. Gilbert\textsuperscript{1,33,38} \and B. Hesford\textsuperscript{1,10,33} \and R. Krieger\textsuperscript{1,13,39} \and K. Lian\textsuperscript{1,7,32} \and E. McKinney\textsuperscript{1,40} \and J. P. Miller\textsuperscript{1,15,41} \and C. Pedersen\textsuperscript{1,7} \and Z. Qu\textsuperscript{1,7,42} \and R. Rozario\textsuperscript{1,7,20} \and M. Rubly\textsuperscript{1,7,43} \and R. Seaton\textsuperscript{1,7} \and A. Subramanian\textsuperscript{1,27,33} \and S. R. Sundin\textsuperscript{1,7,44} \and A. Tan\textsuperscript{1,33,45} \and D. Thomlinson\textsuperscript{1,7,32} \and W. Turner\textsuperscript{1,17,46} \and G. Wing\textsuperscript{1,15,47} \and C. Wong\textsuperscript{1,17,48} \and A. Zarifian\textsuperscript{1,7,10}}


\institute{\at  1 Earth, Planetary, and Space Sciences Department, and Institute of Geophysics and Planetary Physics, University of California, Los Angeles, Los Angeles, CA 90095, USA\\
              \email{vassilis@ucla.edu}
\at 2	Now at: University of Texas at Dallas, Richardson, TX 75080
\at 3    CEA, DAM, DIF, Arpajon, France
\at 4	Atmospheric and Oceanic Sciences Departments, University of California, Los Angeles, CA
\at 5	Now at: Johns Hopkins University Applied Physics Laboratory, Laurel, Maryland
\at 6	Department of Astronomy and Center for Space Physics, Boston University, Boston, MA
\at 7	Mechanical and Aerospace Engineering Department, Henry Samueli School of Engineering, University of California, Los Angeles, CA 90095
\at 8  Now at: School of Mechanical, Industrial, and Manufacturing Engineering, Oregon State University, Corvallis, OR  97331
\at 9	University of St. Petersburg, St. Petersburg, Russia
\at 10	Jet Propulsion Laboratory, California Institute of Technology, Pasadena, CA 91109
\at 11	Space Science Institute, Boulder, CO 80301
\at 12	Sodankylä Geophysical Observatory, University of Oulu, Sodankylä, Finland
\at 13	Materials Science and Engineering Department, Henry Samueli School of Engineering, University of California, Los Angeles, CA 90095
\at 14 Now at: Deloitte Consulting, New York, NY 10112
\at 15 Computer Science Department, Henry Samueli School of Engineering, University of California, Los Angeles, CA 90095
\at 16 Now at: Microsoft, Redmond, WA 98052
\at 17	Physics and Astronomy Department, University of California, Los Angeles, CA 90095
\at 18	Now at: Department of Astronomy and Astrophysics, The University of Chicago, Chicago, IL 60637
\at 19	Now at: Raybeam, Inc., Mountain View, CA 94041
\at 20 Now at: SpaceX, Hawthorne, CA 90250
\at 21	Now at: Reliable Robotics Corporation, Mountain View, CA 94043
\at 22	Now at: Los Alamos National Laboratory, Los Alamos, NM 87545
\at 23 	Mathematics Department, University of California, Los Angeles, CA 90095
\at 24 Now at: Planet Labs, PBC, San Francisco, CA 94107
\at 25	Now at: KSAT, Inc., Denver, CO 80231
\at 26	Now at: Tyvak Nano-Satellite Systems, Inc., Irvine, CA 92618
\at 27	Now at: Northrop Grumman Aerospace Systems, Redondo Beach, CA 90278
\at 28	Now at: Apple, Cupertino, CA 95014
\at 29 Now at: Zipline International, South San Francisco, CA, 94080
\at 30	Now at: Lucid Motors, Newark, CA 94560
\at 31	Now at: College of Engineering and Computer Science, California State University, Fullerton, Fullerton, CA 92831
\at 32	Now at: The Aerospace Corporation, El Segundo, CA 90245
\at 33	Electrical and Computer Engineering Department, Henry Samueli School of Engineering, University of California, Los Angeles, CA 90095
\at 34 Now at: Heliogen, Pasadena, CA 91103
\at 35 Now at: Argo AI, LLC Pittsburgh, PA 15222
\at 36	Now at: Terran Orbital, Irvine, CA 92618
\at 37	Now at: Millenium Space Systems, El Segundo, CA 90245
\at 38 Now at: Department of Electrical Engineering, Stanford University, Stanford, CA 94305
\at 39	Now at: Mercedes-Benz Research and Development North America, Long Beach, CA 90810
\at 40	Now at: Geosyntec Consultants, Inc., Costa Mesa, CA 92626
\at 41	Now at: Juniper Networks Sunnyvale, California, 94089
\at 42 Now at: Niantic Inc., San Francisco, CA 94111
\at 43 Now at: Teledyne Scientific and Imaging, Thousand Oaks, CA 91360
\at 44 Now at: Naval Surface Warfare Center Corona Division, Norco, CA 92860
\at 45 Now at: Epirus Inc., Torrance, CA 90501
\at 46	Now at: Department of Astronomy, Ohio State University, Columbus, OH 43210
\at 47 Now at: Amazon, Seattle, WA 98109
\at 48 Now at: Department of Radiology, University of California, San Francisco, San Francisco, CA 94143
}

\date{Received: date / Accepted: date}

\maketitle

\begin{abstract}
We review comprehensive observations of electromagnetic ion cyclotron (EMIC) wave-driven energetic electron precipitation using data from the energetic electron detector on the Electron Losses and Fields InvestigatioN (ELFIN) mission, two polar-orbiting low-altitude spinning CubeSats, measuring 50-5000 keV electrons with good pitch-angle and energy resolution. EMIC wave-driven precipitation exhibits a distinct signature in energy-spectrograms of the precipitating-to-trapped flux ratio: peaks at $>$0.5 MeV which are abrupt (bursty) (lasting $\sim$17s, or $\Delta L\sim0.56$) with significant substructure (occasionally down to sub-second timescale). We attribute the bursty nature of the precipitation to the spatial extent and structuredness of the wave field at the equator. Multiple ELFIN passes over the same MLT sector allow us to study the spatial and temporal evolution of the EMIC wave - electron interaction region. Case studies employing conjugate ground-based or equatorial observations of the EMIC waves reveal that the energy of moderate and strong precipitation at ELFIN approximately agrees with theoretical expectations for cyclotron resonant interactions in a cold plasma. Using 2 years of ELFIN data uniformly distributed in local time, we assemble a statistical database of $\sim$50 events of strong EMIC wave-driven precipitation. Most reside at $L\sim 5-7$ at dusk, while a smaller subset exists at $L\sim 8-12$ at post-midnight. The energies of the peak-precipitation ratio and of the half-peak precipitation ratio (our proxy for the minimum resonance energy) exhibit an $L$-shell dependence in good agreement with theoretical estimates based on prior statistical observations of EMIC wave power spectra. The precipitation ratio’s spectral shape for the most intense events has an exponential falloff away from the peak (i.e., on either side of $\sim1.45$ MeV). It too agrees well with quasi-linear diffusion theory based on prior statistics of wave spectra. It should be noted though that this diffusive treatment likely includes effects from nonlinear resonant interactions (especially at high energies) and nonresonant effects from sharp wave packet edges (at low energies). Sub-MeV electron precipitation observed concurrently with strong EMIC wave-driven $>$1MeV precipitation has a spectral shape that is consistent with efficient pitch-angle scattering down to $\sim$ 200-300 keV by much less intense higher frequency EMIC waves at dusk (where such waves are most frequent). At $\sim$100 keV, whistler-mode chorus may be implicated in concurrent precipitation. These results confirm the critical role of EMIC waves in driving relativistic electron losses. Nonlinear effects may abound and require further investigation.
\subclass{MSC code1 \and MSC code2 \and more}
\end{abstract}

\section{Introduction}

\subsection{Earth's Radiation Belts}

Relativistic electron fluxes in Earth's radiation belts pose a significant hazard to satellites \citep{Horne13} and astronauts, especially during magnetic storms \citep{Gonzalez94,Baker18}. These fluxes wax and wane in response to upstream solar wind variations \citep{Baker87,Reeves98,Reeves03}, reflecting a delicate competition between acceleration, transport, and loss within the magnetosphere in ways that still defy accurate forecasting \citep{Li&Hudson19}. Additionally, energetic electrons precipitate to the high-latitude mesosphere and lower thermosphere, where they can create $NO_x$ and $HO_x$, ozone-destroying catalysts \citep{Jackman80,Thorne80,Randall05}. Even though $NO_x$ has a short lifetime when in sunlight, during polar winter darkness it can last for days to weeks and can be brought by vertical winds to $30-50$ km altitude, near the stratospheric peak of the ozone layer. Under such conditions it can catalytically convert ozone and contribute significantly to ozone destruction. In particular, relativistic electrons, those at energies of hundreds of keV to several MeV, pose the most significant threat to space assets. These electrons often have both sufficient fluxes and energy to penetrate through spacecraft and space suits, causing deep dielectric charging or high levels of radiation exposure. Additionally, when $>1$MeV electrons are scattered by magnetospheric processes and precipitate to the upper stratosphere, they can produce $NO_x$ directly in the regions of dominant ozone concentration where they can be most destructive \citep{Baker87}.

The transient nature of the high energy electron flux and precipitation complicates forecast efforts. The trapped flux is greatly affected by local acceleration (as opposed to transport) via electron resonant interaction with intense electromagnetic whistler-mode waves \citep{Chen07:NatPh,Thorne13:nature,Li14:storm,Allison&Shprits20}; by heating during injections or during radial diffusion via intense ultra-low-frequency (ULF) waves \citep{Elkington04,Hudson12:simulation, Mann16, Sorathia17}; by magnetopause shadowing at the dayside leading to rapid electron losses \citep{Shprits06,Turner12,Mann16,Olifer18,Sorathia18,Pinto20}; by field-line scattering at the nightside \citep{Sergeev&Tsyganenko82, Artemyev13:angeo:scattering}; and by wave-driven precipitation into the upper atmosphere \citep{Thorne05,Blum15:precipitation,Shprits17}. During active times, relativistic electrons in the radiation belts can be sourced adiabatically by inward diffusion from outer $L$-shells where the electron phase space density is high, due to enhanced electric fields \citep{Dessler&Karplus61,Kim&Chan97} or ULF waves \citep{Elkington03, Mann16}. They can also be sourced from even lower energy, $\sim 10$ keV, perpendicularly anisotropic magnetospheric electrons at larger distances. This can occur by rapid injections (transport and simultaneous acceleration) of the source electrons to the inner magnetosphere, leading to tens to hundreds of keV {\it seed} electrons \citep{Turner15, Gabrielse17}. Waves, in particular whistler-mode chorus, are also excited by these anisotropic, highly unstable source electrons in the magnetosphere. These waves can cause further rapid acceleration of the seed electrons to even higher, relativistic energies \citep{Jaynes15:seedelectrons}. As the energetic electrons are transported into closed drift shells they further interact with waves in the ULF, ELF and VLF range to cause both local acceleration at low L-shells \citep{Elkington99, Horne05Nature, Ma15,Li16:jgr,Thorne10:GRL,Thorne13:nature} and scattering into the loss cone \citep{Thorne06, Millan&Thorne07}. Drift-shell splitting and other dynamic effects such as solar wind compression pulses can cause further wave excitation and scattering of particles into the loss cone \citep{Jin22}.

Acceleration and loss of radiation belt electrons can occur simultaneously, over a wide range of temporal scales (from seconds to weeks) and spatial scales (across different local times, $L$-shells and latitudes). It is evident that precipitating electron fluxes also result from a dynamic competition between acceleration, transport, and loss processes. The two main waves responsible for such lossws are whistler-mode chorus and EMIC waves. The efficacy of such wave-driven losses must then be studied in the geophysical context (geomagnetic activity, plasma environment and location) within which these waves occur. Scattering of $>1$ MeV electrons by chorus waves in a plasma of realistic (on the order of $1-10 $cm$^{-3}$) density requires cyclotron resonance at high magnetic latitudes ($|\lambda|>30^\circ$; see \citep{Thorne05,Shprits06,Summers07:rates,Agapitov18:jgr}) where the average whistler-mode wave intensity is observed to be weak \citep{Meredith12,Agapitov13:jgr,Agapitov18:jgr,Wang&Shprits19:wavemodel}. Thus, local (i.e., unrelated to magnetopause shadowing) and rapid losses of $>2$ MeV electrons are typically attributed to their resonant scattering by EMIC waves \citep{Usanova14, Omura&Zhao12, Mourenas16:grl, Drozdov17}. This interaction is also believed to be a key process controlling the dynamics of the relativistic (as low as hundreds of keV) electron fluxes in Earth’s radiation belts \citep{Thorne&Kennel71}. EMIC waves can be very effective in electron scattering both in the quasi-linear diffusion regime \citep{Ni15,Drozdov17} and in the nonlinear resonant interaction regime \citep{Albert&Bortnik09,Kubota15,Grach&Demekhov20,Bortnik22,Grach22:elfin}. In fact, both event and statistical studies have shown that precipitating fluxes of relativistic electrons at low-altitudes correlate well with equatorial EMIC wave bursts \citep{Blum15:emic,Capannolo19,Zhang21}. However, the transient nature of EMIC wave emissions \citep{Blum16,Blum17} and the effects of hot ions on the EMIC wave dispersion relation \citep{Cao17,Chen19} complicate the evaluation of the relative contribution of EMIC waves to multi-MeV, MeV, and sub-MeV electron losses in the radiation belts.  This is a question that remains open for both observational and theoretical reasons, which we further detail below.

\subsection{Relative impact of EMIC waves on relativistic electron precipitation.}

Trapped fluxes of electrons of relativistic energy, associated with EMIC and chorus wave-driven precipitation, are monotonically decreasing with energy. EMIC waves are expected to be mainly responsible for multi-MeV electron scattering, but exhibit a minimum energy of precipitation potentially extending down to several hundred keV under some specific wave and plasma conditions \cite{Summers&Thorne03, Cao17, Zhang21}. All other wave modes, and in particular whistler-mode chorus waves, are more effective at pitch-angle scattering tens of keV particles, with progressively reduced efficacy at hundreds of keV (except under very special circumstances, such as nonlinear scattering and ducted wave propagation away from the equator \citep{Horne&Thorne03, Miyoshi20, Zhang22Micro}). Thus, the precipitating-to-trapped electron flux ratio, plotted as a function of energy, should be a good indicator of the scattering mechanism. In particular it would be ideal for separating the precipitation induced by EMIC waves from that by other types of waves. However, due to the scarcity of low-altitude satellite data with good pitch-angle and energy resolution in the relevant (sub-MeV to multi-MeV) energy range, this has been difficult in the past, and the relative contribution of the various wave modes to sub-MeV electron losses remains an open question.

This has not been for the lack of trying, as the contribution of energy and pitch-angle scattering by various waves to the overall flux levels and spectra is critically important for modeling and predicting space weather. Theoretical modelling using diffusion theory of EMIC waves, whistler-mode chorus and hiss does a decent job in predicting the high equatorial-pitch-angle flux decay as a function of activity indices, such as $Kp$ or $Dst$, over many days to weeks \citep{Glauert14, Ma15, Drozdov17, Mourenas17, Pinto19}. Similarly, an overall agreement has been found between diffusion theory predictions and observations, from the combination of low-altitude POES satellites and equatorial Van Allen Probes for the flux decay due to chorus waves at low equatorial-pitch-angles \citep{Li13:POES, Reidy21, Mourenas21:jgr:ELFIN}. However, the relative contribution of each mode to relativistic electron acceleration, precipitation, and short-term flux-evolution has been more difficult to pin down. This is particularly true for the relative contribution of EMIC waves to relativistic (hundreds of keV to MeV range) electron scattering. This relative contribution has been surmised recently from Versatile Electron Radiation Belt modeling comparisons with Van Allen Probes data. It was shown that EMIC waves are critical for the flux evolution of $>1$ MeV electrons, but in order to explain these electrons' equatorial-pitch-angle spectra over a broad range of L-shells, a combination of EMIC waves and whistler-mode chorus and hiss waves was needed \citep{Drozdov15, Drozdov17, Drozdov20}. Indeed it has been theoretically and observationally shown that the efficacy of MeV electron precipitation by EMIC waves is enhanced in the presence of whistlers even when the two wave types are operating at different local times, because EMIC waves alone cannot cause precipitation of the most abundant, high pitch-angle electrons \citep{Li07, Shprits09B, Mourenas16:grl, Zhang17}.

Another example of an important unanswered question in the same area is the origin of microburst precipitation. This is a particularly intense, short-lived, electron precipitation phenomenon, lasting on the order of $1$s or less. It is thought to contribute significantly to the overall energetic electron losses at $100$s of keV to several MeV \citep{Blake96, OBrien04, Blum15:precipitation, Greeley19, Hendry19}. While the sub-MeV energy range of this precipitation could be consistent with whistler-mode resonant interactions with electrons, especially since the broad spatial extent of microbursts at dawn overlaps with the typical location of whistler-mode chorus \citep{Douma17, Shumko18, Zhang22Micro}, theoretical studies and observations suggest that intense EMIC waves can also drive relativistic electron microbursts, especially longer duration ones, at dusk \citep{Blum15:emic, ZhangJ16, Kubota&Omura17}. As the two wave modes (chorus and EMIC waves) are both able to scatter electrons in the hundreds of keV to $\sim1$ MeV range, overlap in their spatial distribution, and both can exhibit a bursty nature, it remains unclear which wave mode dominates relativistic microburst precipitation.

Thus, the relative contribution of EMIC waves to scattering loss of sub-MeV to few MeV electrons compared to the contribution by other waves is still an open question. This question is important for accurately modeling short-term variations of both radiation belt fluxes and the atmospheric response to relativistic electron precipitation. Part of the difficulty in addressing this question can be attributed to the previous lack of energy-resolved and pitch-angle resolved spectra of precipitating electrons in the 10s of keV to a few MeV range that would enable a quantitative validation of theoretical models of diffusion. This situation has changed with the recent launch of the ELFIN mission, which provides for the first time such data over a wide range of $L$-shells and local times. ELFIN's multi-year dataset allows us now to accurately compare precipitating-to-trapped electron flux ratios and, thus, to infer electron diffusion rates as a function of energy \citep{Kennel&Petschek66, Li13:POES}. This is especially useful for comparisons of theoretical expectations of such diffusion rates and measurements of such rates using ELFIN observations, especially when they are combined with conjugate, equatorial measurements of waves and plasma parameters at THEMIS, Van Allen Probes, Arase and MMS. This represents a significant improvement over the otherwise massive and previously well-utilized POES dataset  \citep{Rodger10,Yando11,Yahnin16,Capannolo19}, which can only provide limited pitch-angle and integral (high-) energy spectra. It is also an improvement over the dataset from the 1972-076B mission \citep{Imhof77}, which had similar pitch-angle and energy resolution as ELFIN but was not accompanied by conjugate equatorial missions.

ELFIN was proposed as a focused investigation to address the question of whether EMIC waves, the primary candidate for pitch-angle scattering of relativistic electrons, can be definitively proven to be responsible for such scattering using the advantages offered by its new dataset. In this paper we aim to achieve that objective and exemplify the salient features that accompany such scattering. We will address this objective using ELFIN together with its numerous fortuitous conjunctions with equatorial spacecraft and ground observations. We first review, below, the properties of EMIC waves and their interaction with relativistic electrons. We next discuss how chorus waves may be also implicated in the scattering and precipitation of such electrons and how to differentiate the effects of these two wave types. Next, we present the first comprehensive ELFIN measurements of EMIC wave-driven electron precipitation. We discuss the observed features of the precipitating electron fluxes that indicate nonlinear resonant interaction of EMIC waves with electrons, compare precipitating electron energy spectra at high energy-resolution with theoretical expectations, and provide the first statistical distributions of EMIC wave-driven precipitation and its properties.

\section{EMIC waves: generation and effects on relativistic electrons -- present knowledge} \label{sec:background}

\subsection{Generation}

EMIC waves were first postulated to be excited by a low density, high energy population of hot ions which achieve cyclotron resonance with the ion cyclotron wave of a cold, dense ion background by appropriately Doppler-shifting the wave’s frequency in their own frame through streaming along the magnetic field \citep{Cornwall65, Cornwall70}. Such conditions prevail near the plasmapause where drift-shell splitting of ring current ions, or fresh ion injections, or magnetopause compressions of ambient, low density hot plasma may acquire perpendicular anisotropy. Portions of this (hot) ion distribution having a field-aligned streaming velocity that can thus attain cyclotron resonance with the wave can liberate the free energy available in their anisotropy to achieve wave growth \citep{Kennel&Petschek66, Cornwall70, Cornwall71}. This resonance condition is: $\omega-k_\parallel v_{i,\parallel} = n\Omega_{ci}$. Here, $n=+1$ is the relevant harmonic resonance number corresponding to first order resonance; $\Omega_{ci}$ and $v_{i}$ are the ion cyclotron angular frequency and ion velocity, respectively; $\omega$ and $k$ are the wave angular frequency and wave number, respectively; and the parallel symbol denotes components along the ambient magnetic field. Electromagnetic waves of the background (cold, presumed dominant) plasma population propagating opposite to the beam ($k_\parallel v_{i,\parallel}<0$) over a range of frequencies near, say, $0.5\Omega_{ci}$ and with wave vectors satisfying the cold plasma dispersion relation can thus become unstable. The dispersion relation of the cold component for parallel propagation (assuming that ions are protons and that the hot anisotropic component has a sufficiently low density to make a negligible contribution to the plasma dielectric response) is: $ck/\omega = (c/V_A)\left( 1-\omega/\Omega_{ci}\right)^{-1/2}= (c/V_A)\left(1-x\right)^{-1/2}$, or $ck/\Omega_{pi} = x/(1-x)^{1/2}$, with $x=\omega/\Omega_{ci}$, $V_A = B/\sqrt{\mu_0m_iN_i}$ the Alfv\'en speed, $\Omega_{pi}$ the plasma frequency and other symbols having their usual meaning (note that: $c/V_A  \equiv \Omega_{pi}/\Omega_{ci}$). This is a monotonic function of $\omega$, approaching the Alfv\'en wave dispersion relation in the low frequency, MHD limit ($\omega \to 0$). In the high frequency limit,  $\omega \to \Omega_{ci}$ as $k\to \infty$, which means that in this limit the waves are absorbed by the plasma and cannot propagate -- this is the ion cyclotron resonance. At intermediate frequencies, though, when the cyclotron growth provided by the hot component exceeds cyclotron damping by the cold component, the waves can grow. At oblique propagation, the dispersion relation near $\Omega_{ci}$ is only slightly modified, becoming: $ck_\parallel/\omega = (\xi c/V_A) \left(1-x\right)^{-1/2}$, where $\xi=\left[(1+cos^2\theta)/2\right]^{1/2}$.

Using this dispersion relation, the aforementioned cyclotron resonance condition can be recast as: $v_{i,R}/c = v_{i,\parallel}/c = (V_A/\xi c) (1-x)^{3/2}/x$. Maximum growth occurs for parallel propagation since at oblique propagation the resonant velocity decreases and the ion cyclotron damping by the cold component prevails quickly, due to that component's high density and low temperature. It is evident from the above resonance condition that EMIC wave generation depends critically on the ratio $\Omega_{pi}/\Omega_{ci} = c/V_A$ (or equivalently on $f_{pe}/f_{ce}$, the ratio of electron cyclotron and plasma frequencies that we use more commonly below) and on $1-x$, the latter denoting the proximity of the wave frequency to the ion cyclotron frequency. These parameters determine the resonance energy and its proximity to the free energy available in the velocity-distribution's anisotropy. Typical EMIC wave excitation requires that this resonance energy be low enough for the waves to resonate with anisotropic hot magnetospheric ions in the few to 10s of keV range. Hence, the larger the aforementioned frequency ratios are (the smaller $V_A$ is), the easier it is for EMIC waves to resonate with the free energy source of hot ions typically available. Because the Alfv\'en speed increases rapidly away from the equator along a field line, conditions at the geomagnetic equator favor such wave excitation. At high-density equatorial regions that are far enough from Earth so the geomagnetic field is also low, such as near the plasmapause, or within plasmaspheric plumes, $\Omega_{pi}/\Omega_{ci}$ can increase ($V_A$ can decrease) sufficiently for EMIC waves to be excited if anisotropic hot ions are also present.

And indeed, EMIC waves are often excited near the post-noon and pre-midnight sectors where the cold, dense plasmaspheric bulge and plume \citep{Horne&Thorne93} are intersected by the drift-paths of (hot) ring current ions exhibiting velocity space anisotropies. The cold, dense background plasma there is critical for lowering the resonant energy into the energy range where there exists a sufficient number flux of hot ions with high enough anisotropy. This situation occurs in that sector, especially during storm times, according to case studies \citep{Kozyra97, Jordanova98}. However, statistical studies have also revealed that banded, low-frequency electromagnetic waves exist at other local times as well \citep{Anderson92, Erlandson&Ukhorskiy01, Fraser10, Min12, Meredith14, Allen16, Paulson17}. These waves have amplitudes $0.1-10$nT, are typically left-hand polarized and field-aligned near the equator, and can extend from the Alfv\'en mode at low frequencies upwards to the local ion cyclotron frequency \citep{Kersten14}. They too can be identified as EMIC waves. Further supporting this identification is that in the presence of a multi-component plasma, typically $H^+$ with a few percent of either $He^+$, $O^+$, or both, such waves are observed to split into the classical EMIC wave distinct bands (a $H^+$, $He^+$ and $O^+$ band), each between their respective gyrofrequency and the one below it, except that the $O^+$ band extends continuously below the $O^+$ gyrofrequency down to the Alfv\'en branch (e.g., \citep{Cornwall&Schulz71,Young81,Horne&Thorne93}).

While the highest amplitude waves are most frequently observed at the duskside equator in the $He^+$ band with left-hand circular polarization and nearly field-aligned propagation, lower amplitude waves are also routinely observed at the dawnside equator except in the $H^+$ band with linear or elliptical polarization and occasional oblique propagation \citep{Min12}. They are also seen further away from the equator, where they become oblique, likely due to their propagation, and are eventually (at high enough latitudes) Landau damped. Hence off-equatorial waves are seen with lower amplitudes and occurrence rates. However, such waves can occasionally also be ducted. Then they can propagate nearly-field-aligned and evade damping, thus reaching the ionosphere and the ground \citep{Kim10} where they are detected \citep{Engebretson08,Engebretson15,Engebretson18} as {\it continuous} magnetic pulsations of {\it Type 1} (Pc1, 0.2-5Hz) or {\it Type 2} (Pc2, 5-10Hz). Other means of evading damping are mode conversion to the R-mode, and tunneling near the bi-ion frequency, just below the respective ion frequency \citep{Thorne06}. Substorm-related, freshly injected, anisotropic ring current ions drifting duskward from midnight and interacting with the plasmaspheric bulge or plume are most often responsible for exciting the EMIC waves seen at the duskside \citep{Cornwall&Schulz71,Chen10:emic,Morley10}. However, solar wind compressions can also cause hot ions drifting in the inner magnetosphere with pre-existing moderate (marginally stable) anisotropy to attain (through betatron acceleration) an enhanced anisotropy, one that exceeds the threshold for EMIC wave growth. This excitation mechanism is often credited for EMIC wave observations at the dayside, at pre- and post-noon \citep{Anderson&Hamilton93,Arnoldy05,Usanova08,Usanova10}. However, prolonged, quiet time EMIC wave activity over a broad range of local times in the dayside (pre- and post-noon), but over a narrow L-shell range, is attributed to the large anisotropy of freshly supplied ions from the nightside plasmasheet by injections (that can persist at low occurrence rates at large distances even during geomagnetically quiet conditions). Such anisotropy develops at the dayside due to differential drifts at different energies and can excite EMIC waves near an expanded plasmasphere \citep{Anderson96,Engebretson02}.

\subsection{Interaction with electrons}\label{SecINTER}

Relativistic electron pitch-angle scattering due to their resonant interaction with $H^+$ EMIC waves was first considered by \citet{Thorne&Kennel71} and \citet{Lyons&Thorne72}. \citet{Horne&Thorne98} calculated the minimum resonance energies for a multi-ion plasma ($H^+$, $He^+$, $O^+$) inside and outside the plasmapause during storms. \citet{Summers98} addressed relativistic effects, showing that even with such corrections, electrons in gyroresonance with EMIC waves undergo nearly pure pitch-angle (but not much energy) diffusion. \citet{Summers&Thorne03} demonstrated that such interactions can rarely result in scattering of electrons at or below 1 MeV. Such conditions arise only for $\Omega_{pe} /\Omega_{ce}\geq10$ (or equivalently $\Omega_{ci} /\Omega_{pi}  \equiv V_A/c \leq4.3$) which occurs near and just inside the dusk plasmapause, most often at storm times (where $\Omega_{ce}$ is the unsigned electron cyclotron angular frequency). To understand why, we discuss below the fundamental characteristics of this interaction.

Electrons can resonate with an EMIC wave by overtaking (moving in the same direction, but faster than) the wave if they have a sufficiently high (relativistic) speed to Doppler-shift the very low sub-ion-cyclotron wave frequency to the very high, electron cyclotron frequency. The left-hand circularly polarized (in time) EMIC wave electric field vector tip carves a right-handed helical wave-train in space. From the viewpoint (in the frame) of the guiding center of an electron able to overtake the wave's helix crests and valleys, the electric field vector tip rotates now (in time) in a right-handed way, opposite to that in the (ion or plasma) rest frame. This polarization reversal has the potential to put the electron, also gyrating in a right-handed sense, in cyclotron resonance with the EMIC wave. The generic electron cyclotron resonance condition is: $\omega - k_\parallel v_{e,\parallel} = n\Omega_{ce}/\gamma$. Here $v_{e}$ is the velocity of the electron, which in our case has a projection along the magnetic field that is in the same direction as the wave's projection along the field ($k_\parallel v_{e,\parallel} >0$), $\gamma$ is the relativistic correction factor (the Lorentz factor) and $n=-1$, for first order anomalous cyclotron resonance. (Anomalous, because due to the aforementioned overtaking, the sense of polarization experienced by the electron is opposite to that of the wave in the plasma rest frame.) Since $\omega\ll \Omega_{ce}$, the resonance condition becomes simply: $k_\parallel v_{e,\parallel} \sim \Omega_{ce}/\gamma$.

The electron resonance energy obtained from the aforementioned cold plasma dispersion relation of ion cyclotron waves and from the above electron cyclotron resonance condition \citep{Thorne&Kennel71}, simplified for a proton-electron plasma is $E_R/E_0=\gamma-1$, where $E_0$ is the electron rest mass, and (based on the solution of the above two equations) $\gamma$ is given by:
\[
\sqrt {\gamma ^2  - 1}  = \frac{1}{{\xi \cos \alpha }}\frac{{\Omega _{ce} }}{{\Omega _{pi} }}\frac{{\sqrt {1 - x} }}{x}
\]
The minimum resonance energy, $E_{R,\min}$, is obtained for zero pitch angle, $\alpha = 0$, for a given total energy. The most common situation of parallel propagation ($\xi=1$) serves as a case-in-point. Moreover, for a fixed wavenumber, $\theta=0$ also minimizes the resonance energy. We see that $E_{R,\min}$, corresponding to $\sqrt {\gamma ^2  - 1}  = \Omega_{ce}/\Omega_{pi} \sqrt{1-x}/x$, is a monotonic function of $x=\omega/\Omega_{ci}$ and $\Omega_{ce}/\Omega_{pi}$, so the closer $\omega$ gets to $\Omega_{ci}$ the lower the $E_{R,\min}$. This is seen more clearly if the resonant velocity in the resonance condition above can be simply recast as resonance energy: $E_R (k)=E_0\left(\sqrt{1+(\Omega_{ce}/ck)^2}-1\right)$. This is minimum for the maximum unstable wavenumber, which (based on the cold plasma dispersion relation, seen earlier) corresponds to the maximum $\omega/\Omega_{ci}$, closest to 1.

For fixed $\omega/\Omega_{ci}\sim 0.5$, $E_R (k)$ falls off with L-shell as a power law in the plasmasphere, due to the magnetic field decreasing faster than the square root of the density \cite{Sheeley01, Ozhogin12}. At the plasmapause, $E_{R,\min}$ increases abruptly with $L$-shell (outward) by about an order of magnitude, to $\sim 10$ MeV as the density drops by 1-2 orders of magnitude \citep{Cornwall65,Thorne&Kennel71}. Therefore, $E_{R,\min}$ has a local minimum (near $\sim 1$ MeV) just at the interior of the plasmapause. This situation remains true for EMIC resonances with heavy ions, when those are included in the dispersion relation \citep{Summers&Thorne03}.

However, incorporating thermal effects in the cold plasma dispersion relation complicates this picture \citep{Chen11:emic}. When even a fraction of the low-energy ions has a significant temperature ($10$s to $100$s of eV), as is often observed \citep{Lee14}, the dispersion relation is significantly modified: the waves can propagate through their respective cyclotron frequencies and the stop bands can vanish \citep{Silin11, Chen11:emic,Lee12}. While the dispersion relation becomes more complex, and the wave frequency is not limited to just below, or between the ion gyrofrequencies as the case may be, heavy cyclotron damping by the cold species near those frequencies severely limits wave propagation away from the source, even when the waves are, in principle, unstable due to an exceedingly strong anisotropy of the hot ions. These conditions cause excessive wave damping at large wavenumbers, those with $kc/\Omega_{pi}$ higher than $\approx 1$. Yet, the resonance condition, expressed above as the resonance energy as a function of wavenumber, $E_R(k)$, still applies regardless of the dispersion relation, and shows that there is still a lower limit to the minimum resonance energy, the one for the maximum wavenumber permitted for propagation, even with warm plasma effects accounted for. This realization simplifies the analysis: approximating the maximum wavenumber that can be attained under the presence of thermal effects as $kc/\Omega_{pi}\approx 1$, we obtain a similar, monotonic dependence of the resonance energy on $\Omega_{ce} /\Omega_{pi}$ as for the cold plasma approximation: $E_{R,\min}\sim E_0\left(\sqrt{1+(\Omega_{ce}/\Omega_{pi})^2}-1\right)$. We will compare this relationship with data, later in the paper.

A parametric analysis of the instability for multi-species plasmas including warm plasma effects confirms that the maximum unstable wavenumber rarely results in $E_{R,\min}$ below 1 MeV: this only occurs for conditions of large $\Omega_{pe} /\Omega_{ce}$ (15 to 100), and a large hot species anisotropy $A\gtrsim 2$ \citep{Chen11:emic, Chen13:emic}. In those cases, electrons of energy as low as $\sim 500$ keV may be able to resonate with and be scattered by waves of sufficiently high frequency. (Note that diffusion rates still peak at energies higher than $E_{R,\min}$ corresponding to the frequency at peak wave power, that is lower than the maximum observed frequency of wave propagation that corresponds to $E_{R,\min}$.) In particular, $He^+$ EMIC waves which are most easily able to resonate with $\lesssim 1$MeV electrons in cold plasma theory are strongly suppressed by cyclotron absorption; warm plasma effects cause the $H^+$ band to resonate more readily with $0.5-1$ MeV electrons than the $He^+$ band \citep{Chen13:emic}. Thus, even though warm plasma effects allow EMIC wave spectra to reach closer to and even cross the cyclotron frequency, consistent with some observations, at least in the context of quasi-linear theory $E_{R, \min}$ still remains most often above 1 MeV except in rare cases of high density regions such as plumes at high $L$-shells \citep{Ross21} or at low $L$-shells for compressed plasmaspheric conditions during the storm main phase \citep{Cao17}. Observations of precipitating electrons from POES, albeit with instruments of limited energy and pitch-angle resolution \citep{Zhang21}, have shown that $>0.7$ MeV electron precipitation can indeed be observed at POES, preferentially when equatorial spacecraft in close conjunction with POES confirm the existence of plasma conditions favorable for $0.7-1$MeV electron scattering by $H^+$ waves. Note, though, that POES does not have differential energy channels to finely resolve the peak in precipitation as a function of energy, hence these results should be considered suggestive, not conclusive evidence for the operation of EMIC waves. Additionally, many counter-examples were also found (when theoretically expected precipitation from waves was not observed, or vice versa) suggesting that quasi-linear theory alone may not be able to fully explain these observations. Further supporting the latter suggestion is that on occasion, relativistic electron precipitation ($\sim 1$ MeV) events can occur on timescales of a few seconds or less \citep{Imhof92, Lorentzen00, Lorentzen01, OBrien04, Douma17}, whereas the usual timescales of quasi-linear diffusion are on the order of many minutes to hours \citep{Albert03,Li07,Ukhorskiy10, Ni15}. Such short-lived precipitation can often extend down to hundreds of keV. These counter-examples cast doubt on the ability of EMIC waves to fully explain the observations, at least when studied in the quasi-linear regime even when hot plasma effects are incorporated into the theory.

Nonlinear treatments of EMIC wave interaction with relativistic electrons have also resulted in some successes in interpreting observations of rapid sub-MeV electron precipitation. Early work initially showed that nonlinear interaction with moderate amplitude, fixed frequency waves in a dipole field typically leads to scattering towards large pitch angles, away from the loss cone \citep{Albert&Bortnik09}. However, the observed departures of the EMIC waveforms from a constant frequency and the presence of a magnetic field gradient near the equator can cause phase trapping of resonant electrons and result in very rapid pitch-angle scattering and precipitation of MeV and even sometimes sub-MeV energies \citep{Omura&Zhao12, Kubota15, Hendry17, Nakamura19:emic, Grach21:emic}.  This effect can be enhanced by diffusive scattering by large amplitude EMIC waves, which may transport electrons directly into the loss cone from intermediate ($\sim 30^\circ$) pitch-angles \citep{Grach22:elfin}. For realistic EMIC waveforms having sufficiently steep edge effects, or equivalently having a few wave periods in a single packet, even sub-MeV nonresonant electrons can be pitch-angle scattered, when their interaction occurs over a small number of gyroperiods \citep{Chen16:nonresonant,An22}. Additionally, bounce resonance of near-equatorially mirroring, hundreds of keV energy electrons with EMIC waves can also result in moderate pitch-angle scattering and contribute to precipitation at those energies \citep{Cao17:bounce, Blum19}. But since hundreds of keV electrons can also interact with chorus waves, which may occur simultaneously with EMIC waves, either at different local times \citep{Zhang17} or even at the same location when driven by ULF pulsations \citep{Zhang19:jgr:modulation,Bashir22:jgr}, an unambiguous determination of the distinct (let alone independent) EMIC wave contribution to the precipitation can be difficult.

\subsection{Identification in precipitation spectra}

Previous studies have presented suggestive evidence of telltale signatures of EMIC wave-driven relativistic electron precipitation. This was achieved either using in-situ magnetospheric observations of depletion of near-field-aligned flux (in velocity-space) concurrent with EMIC wave enhancements \citep{Usanova14,Zhang16:grl,Bingley19,Adair22}, or by identifying local minima in radial profiles of the phase-space density at $L$-shells consistent with simultaneous ground observations of EMIC waves \citep{Aseev17}, or through observations of simultaneous precipitation of $10$s of keV protons and MeV electrons \citep{Imhof86,Miyoshi08,Hendry17,Capannolo19}. However, chorus waves can also scatter and cause precipitation of electrons of hundreds of keV to MeV \citep{Artemyev16:ssr, Ma16:diffusion, Miyoshi20, Zhang22Micro}. The relative contribution of chorus and EMIC waves was not addressed in those studies (e.g., see discussions in \citep{Zhang21,Zhang17}).

Noting that typical chorus wave scattering is most effective at tens of keV rather than at hundreds of keV, a monotonically decreasing precipitating-to-trapped flux ratio as a function of energy would favor a chorus wave scattering interpretation over an EMIC wave scattering one. Conversely, that ratio increasing with energy, particularly when peaking at 1 MeV or greater, would favor the EMIC wave scattering interpretation, since EMIC waves are most effective scatterers at $>1$ MeV electron energies. However, electron spectra of sufficiently high resolution in energy and pitch-angle to make the above distinction were not available in prior studies, which were mostly based on POES data \citep{Evans&Greer04, Yahnin17,Capannolo18,Capannolo19,Zhang21}. Such high resolution spectra, obtained at a low-altitude (ionospheric) satellite, especially when combined with equatorial or ground-based measurements of the EMIC waves, are critical for determining if such waves are responsible for relativistic electron scattering, and for addressing the physical mechanism of the scattering process (quasi-linear, nonlinear, resonant or nonresonant, etc). Such measurements are needed not only to identify but also to quantify EMIC wave-driven precipitation and its role in radiation belt dynamics and magnetosphere-atmosphere coupling.

EMIC wave resonant interactions with electrons can be attributed to (and studied as) one of two processes: quasi-linear diffusion toward the loss cone \citep{Kennel&Petschek66,Lyons74} and fast nonlinear phase trapping transport toward the loss cone \citep{Albert&Bortnik09,Kubota15,Grach21:emic,Bortnik22,Grach22:elfin}. The relative importance and occurrence rate of these two regimes of wave-particle interaction for EMIC wave scattering has not been addressed yet, even though there is consensus from observations that EMIC waves \citep{Kersten14,Saikin15,Zhang16:grl} are often sufficiently intense to resonate with electrons nonlinearly \citep{Wang17:emic}. The strongest losses associated with quasi-linear diffusion, those in the strong diffusion limit, have (by definition) loss-cone fluxes comparable to trapped fluxes, those next to the loss-cone edge \citep{Kennel&Petschek66}. However, nonlinear electron interaction may exceed the strong diffusion limit and produce loss-cone fluxes higher than trapped fluxes \citep{Grach&Demekhov20}. Distinguishing these two precipitation regimes requires electron flux measurements at fine pitch-angle resolution near and within the loss cone, which is possible with energetic particle detectors of modest angular resolution observing from low altitudes.

The recently launched, ELFIN CubeSat twins, ELFIN A and ELFIN B, provide a new dataset of precipitating electrons that is very helpful for addressing the above questions related to the process and efficiency of EMIC wave resonant scattering of energetic electrons. Their energetic particle detector for electrons (EPDE) measures the full $180^\circ$ pitch-angle distribution of electron fluxes with approximately $22.5^\circ$ resolution, over the energy range $50-5000$ keV sampled at 16 logarithmically-spaced energy channels of width $\Delta E/E < 40$\%. Thus, they can resolve perpendicular (locally trapped), precipitating, and backscattered fluxes with good pitch-angle and energy resolution \citep{Angelopoulos20:elfin}. Due to ELFIN's altitude, 300-450 km, the locally-trapped (perpendicular) flux measured corresponds to particles that are most often in the drift loss cone, i.e., destined to be lost before they complete a full circle around the Earth due to the variation of the geomagnetic field magnitude with geographic longitude. Near the longitude of the south-Atlantic anomaly, in the northern hemisphere the perpendicular fluxes are still inside the bounce loss cone (they will precipitate in the south) but even in that case, intense fluxes generated locally in the same hemisphere above at the equator will still provide valuable information on EMIC wave scattering at a rate faster than a quarter-bounce period, and are therefore valuable to retain. However, at most longitudes the measured perpendicular fluxes still correspond to electrons outside the local maximum bounce loss cone, meaning that such electrons have had a chance to drift in longitude for some time. In this paper we simply refer to perpendicular fluxes as trapped, meaning at least quarter-bounce trapped, or locally trapped.

In Section \ref{sec:casestudies}, below, we present ELFIN examples of EMIC wave-driven electron precipitation. We show the salient features of that precipitation and its difference from whistler-mode precipitation, consistent with the prior discussion in the subsection above. In Section \ref{sec:conjunctions}, we also incorporate in our analysis ancillary observations from other assets, such as conjugate measurements from ground-based stations and equatorial spacecraft. These are providing a regional context for the observed ELFIN precipitation (equatorial density and magnetic field), independent confirmation of the trapped particle fluxes and information on the occurrence and properties of EMIC waves that may be responsible for the observed precipitation. Then, in Section \ref{sec:statstudies}, we take a statistical approach to the study of ELFIN's observed precipitation spectra attributed to EMIC waves. We show that their spectral properties, such as the peak precipitation energy and the slope of precipitating-to-trapped flux ratio as a function of energy, as well as the spatial distribution of the inferred EMIC wave power are all consistent with expectation from theory and equatorial observations of these waves. We also find evidence of nonlinear interactions that can be further explored with the new dataset at hand.

\section{ELFIN examples of EMIC wave-driven electron precipitation}\label{sec:casestudies}
Moving along their low-Earth ($\sim 450$ km altitude), $\sim 90$ min period orbits, the ELFIN CubeSats can, in principle, each record up to four {\it science zones} (covering the near-Earth plasma sheet, outer radiation belt, plasmasphere, and inner belt) during each orbit. However, power and telemetry constraints demand judicious selection of (typically) 4-12 such zones per day per spacecraft. Choice of science zones (planned and scheduled with a weekly cadence) depends on conjunction availability with other missions and ground stations, or uniformity of coverage in time, MLT and hemisphere. For several months after ELFIN's launch in September 2018, there were conjunctions with the near-equatorial, dual satellite, $\sim 12$ hour period Van Allen Probes mission \citep{Mauk13}. During the four years (2018-2022) of ELFIN operations, there have been multiple conjunction periods with the equatorial Time History of Events and Macroscale Interactions during Substorms (THEMIS) mission (three spacecraft, roughly on a string-of-pearls orbital configuration when in the inner magnetosphere) on a $\sim 1$ day orbital period, and $\sim 13R_E$ apogee \citep{Angelopoulos08:ssr}), as well as with the near-equatorial Exploration of energization and Radiation in Geospace (ERG) spacecraft (also known as ARASE; \citep{Miyoshi18:ERG}), and with the near-equatorial, $\sim 3$ day period, four closely-separated satellite Magnetospheric Multiscale (MMS) mission \citep{Burch16}). Additionally there have been very useful ELFIN conjunctions with ground-based magnetometer stations providing magnetic field measurements in the EMIC wave frequency range. Such stations often detect equatorial EMIC waves propagating down to the ionosphere and associated with relativistic electron precipitation \citep{Usanova14,Yahnin17}.

\subsection{EMIC wave-driven versus whistler-mode wave-driven electron precipitation}\label{subsec:event-1}

\begin{figure}
  \centering
  \includegraphics[trim={2.75cm 5cm 2.75cm 2cm}, clip, width=1\textwidth]{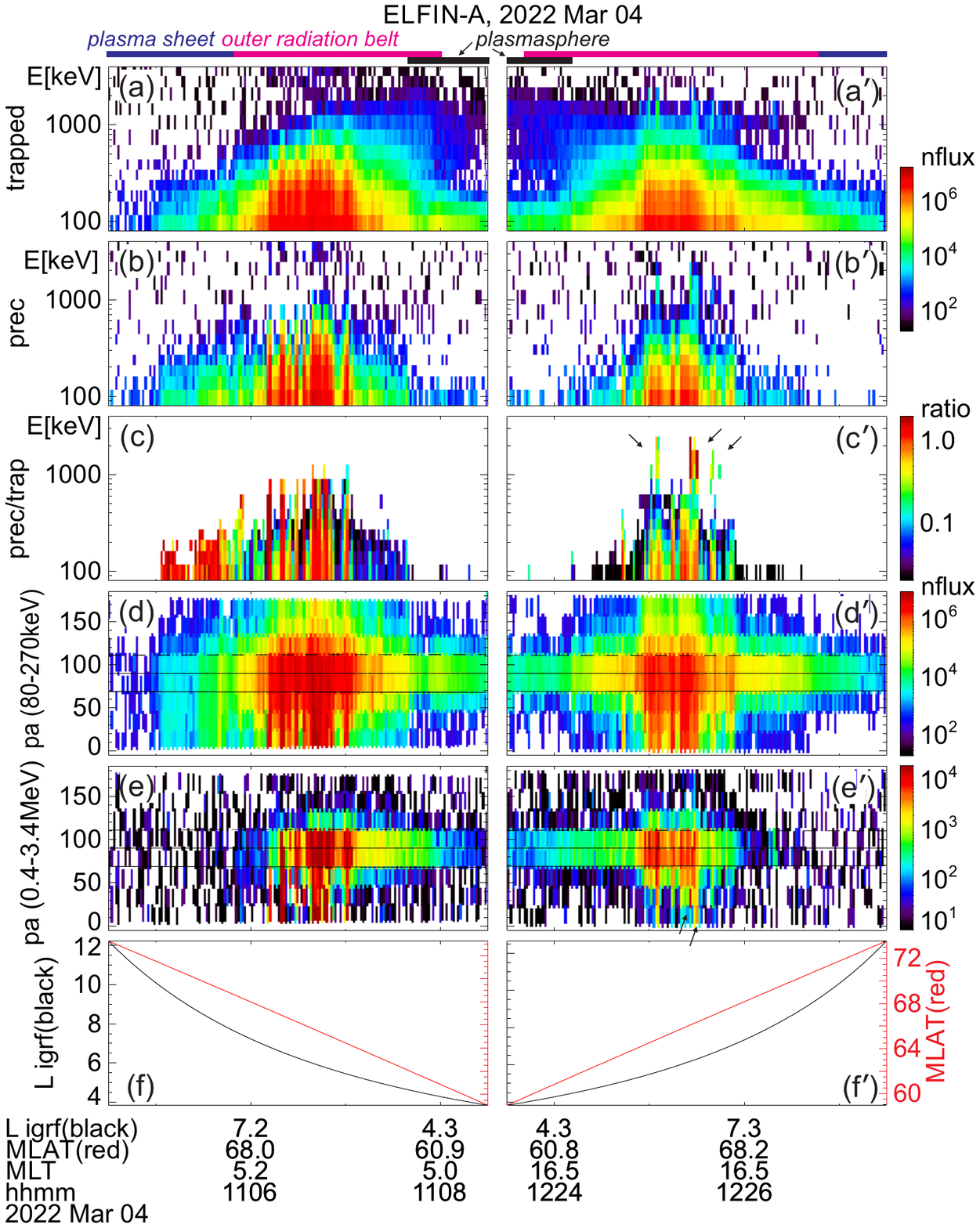}
\caption{Overview of two consecutive ELFIN A science zone crossings: one at the nightside/dawn sector (left) and the other at the dayside/dusk sector (right, primed panel letters). From top to bottom shown are 3 energy-spectrograms (a-c), 2 pitch-angle spectrograms (d-e) and the satellite's L-shell and magnetic latitude (f) computed using the international geophysical reference field (IGRF) model. All spectrograms show products derived from the number-flux of electrons (measured in individual sectors, in units of $1/{\rm cm}^2/{\rm s}/{\rm sr}/{\rm MeV}$) averaged over the selected pitch-angle and energy range as follows: The energy spectrograms in (a) and (a$^\prime$) are for locally trapped electrons (only pitch angles outside the loss cone and anti-loss cone, near perpendicular to the local field line direction, were included); those in (b) and (b$^\prime$) are for precipitating electrons (with pitches in the loss cone); and those in (c) and (c$^\prime$) are precipitating-to-trapped flux spectra ratios formed from the panels right above. The pitch-angle spectrograms in (d-e) and (d$^\prime$-e$^\prime$) are average fluxes in two broad energy ranges: a low energy range, 80-270 keV, and a high energy range, 0.4 - 3.4 MeV. The horizontal lines demarcate 90$\deg$ (vertically centered solid line), the loss cone (the other solid line) and the anti-loss cone (the dashed line). Horizontal color bars above Panel (a) represent magnetospheric regions identified based on the data and discussed in the main text. Arrows in Panels (c$^\prime$) and (e$^\prime$) represent spectral features also discussed in the main text.}
\label{fig:intro}
\end{figure}

Two ELFIN A science zone crossings in Figure \ref{fig:intro}, one at the nightside/dawn flank (left) and the other at dayside/dusk flank (right), depict precipitation patterns representative of the outer radiation belt and auroral zone, under moderately active conditions. (See comprehensive plots of these crossings also in the standard ELFIN overview plots at https://elfin.igpp.ucla.edu under Science $\rightarrow$ Summary Plots. Navigate to the appropriate time using the drop-down menus and then use mouse-clicks to zoom in and out in time on daily-plots and on science zone overview plots respectively.) To orient the reader, we first describe the magnetospheric regions ELFIN A traversed and discuss the predominant scattering mechanisms responsible for sub-MeV precipitation. We then explain how EMIC wave-driven ~MeV electron precipitation can be recognized within this otherwise fairly typical context.

Figure \ref{fig:intro} shows energy-time spectrograms of (locally) trapped fluxes (averaged over near-perpendicular pitch angles, those outside the local bounce loss cone, also referred to as ``trap'' in plots), precipitating fluxes (averaged over pitch-angles inside the local loss cone, and also referred to as ``prec'' in plots), as well as precipitating-to-trapped (or ``prec-to-trap'' in plots) flux ratios. ELFIN A travelled on a post-midnight/dawn meridian from high to low $L$-shells as depicted in the bottom panel of that figure and as demarcated in the annotations at the bottom. It was in the plasma sheet prior to $\sim$11:05:40 UT, and moved to outer radiation belt and plasmasphere soon thereafter.

The plasma sheet is identified by above-background trapped fluxes which do not exceed energies $\sim 200$ keV, concurrent with a precipitating-to-trapped flux ratio of around one. The latter ratio is expected for energetic electrons that have been isotropized by the small field-line curvature radius and the low equatorial field intensity of the plasma sheet \citep{Birmingham84,BZ89,Delcourt94:scattering,Lukin21:pop:df}.
At $\sim$11:05:55 UT (shortly after, but still near the transition region between the plasma sheet and inner magnetosphere), ELFIN A observed a distinct, dispersive feature of the precipitating-to-trapped flux ratio: the lowest energy approaching the ratio of $\sim1$ increased with proximity to Earth (as $L$-shell decreased). This energy versus $L$-shell dispersion is typical of the isotropy (magnetic latitude) boundary. For a given energy, this boundary is the ionospheric magnetic latitude poleward of which electrons of that energy become isotropic (due to field-line scattering) and equatorward of which the same energy electrons are anisotropic (due to the strong intensity and large radius of curvature field in the inner magnetosphere that prevents such scattering; \citep{Sergeev&Tsyganenko82, Dubyagin02,Sergeev12:IB}). At progressively higher energy this isotropy boundary typically resides at progressively lower latitude  (corresponding to a lower $L$-shell), because electrons of higher gyroradius can still be field-line scattered at the stronger and less curved equatorial magnetic field at this lower latitude. This isotropy boundary signature in the precipitating-to-trapped ratio can be used as an additional identifier of the transition from the plasma sheet to the outer radiation belt.

Subsequently, between 11:05:55 UT and 11:08:00 UT ELFIN A was in the outer radiation belt: both the trapped flux magnitude at all energies increased, and the maximum energy of the (statistically significant) trapped fluxes also increased as the $L$-shell decreased. There, a low-level precipitation (ratio $<0.1$) is intermittently punctuated by bursty, intense precipitation (ratio reaching $\sim1$). In this (post-midnight/dawn) science zone crossing, the precipitating-to-trapped flux ratio was highest at the lowest energies. This is expected for electron resonant interactions with whistler-mode (chorus) waves \citep{Kennel&Petschek66}. Such spectra are typical at ELFIN's outer belt crossings at post-midnight or dawn, especially at times of moderate to high geomagnetic activity.

At $\sim$11:07:40 UT, ELFIN A started to enter the plasmasphere, as evidenced by the decrease in the trapped electron fluxes and the simultaneous weakening of the precipitating fluxes. Shortly after that time, at $\sim$11:08:10 UT, the trapped fluxes at $\sim 200-300$ keV decreased below instrument background level. This a distinct feature of the plasmasphere, where 100s of keV electrons are efficiently scattered by plasmaspheric hiss (also whistler-mode) waves. This scattering, occurring at $L\sim 3.5-4.3$, forms a slot region between the outer and inner belts that is nearly devoid of energetic electrons \citep{Lyons&Thorne72, Lyons&Thorne73, Ma16:hiss, Ma17:jgr, Mourenas17}. We thus interpret the ELFIN observations as entry into the slot region at $\sim$11:08:10 UT. The precipitating-to-trapped flux ratio in the plasmasphere (which was seen inside of $L\sim4.5$ in this event) is almost nil because (i) the flux of $>200$ keV electrons near the loss-cone has decreased considerably, and (ii) hiss and lightning-generated whistler-mode wave power at frequencies that can attain cyclotron-resonance with $\sim 50-200$ keV electrons is weak in the 17-24 and 00-05 MLT sectors inside the plasmasphere \citep{Agapitov13:jgr, Agapitov18:jgr, Li15}, where the much higher plasma density compared to outside the plasmasphere further reduces the scattering efficiency of such whistler-mode waves \cite{Mourenas&Ripol12:JGR}.

Figure \ref{fig:intro}, right column, shows ELFIN A post-noon/dusk observations acquired about $1.5$ hours later. In this science zone crossing, ELFIN A moved from low to high $L$-shell (see Panel (f$^\prime$)) and traversed the regions discussed in the previous crossing but in reverse order. The plasmasphere was traversed first, and the plasmapause was encountered at $\sim$12:24:10 UT ($L\sim4.5$) as evidenced by the transition from low to high trapped fluxes and from low to high precipitating-to-trapped flux ratio of $\sim 200-300$ keV electrons. Between 12:23:40 UT and 12:26:40 UT, ELFIN A traversed the outer radiation belt, as evidenced by the significant fluxes of $>200$ keV electrons. In that period it observed whistler-mode (chorus) wave-driven electron scattering, as implied by the bursty nature of the precipitation and by the precipitating-to-trapped flux ratio being largest at the lowest energies. After 12:26:45 UT ELFIN A was likely magnetically conjugate to the equatorial post-noon/dawn plasma sheet (and outside the outer radiation belt) as suggested by the reduction in the flux of trapped electrons of energy $>200$ keV. But because in that local time sector the equatorial magnetic field is both stronger than and not as curved as that at the nightside plasma sheet at a similar L-shell, we do not expect plasma sheet field-line scattering to be significant, explaining the dearth of $<200$ keV precipitation at ELFIN A at that time.

While in the outer radiation belt but near the plasmapause, ELFIN A also observed a distinctly different type of precipitation: bursts of precipitation at relativistic ($E\geq 1$ MeV) energies. These bursts are demarcated by black arrows in the fifth panel (the one depicting the precipitating-to-trapped flux ratio). The most intense bursts occurred at $\sim$12:25:05UT and $\sim$12:25:28UT, but other less intense ones are also evident around the same time period (all occurred within a $\sim$1 min interval, or $\Delta$L$\approx$1.5). The MeV range precipitating-to-trapped flux ratio peaks are clearly separated in energy from the lower-energy peaks of chorus-driven precipitation discussed previously. The MeV range bursts are accompanied by increases in both trapped and precipitating fluxes from near- or below-background levels outside the bursts to above such levels within the bursts. We attribute this type of precipitation to EMIC wave scattering. Aside from the fact that EMIC waves resonate with such high energy electrons and are abundant in this region of space (at or near the post-noon and dusk plasmapause), the spectral properties observed by ELFIN A are also consistent with this interpretation: It is known that equatorial fluxes of relativistic electrons are typically very anisotropic and thus have a low flux level near the loss cone \citep{Gannon07,Chen14:REPAD,Shi16,Zhao18}. Scattering by EMIC waves transports electrons to smaller pitch-angles in velocity-space over a broad range of pitch-angles, increasing both trapped and precipitating fluxes near (and on either side of) the loss cone. Since low-altitude spacecraft, like ELFIN A, measure locally mirroring fluxes that still map to low equatorial pitch angles, they experience EMIC wave scattering as flux increases at both locally trapped (i.e., locally mirroring or perpendicular) and precipitating electrons. In Section \ref{sec:conjunctions}, we provide more examples of such relativistic electron precipitation, using conjunctions with ground-based and equatorial observatories measuring the relevant EMIC waves directly, further supporting our EMIC wave-driver interpretation of these precipitation signatures seen at ELFIN.

\subsection{The latitudinally localized and intense (bursty) nature of EMIC wave-driven electron precipitation}\label{subsec:event0}

\begin{figure}
  \centering
  \includegraphics[trim={2.75cm 5cm 2.75cm 2cm}, clip, width=1\textwidth]{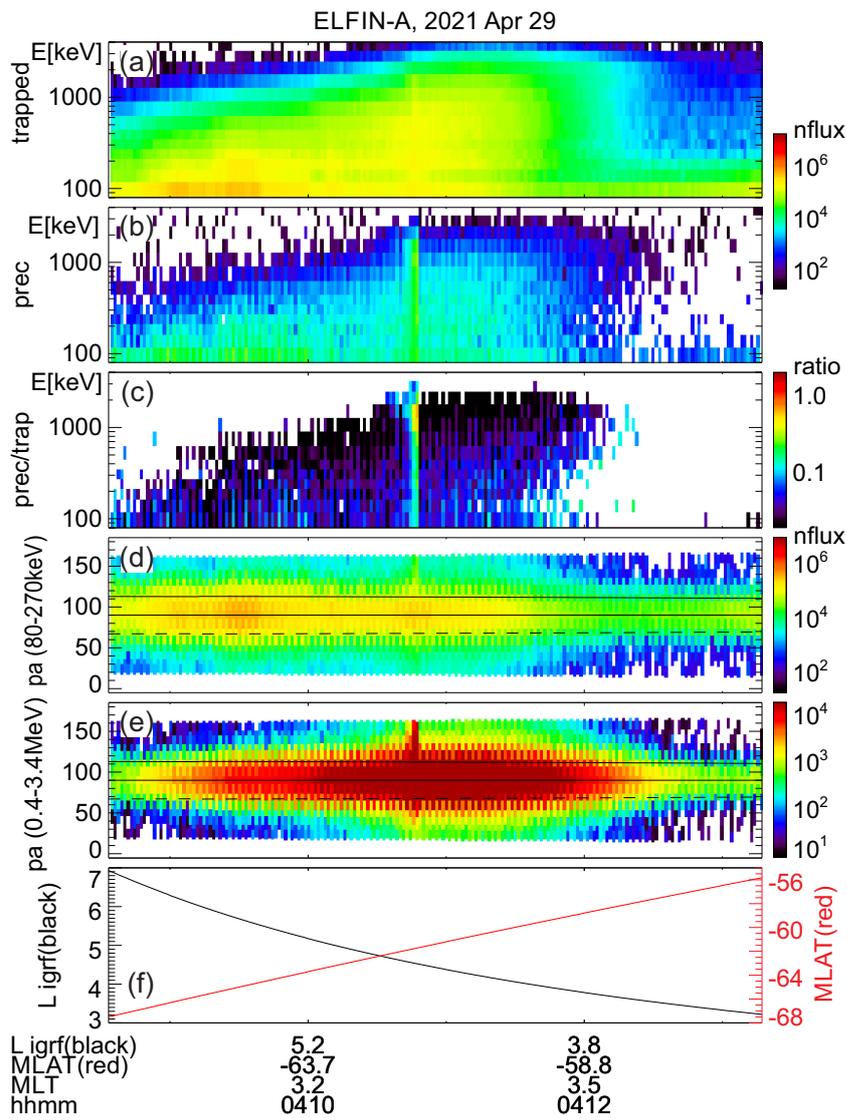}
\caption{Overview of ELFIN A observations for a science zone crossing, exhibiting a typical EMIC wave-driven precipitation signature. Format is identical to that of Figure \ref{fig:intro}}
\label{fig:event0.1}
\end{figure}

The event presented next is a prototypical observation of EMIC wave-driven electron precipitation at ELFIN with accompanying wave measurements on a conjugate platform. Figure \ref{fig:event0.1} shows ELFIN A measurements in the same format as Figure \ref{fig:intro}. The event occurred on 2021-04-29 in the outer radiation belt at dawn (where EMIC waves are also observed quite often, see \citep{Zhang16:grl}). The energy- and time-localized precipitation ratio (Panel (c)) at $\sim$13:21:05 UT marks the relativistic precipitation burst of interest. It lasted anywhere from 7-14s (2.5-5 spins) the range depending on the intensity used for its definition. That ratio peaked in energy at $\sim 1$ MeV, while both precipitating and trapped fluxes increased at the same time. These are all indicative of EMIC wave scattering \citep{Kersten14,Ni15,Mourenas16:grl}. In particular, whistler-mode chorus waves (otherwise also abundant in this region of space) preferentially scatter lower energy electrons (see energy distributions of EMIC and whistler-mode wave scattering rates in \citep{Glauert&Horne05,Summers07:rates,Shprits&Ni09}) and cannot be responsible for these observations. Note that Panels (d) and (e), which show pitch-angle spectrograms during the event, demonstrate that the precipitation was evident far from the edge of the loss cone, close to the downgoing direction (whereas upgoing electons are interpreted as reflected electrons from the atmosphere below). The pitch-angles that enter the precipitating flux energy spectrograms are, by selection, centered at $>$22.5 degrees from the edge of the loss cone, providing clean separation of precipitating and trapped fluxes.

About 35 min prior to these ELFIN A observations at $L\sim5$, the equatorial THEMIS E spacecraft traversed the inner magnetosphere approximately in the radial direction away from Earth at a similar MLT as ELFIN (Figure \ref{fig:event0.2}, Panel (a)) and detected intense wave activity at 0.5 - 1 Hz frequency. The observed waves were propagating near-parallel to the background magnetic field (Panel (b)), had left-hand circular polarization (Panel (c)) and were seen to peak just below the $He^+$ gyrofrequency (dashed lines, calculated using the local magnetic field), allowing us to identify them as $He^+$-band EMIC waves. The waves were seen in a locally measured magnetic field of $(-66, 48, 227) $nT in GSM coordinates, when THEMIS E was about 2000km away from the magnetic equator. They were observed near the plasmapause: a region of density gradient (Panel (d)) exhibiting more than an order of magnitude change throughout the event (from 500 to 15 per cm$^3$) and short-scale density variations of a factor of 2 on timescales as short as tens of seconds. The EMIC wave emission was seen from $L\sim 4.65$ to $L\sim 4.88$, i.e. close to the $L$ shell where ELFIN detected the $\sim 1$ MeV electron precipitation. The narrow $L$ shell range of EMIC observation at THEMIS E, $\Delta L \approx$0.23 (a 6 min duration x 4km/s THEMIS E's radial velocity) is traversed by ELFIN A at 7.9km/s in 9.6sec, or 3.4 spin periods. This is similar to the duration of ELFIN-A's observation of precipitation (7-14s, as discussed earlier), and explains the short lifetime (the bursty apparent nature) of the electron precipitation observed from ELFIN's ionospheric vantage-point. The EMIC wave emission was not only localized in $L$ shell but, in this particular case, was likely also relatively short-lived: THEMIS A and THEMIS D which traversed the same region of space as THEMIS E about 50 and 70 minutes prior, respectively, did not observe the emission. Neither did ELFIN A, which traversed the same region in its prior and subsequent orbits (at 02:40 and 05:40 UT) observe similar type of precipitation. This limits the likely duration of the EMIC wave-driven precipitation to $<$1.5hours. As we shall see in the following subsection, in other instances the precipitation can last a lot longer (many hours), confirming our assumption of the spatial interpretation of the EMIC wave power variability at THEMIS E.

It is instructive to examine several features of the EMIC waves in space that influence the precipitation signatures at ELFIN A. We continue with our assumption that at least the gross features of plasmaspheric density gradients and EMIC wave power last longer than a few minutes (their crossing time by THEMIS E) and are organized by the plasmapause in azimuthal sheets. We also note that the measured flow velocity and ExB velocity in the radial direction at THEMIS E are both ~4km/s consistent with the spacecraft velocity through the plasma. (Corotation, about 2.3km/s in the azimuthal direction, is ignored here as it is both smaller than the satellite velocity and because the structures of interest vary mostly in the radial direction.) Examining the time-series EMIC wave data in one component on the plane of polarization (Panel (e)) we see that the emissions consist of bursts of $<$30s duration ($<$23 wave periods) including transient bursts (such as the one marked by an arrow) as short as 7s (or 5 wave periods). The crossing time of such 7-30 s structures (if spatial at the equator) by ELFIN A in the ionosphere (scaled from the estimates in the previous paragraph) would be 0.19-0.8 s which is much smaller than ELFIN's nominal spin period (2.8 s), and around 1.-4.5 azimuthal spin sectors (each of 16 spin sectors lasts 0.175 s). Examination of the longer waveforms in further detail (Panel (f)) reveals phase skips every 5-15 s, or 4-11 wave periods (marked by the arrows and their separation times), mapping to 0.13-0.40 s, or 0.75-2.3 spin phase sectors at ELFIN. These EMIC wave structures (jumps in coherence and amplitude) from 5-30 s, corresponding to 20 - 120km at THEMIS, are smaller than the $He^+$ inertial length ($d_{He+} \approx$124-264 km in this case for $N_{e}\approx$ 50-100/cc and a $He^+$ number density $\sim$ $N_{e}$/16) and the thermal proton gyroradius (100 km for 60keV $H^+$) and much smaller than the expected (parallel) EMIC wavelength ($\lambda \approx 4\pi d_{He+}$). They may be due to oblique propagation and interference of waves from different source locations in the presence of plasma density inhomogeneity of a comparable gradient scale (Panel (d)). As bouncing and drifting energetic electrons near cyclotron resonance encounter these EMIC wave structures, both along and across the field, they should experience short-lived coherent interactions with an ensemble of waves of a range of wavelengths, amplitudes and phases that, for small amplitudes, would appear like turbulence. (Even if wave-field temporal variations occur and are partly responsible for the wave observations, above, the energetic electrons stream and drift so fast across them that they are effectively stationary in the Earth frame.) Their resultant interaction with electrons would then be describable by quasi-linear theory. Field-aligned electrons are organized on drift shells with equatorial cross-sections encircling Earth but centered towards dawn, since field lines are stretched out further from Earth at post-midnight. Conversely, waves excited due to ion cyclotron resonance near density gradients are (like the plasmapause) arranged on distorted drift shells with circular equatorial cross-sections having centers near Earth but displaced towards dusk due to the plasmasphere buldging toward dusk. The intersection of resonant electron drift-shells and plasmaspheric density enhancements filled with EMIC waves would be sheet-like structures: longer in azimuth and thin in radial distance. A good fraction of that layer of waves could be also interacting with resonant drifting electrons, resulting in precipitation over several spins. We would expect the EMIC wave-driven electron precipitation to be organized in thin azimuthal layers of thickness roughly consistent with the aforementioned variations in wave power at THEMIS E over a 6min in duration, i.e., over a $\Delta L \approx$0.23 as discussed above. The significant variations in power on spatial scales 7-30 s at THEMIS E could result in abrupt enhancements in near-field-aligned equatorial fluxes on that time scale. Mapping at ELFIN A to 1-4.5 spin sectors sectors, they can be either inside the loss cone or outside it (both are quite close to the equatorial loss cone), depending on ELFIN's spin phase and look direction. Hence the perpendicular-to-trapped flux ratio would be time-aliased at ELFIN, exhibiting large increases (even above unity) or decreases due to abrupt changes in equatorial wave power, even when that power is consistent with the quasi-linear regime.  Two arrows in Figure \ref{fig:intro}(e$^\prime$), of the event discussed in the previous subsection, show instances when the average flux in the loss-cone dominates the precitating-to-trapped flux ratio (the ratio exceeds one) due to flux enhancements in a single sector -- clearly aliased. In fact, because there are more spin-phase sectors in the loss cone than outside (typically 6 versus 4 in each full spin) a random distribution (in time) of 1-5 sector-long flux enhancements would result in more flux ratios exceeding one than below one, statistically. That bias can be normalized away, however, in statistical studies. Case studies can rely on the consistency between consecutive trapped fluxes, twice per spin, to ensure aliasing has been minimized (e.g., Figure 1 in \citet{Zhang22Micro}).

Figure \ref{fig:event0.3} shows energy-spectra of the number flux of precipitating and trapped electrons, averaged over the 6s (four ELFIN A half-spins) when the dominant precipitation attributed to EMIC waves occurred, in the event of Figure \ref{fig:event0.1} that was discussed earlier in this subsection. The measured spectrum of the precipitating flux (dotted thin red line) shows a peak near 1 MeV. This peak is even more pronounced when the average of the precipitation 6s before and 6s after the dominant precipitation interval (the trend, dashed thin red line) is removed, revealing the net contribution to the precipitation from just the EMIC waves (solid thick red line). (We interpret the trend as most likely due to low-level hiss waves.) The ratio of the average measured precipitating-to-trapped flux (not shown) also peaks near 1 MeV at a high value $\sim 0.4$, consistent with the color spectrogram in Figure \ref{fig:event0.1}(c), that depicts this ratio for individual half-spins peaking at $\sim 0.7$. After detrending the average of the precipitating flux in Figure \ref{fig:event0.3}, the detrended precipitating flux has an even more clear peak near 1 MeV (see the solid thick red line), corresponding also to a stronger peak near 1 MeV for the detrended precipitating to un-detrended trapped flux ratio appropriate for comparisons with quasi-linear theory \citep{Kennel&Petschek66}. The detrended precipitating and trapped fluxes have a very similar energy spectrum peaked near 1 MeV (compare solid thick red and blue curves), suggesting that EMIC waves may be responsible for both flux increases compared to the trend, possibly through nonlinear transport from higher pitch-angles \citep{Grach22:elfin}. As we will see later, in about 50\% of the time in our database of EMIC events the un-detrended flux ratio actually peaks and exceeds one above 1 MeV. However, given the significant spatial variability of the EMIC wave field and its electron interaction region in the magnetosphere, the ratio exceeding unity must be viewed with caution, because of the temporal aliasing effects arising from latitudinally narrow regions of precipitation lasting a few spin sectors, which prevents the precipitating and trapped fluxes to be measured simultaneously.

Nevertheless, it is evident from the above discussion that the precipitating flux during EMIC events  exhibits a strong peak and a high precipitating-to-trapped flux ratio in the $0.5 - 1.5$ MeV energy range, making this a hallmark of EMIC wave-driven precipitation in spectra that are well-resolved in energy and pitch-angle.

Assuming that there exist cases when time aliasing does not affect the precipitating-to-trapped flux ratio, that ratio exceeding one would signify the presence of nonlinear EMIC wave-relativistic electron interactions. This is because a ratio $>$1 cannot be explained by quasi-linear diffusion, which has an upper limit of precipitation, the {\it strong diffusion} limit (see \citep{Kennel&Petschek66}), that necessitates that precipitating and trapped fluxes be equal. However, nonlinear resonant interactions can indeed result in loss cone fluxes that are greater than permitted by quasi-linear theory. This could be due to the phase trapping effect \citep{Kubota15,Kubota&Omura17,Grach&Demekhov20,Grach21:emic}.
In this mechanism, very intense EMIC waves can interact resonantly with electrons initially located well above the loss-cone edge and transport them in phase space directly into the loss cone. Such nonlinear trapping results in a large pitch-angle change, $20^\circ-40^\circ$, during a single resonant interaction. Therefore, electrons with large equatorial pitch-angles exhibiting increasingly larger flux at fixed energy (due to the typically strong perpendicular anisotropy of relativistic electrons \citep{Ni15:flux}) can be transported all the way into the loss cone directly and without simultaneously enhancing the trapped flux near the edge of the equatorial loss-cone that corresponds to the only trapped electron population visible at ELFIN's altitude. This has been already demonstrated for a case study of EMIC wave-driven precipitation by \citet{Grach22:elfin} who used simultaneous equatorial observations of the EMIC waves and modeling of the wave-particle interactions to demonstrate the precipitation ratio should exceed unity, as was indeed observed on ELFIN. In that case the ratio exceeded unity for 3 consecutive spins (albeit at different energies) which bolsters the case for nonlinear scattering. Similar case-by-case studies of the details of the ELFIN particle distributions are needed, hand-in-hand with modeling, to verify the presence of nonlinear effects and separate them from temporal aliasing. Statistical studies of the problem can either rely on multiple consecutive spins with similar signatures, or probabilistic analysis of the cluster of strong precipitation events (ratio $>$1) after removal of trend and biases. Thus, the fine energy and pitch-angle resolution of ELFIN's  energetic electron measurements from low altitude can allow us to also identify and study the properties of nonlinear interactions of EMIC waves and relativistic electrons, which are likely important at times of intense EMIC waves.

\subsection{Evolution of long-lasting EMIC wave-driven electron precipitation}\label{subsec:event1}

As we saw in the previous section, relativistic electron precipitation driven by EMIC waves is often localized in $L$-shell due to the localization of the EMIC wave excitation and distribution in the magnetosphere \citep{Blum16,Blum17}. From ELFIN's low-altitude vantage point, such a spatial localization is evidenced as temporally localized, transient or bursty precipitation (as in Fig. \ref{fig:event0.1} and in events to be discussed subsequently in Sections \ref{subsec:event2}, \ref{subsec:event3}). However, EMIC waves can occasionally persist for hours, as revealed in prior studies from combinations of ground-based and multi-spacecraft observations \citep{Engebretson15,Blum20}. From its low-altitude, $\sim$ 90 min period orbit, ELFIN can traverse the same $MLT$ and $L$-shell region repeatedly, and thus it too can identify long-lasting EMIC waves, as well as monitor and study their gradual evolution in intensity and latitude using their clear electron precipitation signatures. Figure \ref{fig:long} shows such an event over three ELFIN B orbits (lasting $\sim 3$ hours).

At first (Figure \ref{fig:long}, Panels (a-c)), ELFIN B observed EMIC wave-driven electron precipitation at 02:42:50 UT, at $L$-shell$\sim 5.4$ (see Panel (j)), lasting $\sim 8.4$ seconds ($\Delta L \approx 0.1$). This is evident based on the precipitating-to-trapped flux ratio, showing that the most efficient precipitation lies in the range $[0.5, 2]$ MeV, where that ratio is $\sim$1 -- precipitating fluxes at these energies are comparable to trapped. An orbit later (Figure \ref{fig:long}, Panels (d-f)), ELFIN B passed over approximately the same MLT (at a distance of $\Delta MLT \approx 0.1$) and observed again strong precipitation (at 04:15:25--04:15:35 UT) at $L$-shell$\sim 5.2$, over a somewhat broader energy range, now $[0.4, 2]$ MeV, and lasting much longer, $\sim 22.4$ seconds ($\Delta L \approx 0.25$). The third time around (Figure \ref{fig:long}, Panels (g-i)), EMIC wave-driven precipitating fluxes were seen again at ELFIN B (at 05:48:20--05:48:35 UT) at $L$-shell$\sim 5.0$. They were comparable to trapped fluxes and extended over an even greater range in energy, $[0.3.,1.5]$ MeV, and L-shell ($\Delta L \approx 0.5$). There is a small $MLT$ evolution between the three orbits (from the first to the third crossing the MLT of the precipitation event changed by $\Delta MLT \approx 0.35$), but this is well within the expected EMIC wave azimuthal extent of many hours in MLT in the equatorial magnetosphere \citep{Blum16,Blum17}. The location of the center-time of the emissions moves closer to Earth in each encounter (from $L \approx$ 5.4, to 5.2 to 5.0.) Given their similarity, we conclude that these observations are likely due to continuous EMIC wave activity from the roughly same region in space, where either the EMIC wave free energy source (e.g. drifting ions) or the density gradient that enables resonant interactions with that source was evolving in time (moving closer to Earth and expanding in radial extent).
Therefore, using ELFIN's repeated passes over a long-lasting ($\sim 3$ hours) relativistic electron precipitation event also allows us to infer characteristics of the EMIC wave spatial location, intensity, extent, and temporal evolution at timescales of an orbit period or, on occasion, even faster (due to the occasional availability of data from two satellites or from two science zones at the same MLT, in the north and south hemispheres, on each orbit).

\section{Studies of EMIC wave-driven precipitation with ELFIN and its conjunctions with ancillary datasets}\label{sec:conjunctions}

\subsection{Confirming $He^+$-band EMIC wave resonance with electrons}\label{subsec:event2}

Figure \ref{fig:event1.1} shows an overview of ELFIN A observations on 2 November 2020. There is a clear peak of electron precipitation above 1 MeV at $\sim$15:19:00 UT. The precipitating-to-trapped flux ratio is $\sim$ 1 for energies $\sim 1-3$ MeV and decreases with decreasing energy at energies $<1$ MeV. Enhanced precipitation is observed during 6 consecutive ELFIN half-spins ($\sim 8.5$ seconds), the temporal localization again likely being due to ELFIN A’s crossing of flux-tubes mapping to a spatially localized region of EMIC waves near the equator at $L$-shell$\sim 5.5$. The typical scale of such wave emissions in the radial direction at the equator is about $0.5RE$ \citep{Blum16,Blum17}, consistent with the observed L-shell extent in this event of $\Delta L \approx 0.4$.

ELFIN A's trajectory projections to the north and south hemisphere, shown in Fig. \ref{fig:event1.2}, demonstrate the presence of several nearby ground-based magnetometer stations. These could potentially provide high-resolution magnetic field measurements that can reveal EMIC waves of interest during the time of interest. (The time of strong precipitation at ELFIN A is denoted by a thick trace superimposed on its projections in that figure.) Thus, although there is no equatorial spacecraft conjugate to ELFIN A at this time, it is still possible to obtain information on the presence and properties of the EMIC waves associated with the observed precipitation using such stations \citep{Usanova14}, if the EMIC waves managed to propagate to the ground. And indeed, stations PG3, PG4, and SPA measured low frequency, banded emissions consistent with EMIC waves at this time. We elect to work with data from SPA (Fig. \ref{fig:event1.3}) which was closest to ELFIN A at the time of interest. Using the T89 magnetic field model \cite{Tsyganenko89} we plot the equatorial helium gyrofrequency for the $L$-shells pertaining to these stations. The observed wave emission, exhibiting an upper limit $f_{\max}$ just below the $He^+$ gyrofrequency, $f_{cHe}$, evidently corresponds to helium band EMIC waves.

To further confirm the resonance condition for these waves, we use an empirical plasma density model \citep{Sheeley01} to estimate the equatorial plasma frequency $f_{pe}$; its ratio to the electron cyclotron $f_{ce}$ frequency (from the aforementioned use of T89) is about $8.5-9$. This is typical of the plasma trough, where ELFIN A was located during the subject precipitation (between the plasmasphere and the inner edge of the plasma sheet, marked atop the trapped electron energy spectrogram in Fig. \ref{fig:event1.2}(a) using the same criteria as explained earlier for the events of Fig. \ref{fig:intro}). For reasonable $He^+$ concentrations, $\approx 5-10\%$, the above $f_{pe}/f_{ce}$ ratio and the observed value of $f_{\max}/f_{cHe} \approx 0.97 - 0.99$ (Fig. \ref{fig:event1.3}(a)) can result in a theoretical estimate of the minimum resonance energy of EMIC waves. To demonstrate this, we plot the theoretical minimum resonance energy for a wide range of $f_{\max}/f_{cHe}$ (vertical axis) and $f_{pe}/f_{ce}$ (horizontal axis) in Fig. \ref{fig:event1.4}. We then denote the  range of the latter two parameters expected from observations as the gray areas between two horizontal and two vertical lines. The intersection of those two gray areas defines the region in this two-parameter space that corresponds to the expected range of parameters and solutions for the resonance energy as depicted by the plot's contours. This resonance energy estimate, between 1.5 and 2.5 MeV, is quite close to the moderate and strong electron precipitation energies observed at ELFIN A.

To make the last point point clear, we marked in yellow and red in Fig. \ref{fig:event1.4} the energies where moderate and strong precipitation was observed, respectively. This determination was based on the ratio $R$, depicted in Panel (a), being $0.5>R>0.3$ and $R>0.5$, for moderate (yellow) and strong (red) precipitation, respectively. Transferring the energy range of these two categories onto the contours of Panels (b) and (c) immediately depicts our observational assessment of moderate and strong precipitation within the resonance energy contours. We can thus see that the expected resonance energy based on the most likely values of the two aforementioned parameters, $f_{\max}/f_{cHe}$ and  $f_{pe}/f_{ce}$ (the intersection of their respective grayed area bounds), overlaps with the observed moderate and strong precipitation (yellow and red highlighted contours), as one would expect from quasi-linear theory.

\subsection{Confirming $H^+$-band EMIC wave resonance with electrons}\label{subsec:event3}

Figure \ref{fig:event2.1} shows an overview of ELFIN A observations on 6 Dec 2020 exhibiting a clear peak of electron precipitation at $E>1$ MeV around 20:20:40 UT (marked by a left-leaning arrow), a putative EMIC wave scattering event. At that time, the precipitating-to-trapped flux ratio peaked (at $\gtrsim1$) near $\sim1-2$ MeV, and decreased with decreasing energy at $<1$MeV. Strong relativistic electron precipitation at energies $\sim1$ MeV was also seen at four earlier times in the same ELFIN science zone crossing (denoted by down-pointing arrows). Each of the five instances of intense precipitation was transient, lasting for a few (2-8) spins ($\sim5-20$ seconds), had a peak ratio $\gtrsim1$ at $\sim0.5-2$ MeV and exhibited a decreasing ratio with decreasing energy below its peak.

As in the event we examined previously, ELFIN A's magnetic projections to the north and south hemispheres (Fig. \ref{fig:event2.2}) reveal that several ground-based stations were magnetically conjugate to ELFIN A, and might be able to provide magnetic field data that can explore the presence and properties of EMIC waves during this interval. And indeed, stations KEV, OUL, and SOD of the Finnish pulsation magnetometer network do show evidence for narrow-banded waves, likely EMIC waves, that started abruptly with a broadband burst, probably due to a nightside injection. Here we elect to show in Fig. \ref{fig:event2.3}) only data from SOD which was closest to ELFIN A during the 20:20:40 UT burst. Although the wave emission was not well defined spectrally on the ground at the time of ELFIN A's science zone crossing (demarcated by the two vertical magenta lines in that figure), it became clearly defined a few minutes after (at $\sim$20:25:00 UT): it had peak-power and maximum frequencies ($f_{peak}$ and $f_{max}$), that varied (rose) considerably over the next 2 hours, both within the range 0.2-0.7 Hz. T89 magnetic field model-based $O^+$, $He^+$ and $H^+$ gyrofrequencies at the equatorial location conjugate to SOD are over-plotted in the above figure. They reveal that the observed wave intensity peaked below the expected equatorial $H^+$ gyrofrequency ($\approx 0.7Hz$) and was suppressed as that frequency was approached from below. The lack of a local power minimum (a stop band) near the helium gyrofrequency and just above it leads us to surmise that there is likely negligible helium concentration in the wave source region \citep{Lee12,Chen19}, and that the observed emission is likely due to an equatorial source of $H^+$ EMIC waves. We take note of the delay in appearance of the EMIC wave on the ground relative to the broadband burst that initiated the wave activity and also relative to the overhead passage of ELFIN A. We interpret this delay as due to the fact that it takes some time for the duct that facilitates unimpeded EMIC wave propagation to the ground to establish itself along the entire flux tube. The broadband waves, in the Pi2 and Pi1 range, are (at least part of the way) compressional and Alfv\'enic and arrive to the ionosphere from the equatorial magnetosphere without the need for ducting. Thus, due to the need for duct formation for EMIC waves to propagate far from the equator, ground EMIC wave detection is delayed relative to the equatorial injection that likely initiated the hot anisotropic ions, the equatorial EMIC wave emission, the associated electron precipitation, the initial broadband waves on the ground, and the density perturbation that eventually led to the duct.

Equatorial observations of spacecraft potential from THEMIS-A, -D and -E \citep{Angelopoulos08:ssr}, which crossed the same L-shell as ELFIN A two hours later and a few MLT hours away (at $\sim$20 MLT), provide estimates of the equatorial plasma density (and plasma frequency). At the times when the locally measured magnetic field at THEMIS was $\sim 45-50$nT, consistent with the aforementioned $H^+$ gyrofrequency, the inferred equatorial density from the spacecraft potential was 2.5-3.0 cm$^{-3}$, such that the inferred equatorial $f_{pe}/f_{ce}$ ratio was about $10-12$. Together with the observed value of $f_{\max}/f_{cp} \sim0.6-0.9$ from the ground-observatories, this gives a theoretical estimate of the wave minimum resonance energy for electrons of $\approx$ 0.3 - 2 MeV as marked by the cross-section of the horizontal and vertical highlighted regions in Fig. \ref{fig:event2.4}. This estimate is quite close to the energies of moderate and strong electron precipitation observed by ELFIN, depicted in the same figure with red and yellow areas within the wider $f_{\max}/f_{cp}$ versus $f_{pe}/f_{ce}$ parameter space. Our ELFIN observations of peak precipitation combined with ground based estimates of the wave frequency and equatorial spacecraft estimates of plasma density are therefore consistent with a $H^+$ band EMIC wave-driven interpretation in the context of quasi-linear theory.

\subsection{EMIC wave-driven precipitation in the context of TEC maps}\label{subsec:event4}

As discussed above, the equatorial electron density is one of the most important parameters affecting EMIC wave generation and resonance with relativistic electrons. When that density is large, as in the case of plasmaspheric plumes \citep{Fraser05,Usanova13,Halford15}, the EMIC wave resonant excitation by hot anisotropic ions is favored \citep{Horne&Thorne93,Chen09:EMIC&plumes, Chen10:emic}, while the energy of electrons resonant with EMIC waves also decreases down to energies of peak flux in the outer radiation belt, 0.5-2MeV, leading to intense precipitation of such relativistic electrons \citep{Summers&Thorne03, Summers07:rates}. Thus, magnetospheric density enhancements, collocated with azimuthally drifting energetic ions, have been commonly reported as favorable sites for EMIC wave growth and relativistic electron precipitation. Such sites are expected around boundaries between the nominal ring current region (filled by hot ions injected from the plasma sheet \citep{Gkioulidou15,Runov15}) and the plasmasphere \citep{Thorne&Kennel71, Horne&Thorne93, Chen10:emic}. It is challenging, however, to definitively establish the connection between EMIC wave-driven electron precipitation and density enhancements solely based on fortuitous conjunctions between low-altitude and equatorial spacecraft. This is because density ducts (spatially limited enhancements) can be highly-localized and the equatorial spacecraft in those fortuitous conjunctions are not always in the optimal location to reveal the spatial profile of the pertinent density enhancement responsible for wave excitation. Mapping uncertainties due to imperfect magnetic field models further complicate such conjunction-based studies.

An alternate approach to exploring the density variations at play during ELFIN measurements of relativistic electron precipitation is to use total electron content (TEC) measurements of the ionosphere. These are obtained using the phase delay of radio signals transmitted from Global Navigation Satellite Systems (GNSS) satellites (moving along circular orbits at an altitude of $\sim 20,000$ km) to ground-based receivers. That phase delay allows estimates of the altitude-integrated electron density \citep{Davies65,Foster01:tec} along the line-of-sight. Data from multiple propagation rays during a finite time-step (from tens of seconds to minutes) at a wide range of propagation angles are assimilated through a tomographic reconstruction process to produce TEC maps over a wide area on the ground. TEC maps reveal well the spatial and temporal variations of plasmaspheric density  \citep{Heise02, Belehaki04,Lee13:tec}, including those at adjacent density enhancement structures, like plasmaspheric plumes \citep{Foster02:tec, Walsh14:tec&reconnection}. Moreover, TEC data also reveal quite well the nightside region of enhanced hot (ring-current energy) ions, which, is known to the ionospheric community as the mid-latitude ionospheric trough (MIT; see discussion on its formation in \citet{Aa20}). This topside ionosphere,  subauroral latitude region is recognized in satellite data by its enhanced ionospheric electron temperature and its prominent density reduction. The latter is also commonly captured in TEC data (see: \citep{Yizengaw&Moldwin05,Weygand21:MITPP}). Because ring current ions can provide free energy for EMIC wave generation in the pre-midnight sector, a significant fraction of EMIC wave-driven precipitation is observed in the nightside region \citep{Capannolo22,Carson13}, precisely where the MIT develops. Using TEC data to study the correlation between the TEC-derived density gradients with relativistic electron precipitation can be quite advantageous: The magnetic projection of the low-altitude ELFIN satellites downwards, onto ionospheric TEC maps is highly accurate compared to upwards, to the magnetic equator. Moreover, the large availability and wide spatial coverage (in MLT and $L$-shell) of TEC observations provides a large dataset with which a correlation between precipitation events and plasma density boundaries can be investigated. In the following, we present first results from such studies, while also exemplifying advances that can be made in the future using a similar approach.

Figure \ref{fig:event4.1} shows two relativistic electron precipitation events on successive orbits of ELFIN A ($\sim$ 90 min apart, at the same MLT). (Note that Panel (e), showing MLAT and L-shell, applies to both events.) The events show clear signatures of EMIC wave-driven scattering at around 11:30:00 UT and 13:02:30 UT, respectively. At those times, ELFIN observed strong, transient increases in trapped and precipitating electron fluxes at an L-shell $\sim$ 6.5. The precipitating-to-trapped flux ratio peaked at relativistic energies ($\sim 1$MeV). Both bursts were observed between the inner edge of the plasma sheet (the earthward boundary of isotropic plasma sheet fluxes in the energy range of $<200$ keV, at $\sim$11:29:23 UT and $\sim$13:02:05 UT in the two crossings, respectively) and the plasmasphere (where trapped fluxes of energy $\sim$200 keV fall below those at $>$500 keV energy due to effective scattering of  $<$500keV electrons by plasmasheric hiss \citep{Ma16:hiss, Mourenas17}, at and after 11:30:55 UT and 13:03:25 UT in the two crossings, respectively). The precipitation was likely spatially localized ($\Delta L \sim 0.4$ for both events). The similarity of the EMIC-driven precipitation signatures in two ELFIN orbits suggests here too that EMIC wave generation persists for at least 1.5 hours (as in the cases presented earlier, in Fig. \ref{fig:long}, and as was previously reported by \citet{Engebretson15, Blum20}).

Figure \ref{fig:event4.2} shows projections of ELFIN A orbits onto TEC maps in the northern hemisphere (provided by MIT Haystack via the Madrigal database \citep{Rideout&Coster06,Coster13,Vierinen15}). The spatial resolution of the TEC map is 1 by 1 degree. The images are geographic projections but the magnetic longitude is denoted by the two overlaid MLT meridians, for $MLT=0$ and $MLT=6$ respectively. The ELFIN A orbits map magnetically near the midnight meridian, slightly post-midnight (MLT $\approx$ 0-1). This is to the west of the pre-midnight boundary of the MIT, which can be identified as the blue regions of TEC depletion. The intense TEC values at the dayside are due to enhanced TEC due to the sunlight. Note that during the second ELFIN A orbit, at $\sim$13:02:00 UT, the western (pre-midnight) boundary of the MIT is not clearly visible due to statistical uncertainties from lack of high-latitude TEC stations in the Northern Pacific. This TEC quality degradation occurs as the Earth rotates counterclockwise, viewed from the north, by 22.5 degrees between the two successive orbits, and the American sector moves eastward. The ELFIN A orbit and the sunlit-enhanced TEC at the dayside, which are both nearly-fixed in magnetic local time coordinates, appear to rotate clockwise (westward) by the same amount in the geographic coordinate system of this figure.

The MIT region's western edge evolves as the season changes between the summer solstice and the autumnal equinox, moving from dawn to dusk, according to previous statistical studies \citep{Aa20}. As expected for the season of our event, close to autumnal equinox, the MIT is observed in Fig. \ref{fig:event4.2} preferentially at post-midnight to dawn ($MLT\in[0,6]$). It has also been previously established through statistical investigations of the MIT in conjunction with near-equatorial spacecraft measurements that at each longitude the minimum of TEC viewed as a function of latitude within the MIT corresponds to the equatorial plasmapause location \citep{Shinbori21, Heilig22}, whereas the pre-midnight MIT boundary (roughly along a magnetic meridian) is magnetically conjugate to the plasmaspheric plume \citep{Heilig22}. Thus, both ELFIN A orbits map magnetically to the pre-midnight MIT boundary. This boundary should be interpreted as the overlap between ring current ions (resulting in the MIT's formation in the first place, \citep{Aa20}) and the cold plasma density region at pre-midnight (associated with the plasmaspheric plume \citep{Heilig22}). Finally, the poleward TEC boundary of the MIT maps to the inner edge of the plasma sheet, as it is attributed to local, ionospheric electron enhancements brought-about by plasma sheet electron precipitation \citep{Aa20}.

During its 6-min-long science-zone crossing in each orbit, ELFIN A was moving  equatorward and was initially in a region of localized TEC enhancement associated with plasma sheet electron precipitation, as also evident by the electron spectrograms in Figure \ref{fig:event4.1}, Panels (a) and (a$^\prime$) and denoted by horizontal black bars above those panels. It traversed the inner edge of the plasma sheet at approximately 11:29:23 UT and 13:02:05 UT, respectively (the times when the precipitating-to-trapped flux ratio stopped being isotropic and trapped fluxes exceeding background levels were still below $<200keV$ in Fig. \ref{fig:event4.1}). Observations of EMIC wave-driven precipitation, which ensued and are demarcated along its orbit projections in Figure \ref{fig:event4.2} by thick magenta lines, appear to map into the {\it green or blue} TEC region, between the poleward and equatorward MIT boundaries and near the west MIT boundary. The TEC values along the satellite tracks on Figure \ref{fig:event4.2} are transferred as line plots into Figure \ref{fig:event4.1} (see Panels (d) and (d$^\prime$)) to facilitate comparison with the ELFIN A precipitation signatures in the panel stacks right above them and with region identifications depicted by the horizontal color bars at the top of each stack. At $\sim$11:31:20 UT and $\sim$13:03:55 UT, the times of ELFIN A's crossings of the plasmapause, as inferred from in-situ measurements in Figure \ref{fig:event4.1} and shown by the black color bars, ELFIN A was located on the poleward edge of, but near the MIT's TEC minima in Figure \ref{fig:event4.1}(d) and (d$^\prime$) and also evident in Figure \ref{fig:event4.2}. These minima are known to map to the equatorial plasmapause location \citep{Shinbori21, Heilig22}) and therefore are consistent with our approximate identification of the plasmapause from in-situ energetic electron measurements. Therefore, comparison of TEC and ELFIN A-measured electron precipitation demonstrates that regions of EMIC wave-driven electron precipitation indeed correlate well with the TEC minimum and its poleward gradient within the MIT region i.e., they are located just outside the plasmapause. Similar comparisons with TEC maps can provide two-dimensional ionospheric and plasmaspheric context information for ELFIN's locally trapped and precipitating flux measurements.

\section{Statistical results}\label{sec:statstudies}

To analyze the spatial and spectral properties of EMIC wave-driven precipitation statistically, we examined all ELFIN A\&B observations from 2019-2021. Based on the aforementioned telltale signatures of EMIC wave-driven precipitation, we applied an operational definition for such a precipitation event to be a peak in the precipitating-to-trapped flux ratio at $>0.5$ MeV. We thus identified $\sim 50$ relativistic electron precipitation events similar to those discussed in the previous sections. We specifically excluded from the event database those which were within the expected location, or were consistent with, the electron isotropy boundary. The latter is distinctly evident by its location at the inner edge of the plasma sheet and the characteristic energy dispersion exhibited by the precipitating-to-trapped flux ratio \citep{Imhof77,Imhof79,Yahnin97}. The putative EMIC wave-driven precipitation events thus selected correspond to a total of $\sim 310$ spins, with an average event duration of $\sim6$ spins or $\sim17$ s. They are most often located at L$\sim$6 and have typical $\Delta L \sim$ 0.56, consistent with prior reports of the radial extent of the EMIC waves at the equator \citep{Blum16,Blum17}. This is in accordance with our earlier assertion (in Section \ref{subsec:event0}) that the bursty nature of the subject precipitation at ELFIN is due to the radial localization of the EMIC waves and of their electron-interaction region in the magnetosphere.

\subsection{Spatial distribution of all EMIC wave-driven events} \label{subsec:spatialstats}

Figure \ref{fig:statistics}(a) shows that most events occurred at around $L\sim 5-7$ consistent with their expected location within the outer radiation belt. The events occurred predominantly at the midnight and dusk sectors, even though the ELFIN database of science zone crossings has uniform coverage in local time. Their probability distribution in MLT - L space, shown in Figure \ref{fig:statistics}(b), reveals that while events are indeed most probable at pre-midnight and dusk at $L\sim 5-7$, a second class of events is present at midnight, post-midnight and dawn at higher L-shells, $L\sim 8-12$. This bares close resemblance to the distribution of EMIC waves in the magnetosphere (see e.g., \citep{Min12}).


All events have a large precipitating-to-trapped flux ratio, near one (Panel (c)). In fact, for almost half of the events that ratio exceeds unity. Full-spin resolution data have been used (thus trapped fluxes represent an average of flux accumulations from two look directions, one before and one after the precipitation measurement). Only points with relative error of the ratio, $R$, $\Delta R/R<$ 50\% were included herein. We thus find statistically that the above ratio exceeds unity for a very considerable fraction of all events. This suggests that such nonlinear effects may be considerable in our database of EMIC wave-driven electron scattering process. However, because transient precipitation on a sub-spin resolution, lasting only a few spin sectors, may be common, careful consideration and removal of aliasing needs to take place in a statistical or event-by-event analysis of this database in a future study to definitively identify events that are due to nonlinear scattering.

\begin{figure}
  \centering
  \includegraphics[trim={0cm 0cm 0cm 0cm}, clip,width=0.9\textwidth]{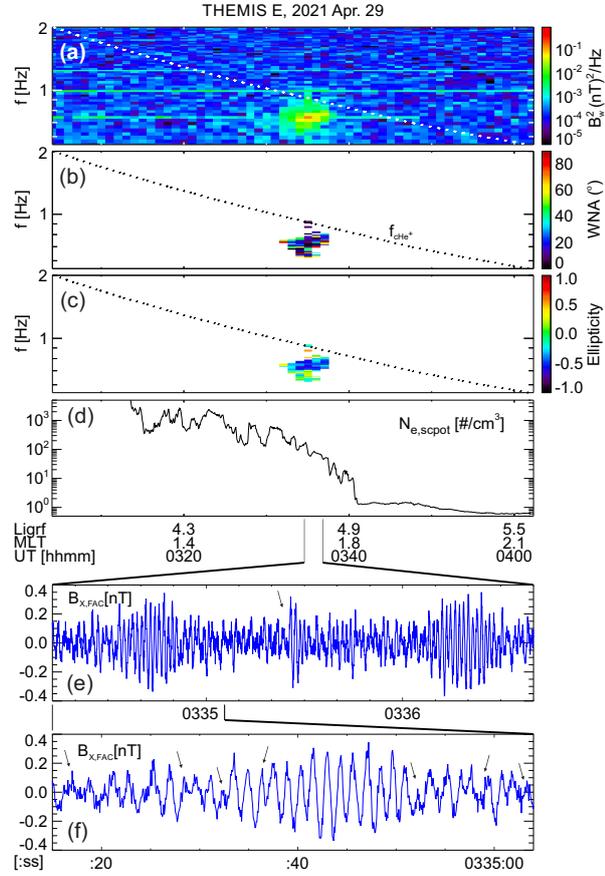}
\caption{Observations from THEMIS E at the equator at an MLT and UT near those of the science zone crossing by ELFIN A depicted in Figure \ref{fig:event0.1}. Panels (a-d) show about an hour of data centered around an EMIC wave emission: power spectral density of the magnetic field measured by the fluxgate magnetometer, FGM, instrument (a), wave normal angle (b), ellipticity (c), electron density inferred from the spacecraft potential computed on-board by the electric field instrument, EFI, and processed on the ground using the measured electron temperature by the electrostatic analyzer, ESA, instrument (d). The black dashed line is the $He^+$ gyrofrequency. Panel (e) shows $\sim$2.5 min of data from a single magnetic field component in a field-aligned-coordinate system (FAC), $B_{X,FAC}$. It is oriented perpendicular to the average magnetic field direction (hence, near the plane of polarization) and lies on a plane also containing the sunward direction. Panel (f) is an expanded, $\sim$1 min long, view of the same quantity as above it. Arrows in Panels (e) and (f) are discussed in the main text.}
\label{fig:event0.2}
\end{figure}

\begin{figure}
  \centering
  \includegraphics[trim={3cm 11.5cm 5cm 5cm}, clip,width=1\textwidth]{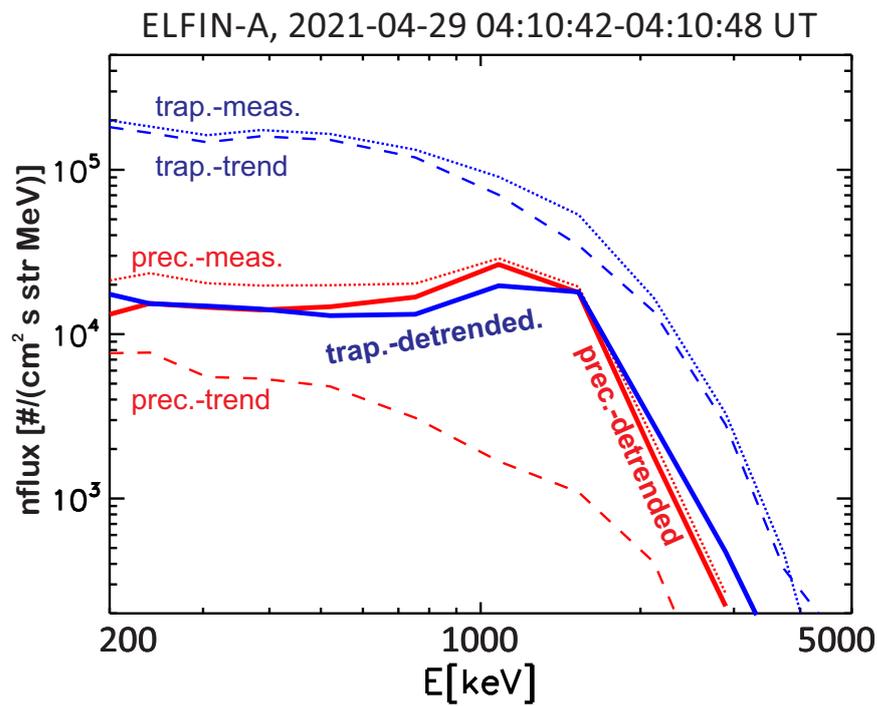}
\caption{Average spectra of precipitating (red) and trapped (blue) electrons at the moment of the strongest precipitation (04:10:42-04:10:48 UT) at ELFIN A for the event of Figure \ref{fig:event0.1}. Detrended fluxes (solid, thick lines) are
measured averages (dotted, thin lines) minus the trends (dashed, thin lines). Trends are average fluxes from 6s immediately before and 6s immediately after the strongest precipitation interval.
}
\label{fig:event0.3}
\end{figure}

\begin{figure}
  \centering
  \includegraphics[trim={2.75cm 1cm 2.75cm 1cm}, clip, width=1\textwidth]{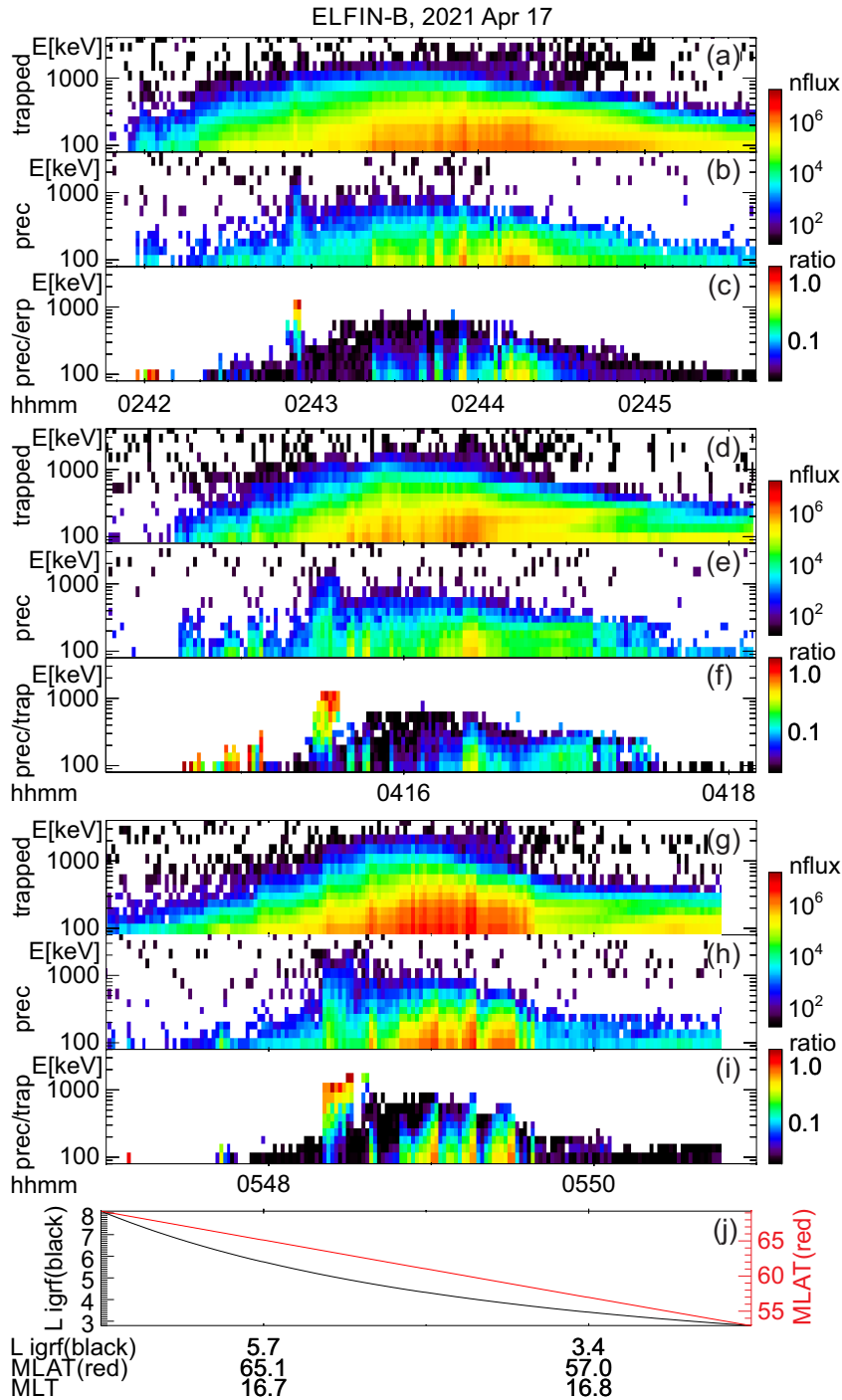}
\caption{An example of prolonged relativistic electron precipitation, presumably due to the long-lasting presence of EMIC waves in the equatorial magnetosphere. Three consecutive northern hemisphere, equatorward science zone crossings by ELFIN A at post-noon/dusk (MLT $\sim$16.75) are depicted in three panels per crossing (a-c; d-f; g-i), arranged in time to have a common L-shell and magnetic latitude, shown in Panel (j). Each crossing's three panels have the same format as the top three panels in Figures \ref{fig:intro} and \ref{fig:event0.1}, i.e., they are energy spectrograms of trapped fluxes, precipitating fluxes and precipitating-to-trapped flux ratio.}
\label{fig:long}
\end{figure}

\begin{figure}
  \centering
  \includegraphics[trim={2.75cm 11cm 2.75cm 1cm}, clip, width=1\textwidth]{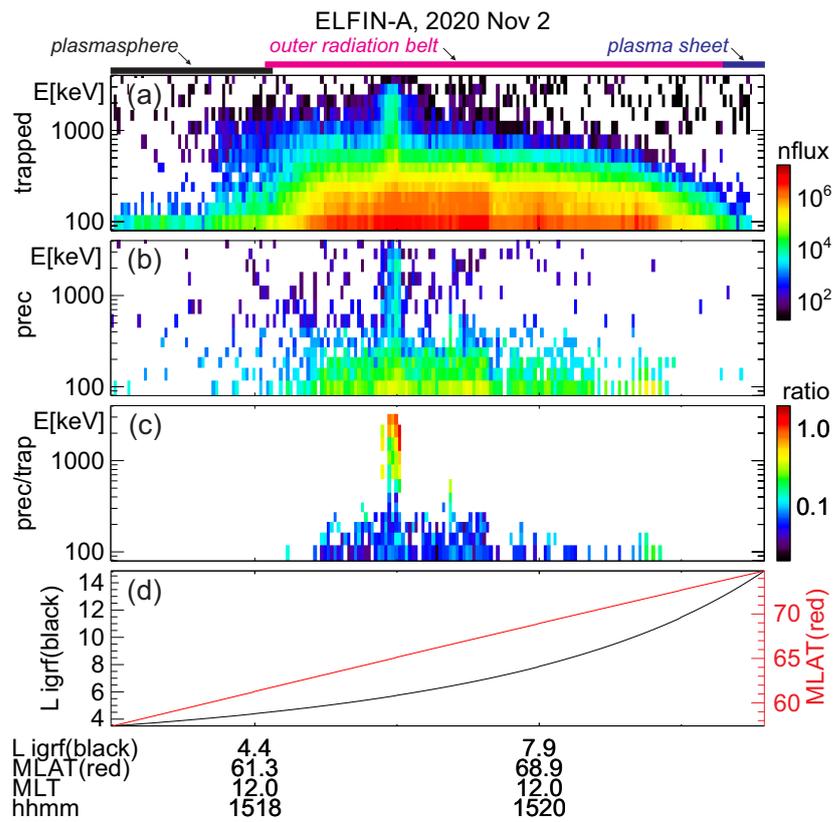}
\caption{Overview of ELFIN A observations during a $He^+$ band EMIC wave-driven event, on 02 Nov 2020, in a format similar to that of Figure \ref{fig:intro}.}
\label{fig:event1.1}
\end{figure}

\begin{figure}
  \centering
  \includegraphics[width=0.95\textwidth]{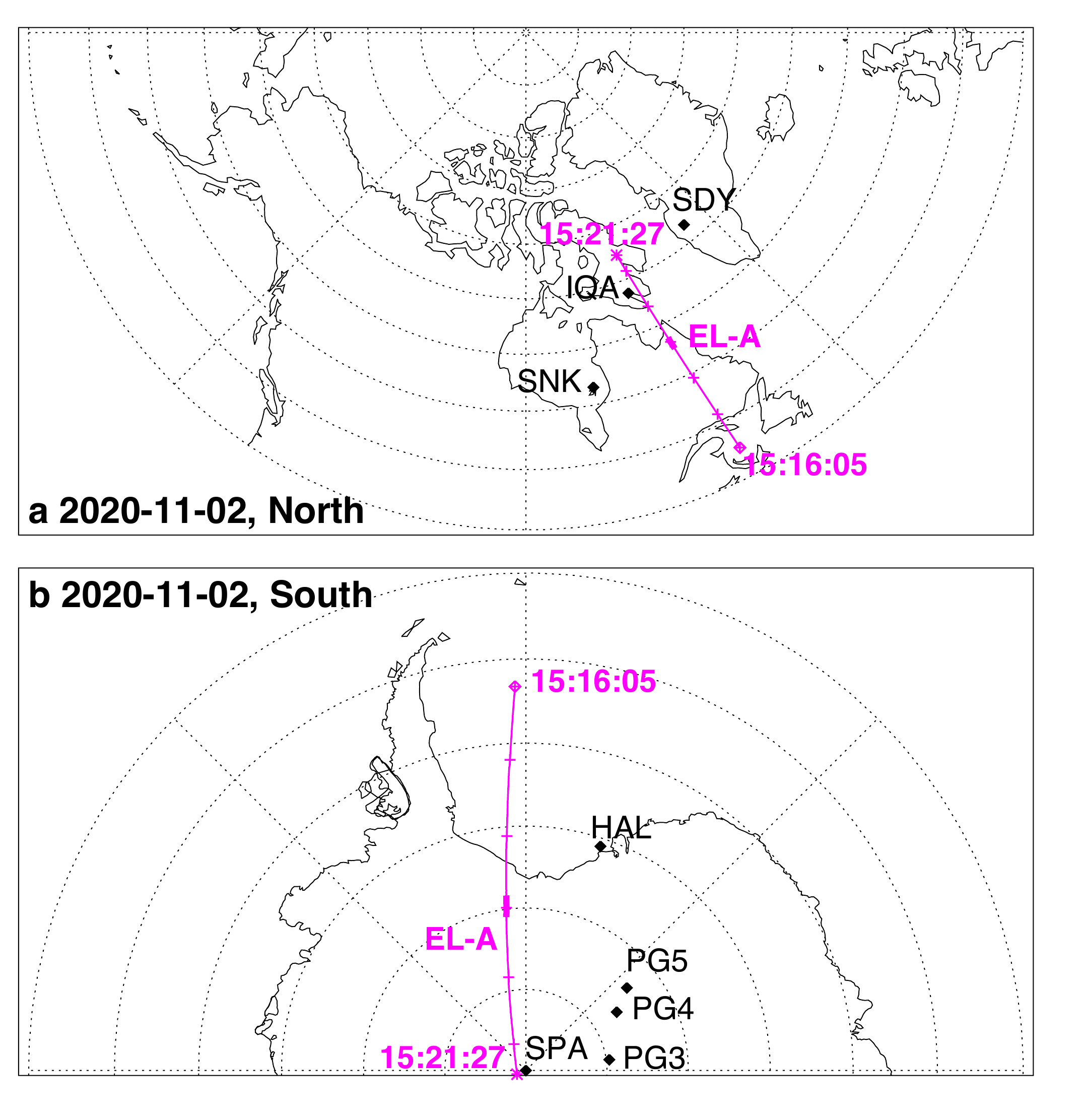}
\caption{ELFIN A projections to the ionosphere in the north and south for the event of Fig. \ref{fig:event1.1}. Diamonds and asterisks mark the start and end times of the trajectories; crosses are 1 min tickmarks; thick traces denote times of intense relativistic electron precipitation identified from Fig. \ref{fig:event1.1}(c) as a putative EMIC wave-driven precipitation event.}
\label{fig:event1.2}
\end{figure}

\begin{figure}
  \centering
  \includegraphics[width=1.0\textwidth]{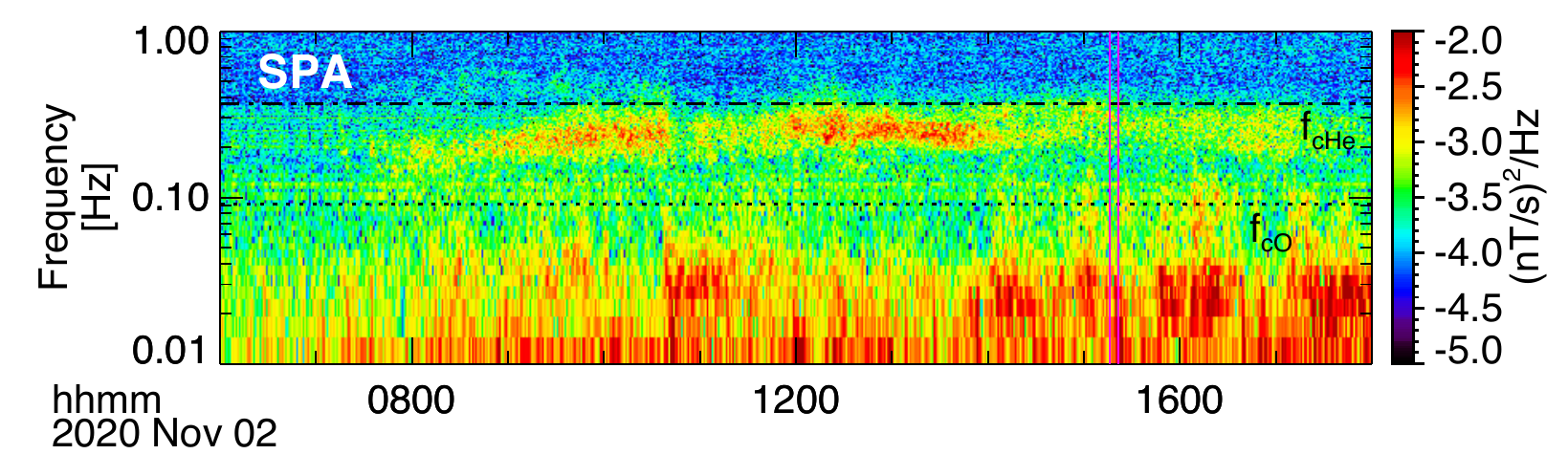}
\caption{Magnetic field spectra from the ground-based station at SPA in the Antarctic. As shown in Fig. \ref{fig:event1.2}, SPA is in close conjunction with ELFIN-A during this event. Superimposed in the spectra are $He^+$ and $O^+$ equatorial gyrofrequencies (horizontal lines) using the magnetic field at their equatorial projection, inferred from the T89 model. The vertical magenta lines mark the time interval of ELFIN A' science zone crossing during this event.}
\label{fig:event1.3}
\end{figure}

\begin{figure}
  \centering
  \includegraphics[width=0.8\textwidth]{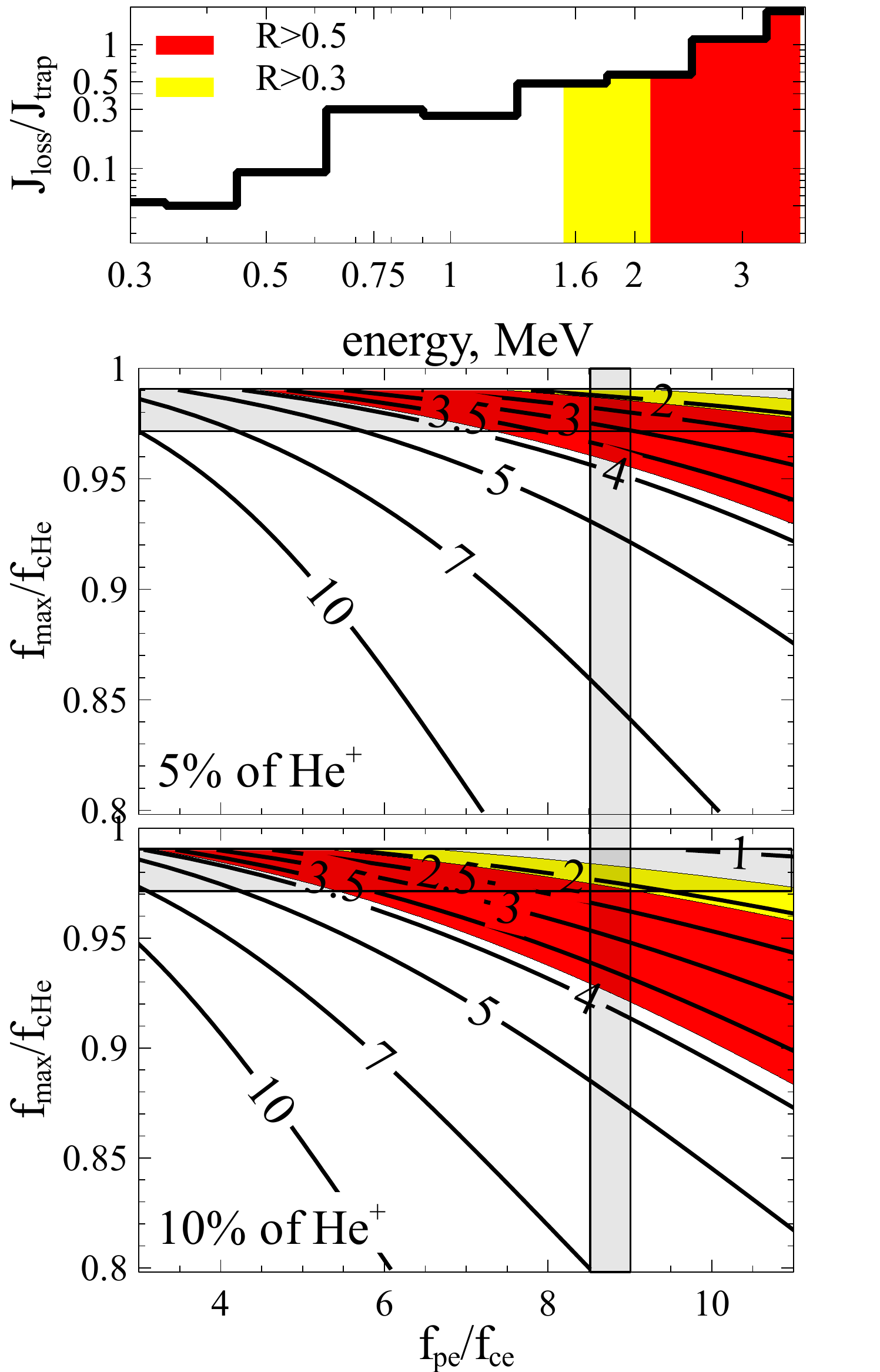}
\caption{Theoretical estimates of minimum resonance energy and comparison with observations for the event of Fig. \ref{fig:event1.1}. From top to bottom: average precipitating-to-trapped flux ratio during the time of peak precipitation (15:18:51-15:19:03UT); contour plot of minimum resonance energy (in MeV) as a function of the maximum unstable frequency $f_{\max}$ (normalized to the relevant ion cyclotron gyrofrequency $f_{cHe}$) and of the $f_{pe}/f_{ce}$ ratio for a $He^+$ ion concentration of 5$\%$; and same as the panel above, but for a 10$\%$ $He^+$ concentration. Red and yellow colors depict the electron energy ranges for which ELFIN A measured strong precipitation and moderate precipitation (R$>$0.5 and 0.5$>$R$>$0.3, respectively). The plot shows that resonance energies exhibiting moderate and strong precipitation at ELFIN are consistent with the range of parameters $f_{\max}/f_{cHe}$, $f_{pe}/f_{ce}$ inferred from in-situ measurements at conjugate platforms (at the intersection of the corresponding grayed areas), for a reasonable range of $He^+$ concentrations.}
\label{fig:event1.4}
\end{figure}

\begin{figure}
  \centering
  \includegraphics[trim={2.75cm 11cm 2.75cm 1cm}, clip, width=1\textwidth]{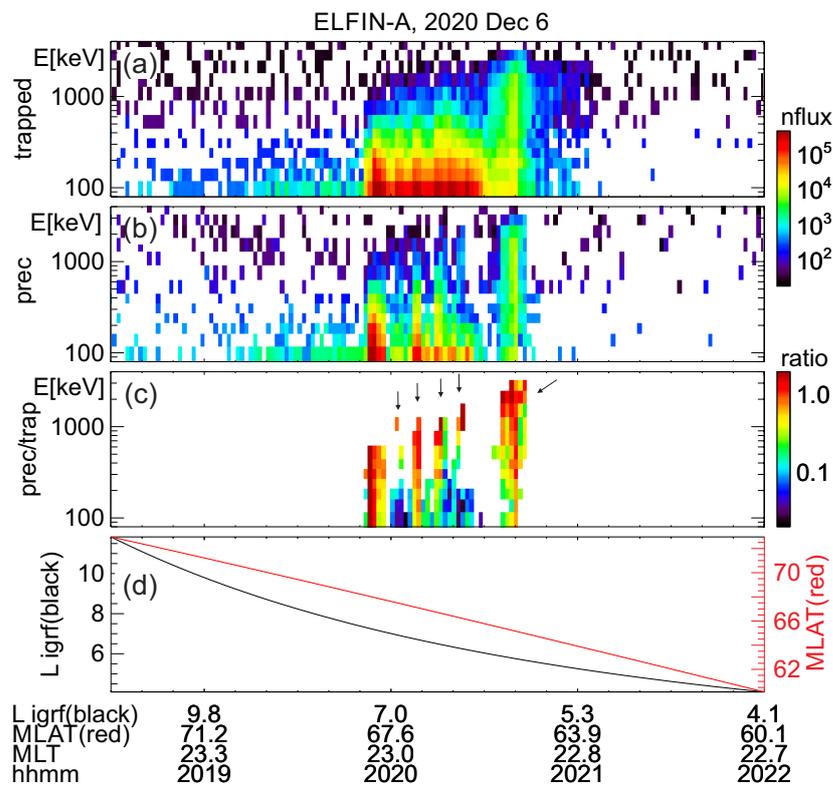}
\caption{Overview of ELFIN A observations during a $H^+$ band EMIC wave-driven event, on 6 December 2020, in a format similar to that of Figure \ref{fig:intro}.}
\label{fig:event2.1}
\end{figure}

\begin{figure}
  \centering
  \includegraphics[width=0.95\textwidth]{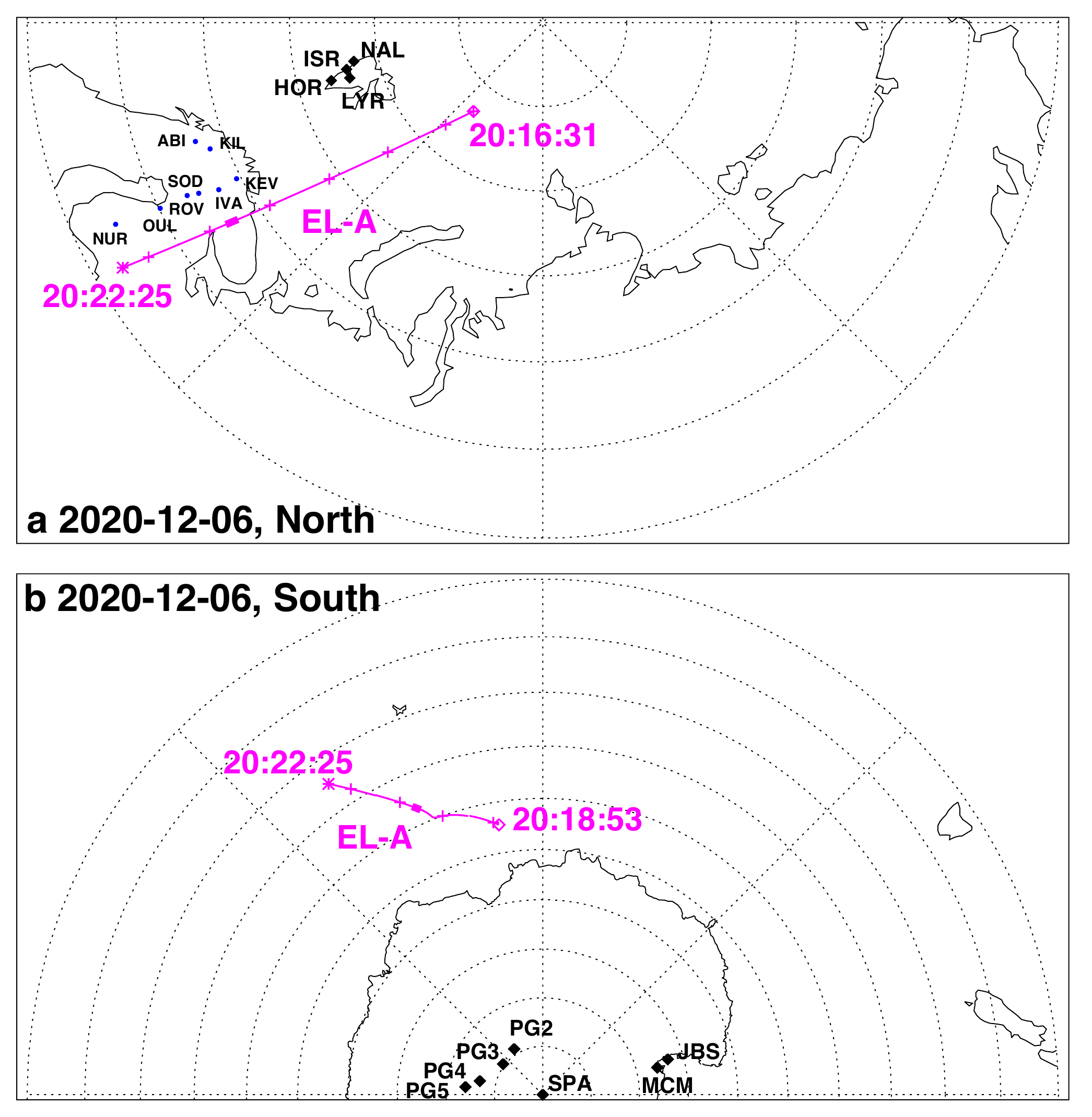}
\caption{ELFIN A projections to the ionosphere in the north and south for the event of Fig. \ref{fig:event2.1}, in a format similar to that of Figure \ref{fig:event1.2}.}
\label{fig:event2.2}
\end{figure}

\begin{figure}
  \centering
  \includegraphics[width=1.0\textwidth]{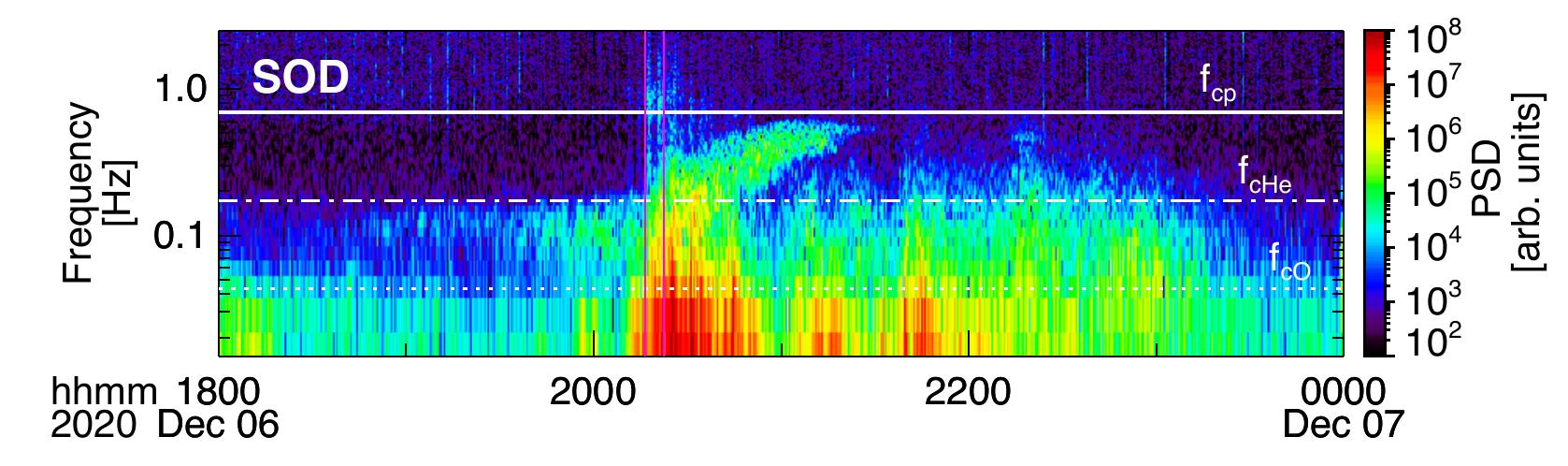}
\caption{Magnetic field power spectral density (arbitrary units) from the Finland ground-based station at SOD, located as shown in Fig. \ref{fig:event2.2}. Superimposed $H^+$, $He^+$ and $O^+$ equatorial gyrofrequencies (horizontal lines) using the equatorial  magnetic field conjugate to these stations. The magenta vertical lines bracket the time interval of ELFIN A's science zone crossing during this event.}
\label{fig:event2.3}
\end{figure}

\begin{figure}
  \centering
  \includegraphics[width=0.8\textwidth]{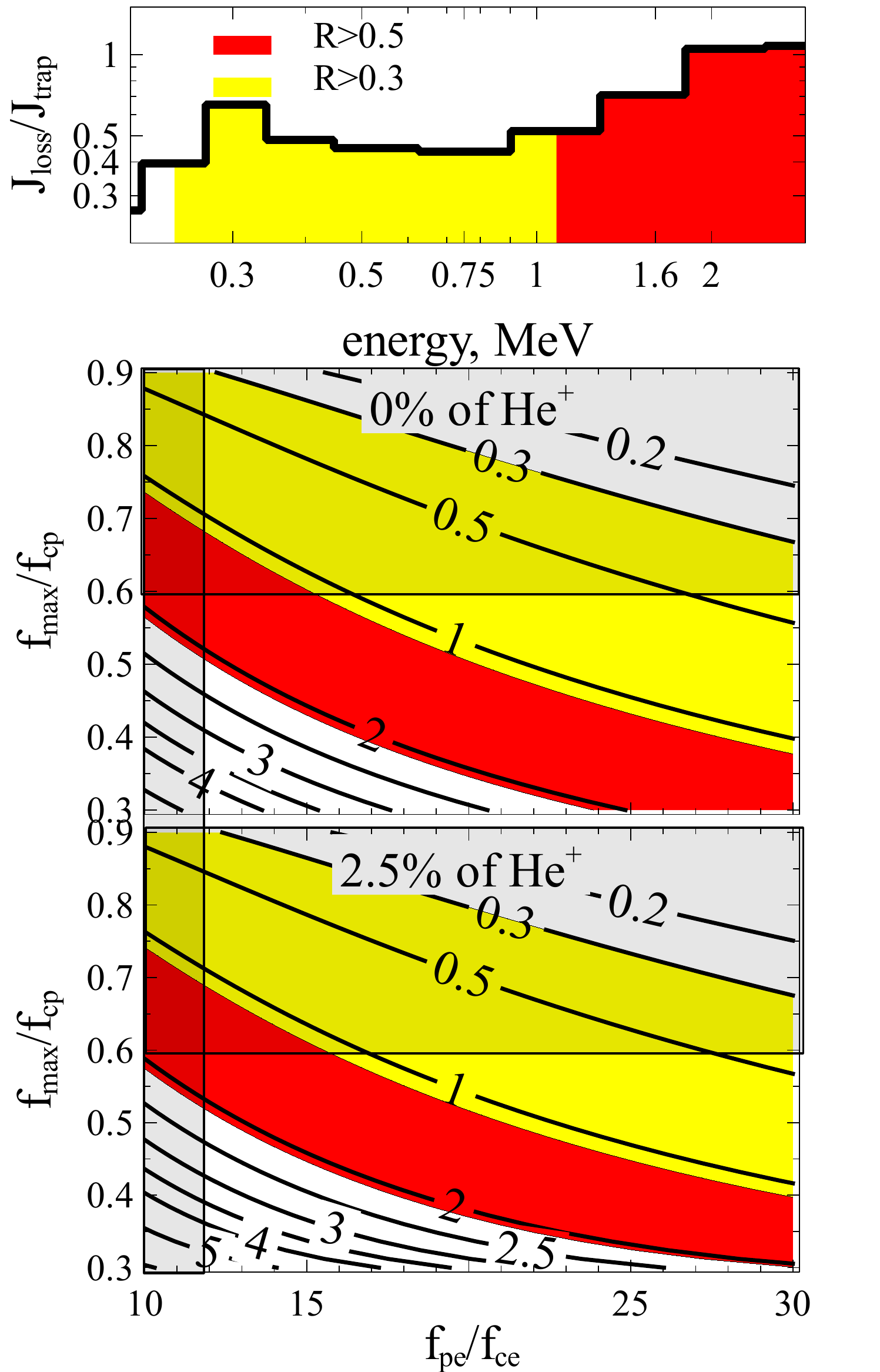}
\caption{Theoretical estimates of minimum resonance energy and comparison with observations for the event of Fig. \ref{fig:event2.1}. From top to bottom: average precipitating-to-trapped flux ratio during the time of peak precipitation (20:20:33 - 20:20:48 UT); contour plot of minimum resonance energy (in MeV) as a function of the maximum unstable frequency $f_{\max}$ (normalized to the relevant ion cyclotron frequency, $f_{cp}$), and of the $f_{pe}/f_{ce}$ ratio for $0\%$ helium concentration; and same as the panel above but for a $2.5\% He^+$ concentration. Red and yellow colors depict the electron energy ranges for which ELFIN A measured strong and moderate precipitation (R$>$0.5 and 0.5$>$R$>$0.3, respectively). The plot shows that resonance energies exhibiting moderate and strong precipitation at ELFIN are consistent with the range of parameters $f_{\max}$/$f_{cp}$ and $f_{pe}/f_{ce}$ inferred from in-situ observations at conjugate platforms ($f_{\max}/f_{cp} \sim$0.6-0.9 and  $f_{pe}/f_{ce} \sim$10-12) for a reasonable range of $He^+$ concentrations.}
\label{fig:event2.4}
\end{figure}

\begin{figure}
  \centering
  \includegraphics[trim={2.75cm 2cm 2.75cm 1cm}, clip, width=1\textwidth]{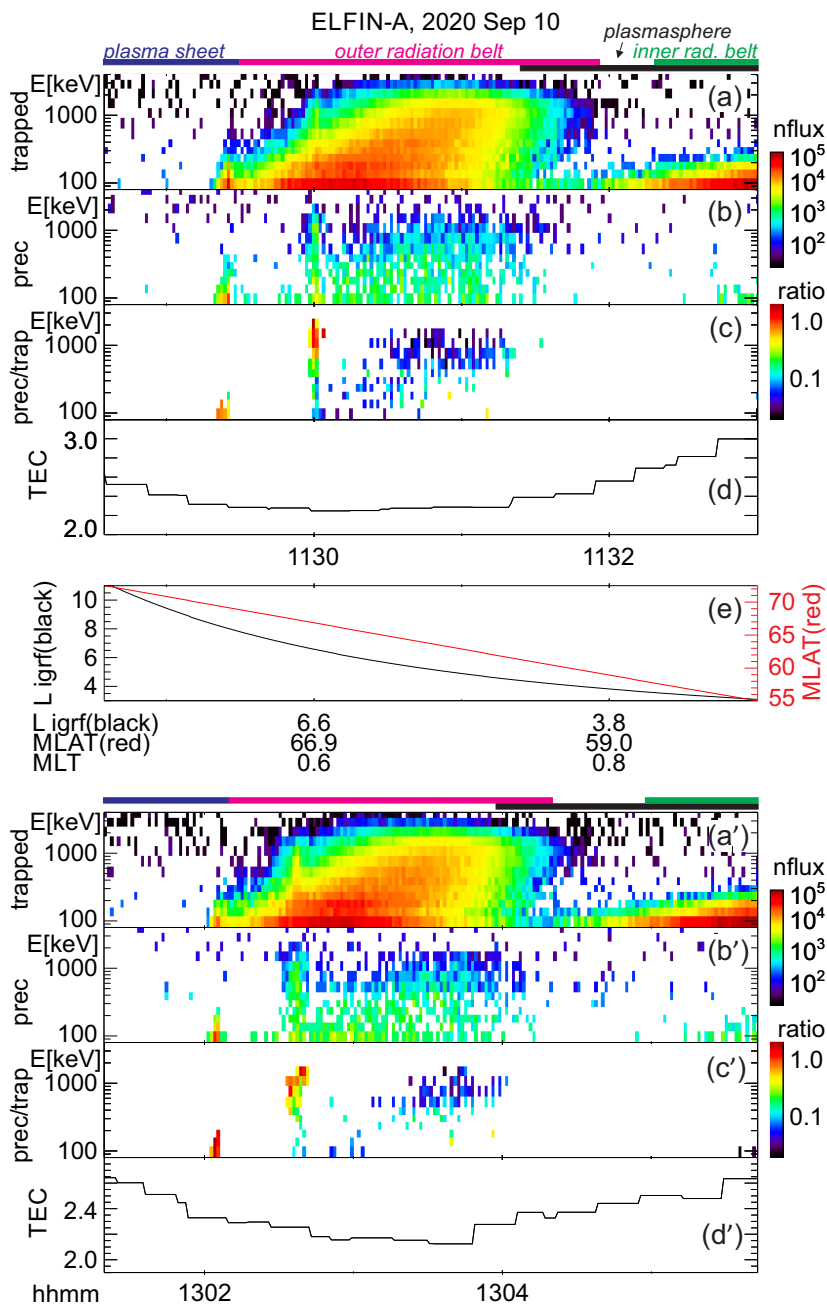}
\caption{Overview of ELFIN A observations on 10 September 2020, which is being examined together with concurrent TEC observations. Two consecutive northern hemisphere, equatorward science zone crossings at midnight/pre-midnight (MLT $\sim$0.75) are depicted in three panels per crossing (a-c; and f-g), arranged in time to have a common L-shell and magnetic latitude, shown in Panel (e). Each crossing's three panels have the same format as the top three panels in Figures \ref{fig:intro} and \ref{fig:event0.1}, i.e., they show energy spectrograms of trapped fluxes, precipitating fluxes and precipitating-to-trapped flux ratio. A fourth panel for each crossing (Panels (d) and (h), respectively) show the TEC values at the satellite ionospheric projection (after averaging over 3 degrees each in latitude and longitude, and over 20min in time).}
\label{fig:event4.1}
\end{figure}

\begin{figure}
  \centering
    \includegraphics[width=0.95\textwidth]{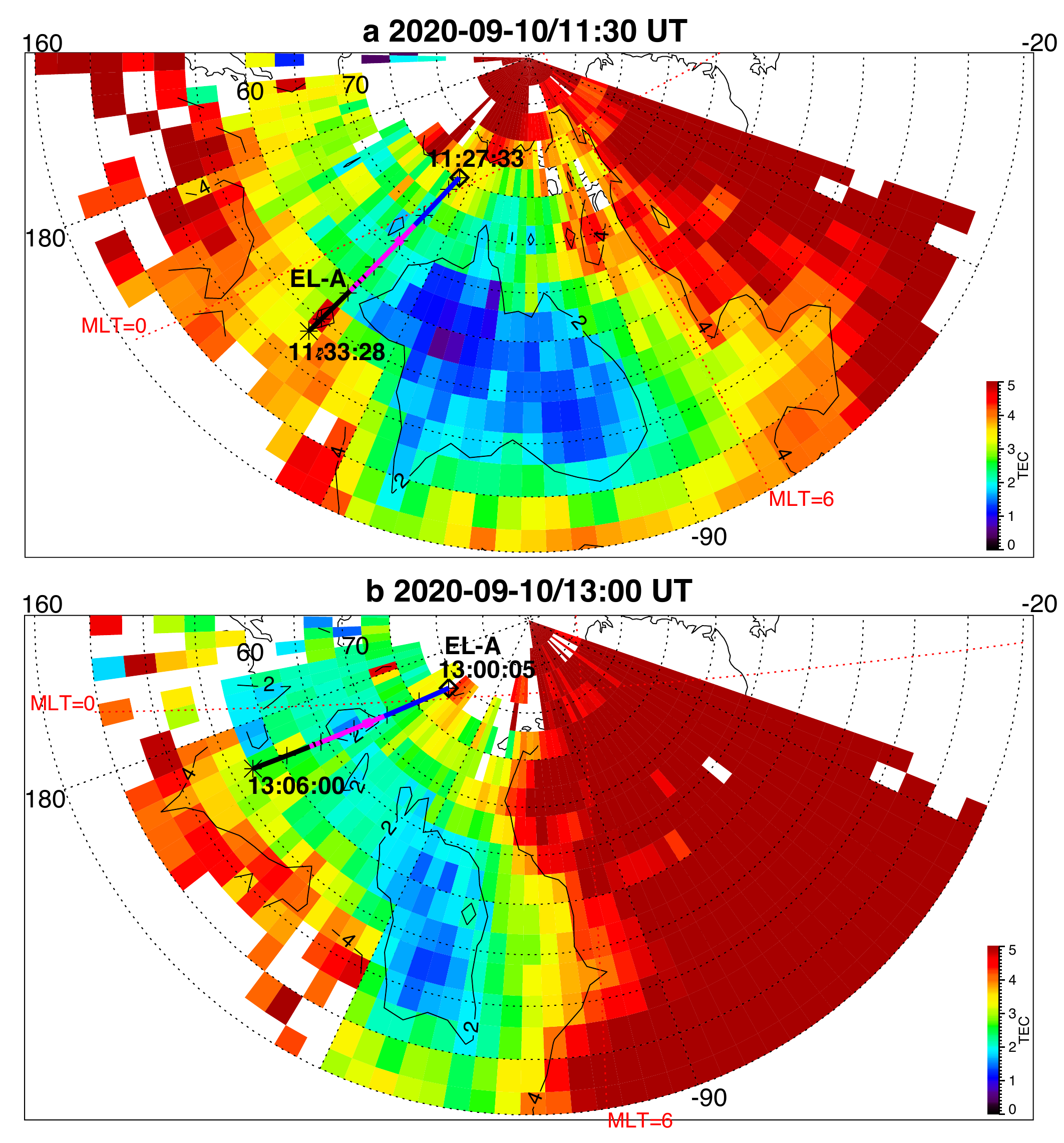} 
\caption{Projections of ELFIN A science zone crossings of Figure \ref{fig:event4.1} onto the Madrigal TEC maps at 11:30:00 UT (top panel) and 13:00:00 UT (bottom panel) in the northern hemisphere: start and end times of ELFIN A trajectories are marked by a diamond and an asterisk, respectively; crosses are 1 min marks. Regions of interest along the ELFIN trajectory are denoted with same colors as at the top of Figure \ref{fig:event4.1} (blue: plasma sheet, magenta: outer radiation belt, black: plasmasphere). The thick magenta line sections depict the times of relativistic electron precipitation related to EMIC wave scattering identified from Figure \ref{fig:event4.1}, Panels (c) and (c$*\prime$), respectively. Numbered black traces in the TEC maps show contours of TEC.}
\label{fig:event4.2}
\end{figure}

\begin{figure}
  \centering
  \includegraphics[width=0.9\textwidth]{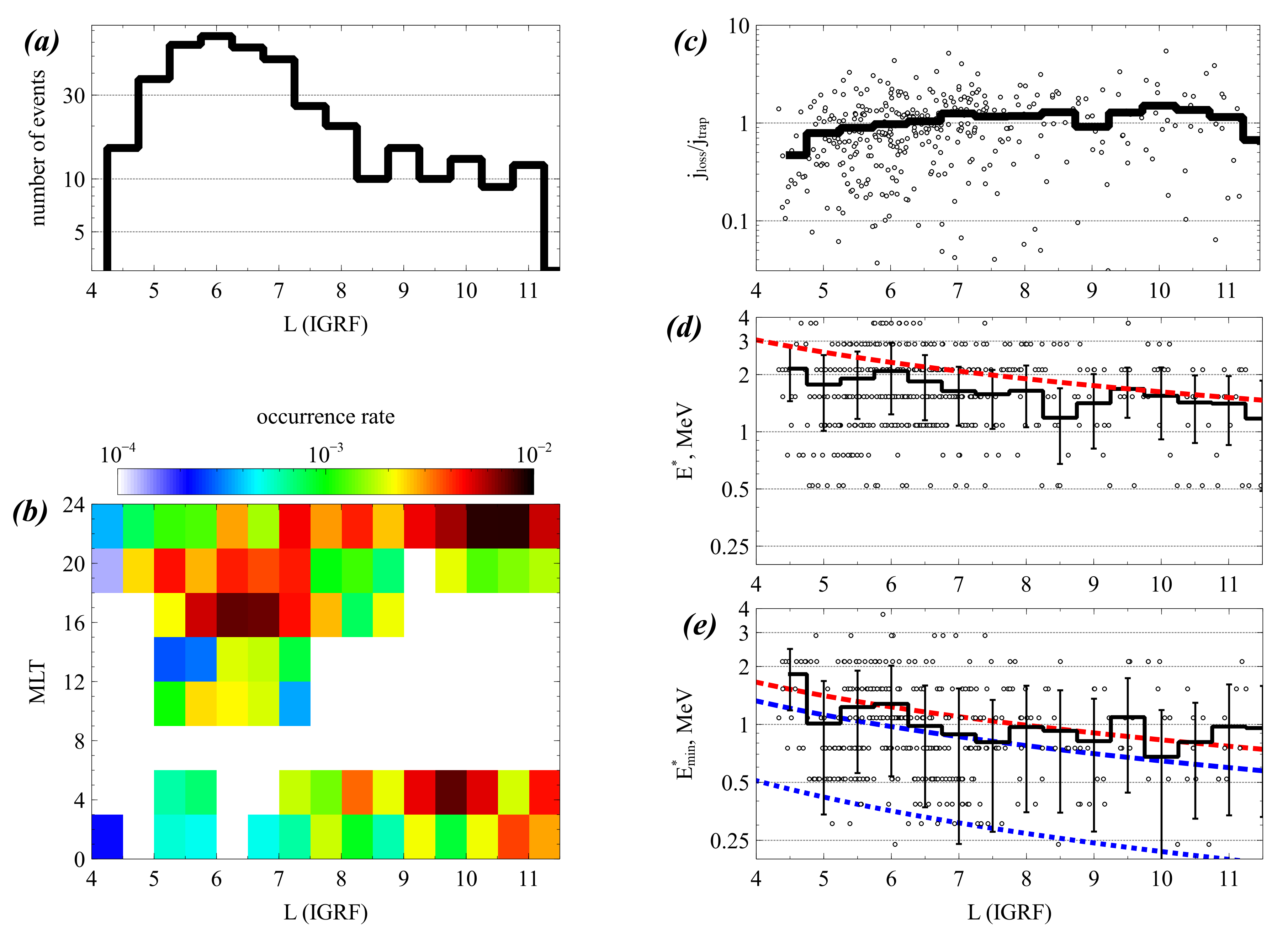}
\caption{Properties of relativistic electron precipitation inferred from $\sim 50$ EMIC wave-driven events (comprising $\sim 310$ individual spin samples) measured by ELFIN A\&B. (a) Ensemble spatial distribution versus $L$-shell. (b) Occurrence rate of events (total event duration normalized by the ELFIN residence time in each bin) in $(L,MLT)$ space. (c) Peak precipitating-to-trapped flux ratio at each spin sample (circle) and average of that ratio within each $L$-shell bin (solid curve), both shown as a function of $L$-shell. (d) Energy of the peak precipitating-to-trapped electron flux ratio, $E^*$, at each spin sample as a function of $L$. (e) Energy of half-peak (below the peak) of the precipitating-to-trapped electron flux ratio, denoted by $E^*_{min}$ at each spin sample, shown as a function of $L$. Red dashed lines in Panels (d) and (e) depict theoretical estimates of the respective quantities based on statistical wave spectra (derived as discussed in text). Dashed and dotted blue lines in Panel (e) depict theoretical lower limits of the resonance energy estimated based on wave cyclotron damping at $kc/\Omega_{pi}\sim 1$ and $kc/\Omega_{pi}\sim 2$, respectively.}
\label{fig:statistics}
\end{figure}

It is possible that few, some, or many of our events are due to nonlinear scattering due to spatially-localized or short-coherence scale, high-amplitude wave emissions. In fact, it has been known for a while that individual EMIC wave-packets often reach sufficient amplitudes to allow faster, nonlinear wave-particle interactions \citep{Engebretson08, Albert&Bortnik09, Pickett10, Omura&Zhao12, Grach21:emic}. However, despite the presence of such nonlinearly scattered, bursty precipitation events in our database, it is still quite likely that such events statistically conform to a diffusive treatment. The argument for this is an analogy with whistler-mode chorus waves. Chorus wave-packets having large amplitude (thus, in the nonlinear regime), can have short wave-packet lengths or exhibit strong, random phase jumps, or both, resulting in phase decoherence. This allows a diffusive description of the resulting
electron phase space density evolution, despite the large wave amplitudes at play \citep{Zhang18:jgr:intensewaves, Zhang20:grl:phase, Artemyev21:pre, Artemyev22:jgr:QLNL, Mourenas22:jgr:elfin}. And for EMIC waves too, it is likely that a prevalence of short EMIC wave-packets exhibiting strong and random wave phase jumps across or within them (e.g., see various examples of short packets in \citep{Usanova10} and \citep{An22}; some examples have also been seen in Figure \ref{fig:event0.2} and discussed in Section \ref{subsec:event0}) can permit a diffusive description of the scattering process despite the nonlinear nature of the scattering. In this approach the average precipitation fluxes and the average wave amplitudes incorporate the combined effects of linear and nonlinear resonant interaction regimes. We therefore proceed in our analysis of the precipitation energy spectra with a diffusive paradigm in mind, and study the minimum resonance energy, peak precipitation energy and the typical wave amplitude variation with frequency that corresponds to the energy-spectrum of the precipitation, next.

Figure \ref{fig:statistics}(d) shows that the energy of the peak precipitating-to-trapped flux ratio $E^{*}$ lies typically in the range 1 to 3 MeV. We interpret $E^{*}$ as the energy of electrons resonating with the most intense EMIC waves. This is because the typical spectrum of EMIC waves on the ground or in space (as seen for example in Figure \ref{fig:event0.2}(a) or Figure \ref{fig:event1.3}) has a peak intensity at a certain frequency, $f_{peak}$. Then, electrons of energy $E_{peak}$ resonating with the most intense EMIC waves at that frequency should also exhibit a peak in their precipitating-to-trapped flux ratio. The higher frequency part of the EMIC wave spectrum (at $f > f_{peak}$), has a lower wave intensity, but it will still lead to electron precipitation at energies below $E_{peak}$, down to the minimum resonance energy $E_{R,min}$ corresponding to $f_{max}$, the maximum wave frequency of appreciable wave power.

To better approximate the minimum resonant energy for the purpose of statistical studies of its spatial distribution, we define $E^*_{\min}$ as the half-way point in precipitation intensity below its peak. $E^*_{\min}$ is thus the energy at half-peak of the measured precipitating-to-trapped electron flux ratio at energies lower than that of peak precipitation ($E^*_{\min}<E^*$). (For some EMIC spectra without appreciable wave power at frequencies higher than their peak intensity, it would be $E^*_{\min} = E^*$, but that is a rare occasion.) We note that warm plasma effects may limit scattering to frequencies below $f_{max}$ (energies above $E^*_{\min}$) but also that such effects may also limit propagation and suppress the average wave power to below its $f_{max}$ attainable in each individual event time series. Thus the half-way point appears as a reasonable energy selection in precipitation data to represent the theoretical $E_{R,\min}$ expected from a statistical average of $f_{max}$ in available datasets.

In summary, $E^{*}$ is our measured estimate for the theoretical $E_{peak}$ corresponding to the point of maximum precipitation by EMIC wave-driven waves of peak wave power at $f_{peak}$, and $E^{*}_{min}$ is our measured proxy for the theoretical minimum resonance energy $E_{R,\min}$ corresponding to the maximum frequency of appreciable wave power $f_{max}$ in statistical averages of such wave power observations. In other words, $E^{*}_{min}$ is an estimate of the minimum resonance energy $E_{R,\min}$ corresponding to significant wave-driven electron scattering toward the loss-cone. Figure \ref{fig:statistics}(e) shows $E^*_{\min}$ for each spin in our database of events. Evidently, its average lies in the range of 0.5 to 1.5 MeV.

Both $E^{*}$ and $E^{*}_{min}$ tend to decrease, on average, as $L$ increases (despite the large scatter, which is in part due to uncertainties in $L$-shell determination). This is expected, due to the decrease of the minimum resonance energy for cyclotron resonance with distance from Earth: Since the dipole field intensity fall-off with distance ($\sim1/L^3$) is faster than that of the square-root of the density \cite{Sheeley01}, $f_{pe}/f_{ce}$ increases with distance, and the minimum resonance energy for electrons (which is monotonically and inversely dependent on $f_{pe}/f_{ce}$, as discussed in Section 2.2) decreases with distance. We can see this behavior even in case studies with multiple EMIC wave-driven bursts, such as that of Figure\ref{fig:event2.1}: the energy of peak precipitating-to-trapped ratio, i.e., the energy of the most efficient precipitation, decreases at progressively larger $L$-shells.

To assess the trends revealed in the averages shown, we compare them with estimates for $E^*$ and $E^*_{min}$ determined independently, from published statistical averages of EMIC wave spectra providing peak-power and maximum frequency \citep{Zhang16:grl} and from a model-based estimation of the $f_{pe}/f_{ce}$ ratio \citep{Summers&Thorne03}. Our independent estimates resulted in the dashed red lines in Panels (d) and (e). Specifically, our procedure was as follows: Using an empirical model for the plasmaspheric plume density $n_e\sim 1300(3/L)^{4.83}$ cm$^{-3}$ at $L>4$ \citep{Sheeley01}, as appropriate for the dusk sector where most of our events were observed, and assuming that the resonance is mainly with hydrogen band EMIC waves (based on previous statistical observations \citep{Kersten14,Mourenas16:grl,Mourenas17,Zhang16:grl,Zhang17,Zhang21}), we get for the resonance energy of precipitating fluxes $E[{\rm MeV}] = [(1 +C/L^{1.17})^{1/2} -1]/2$. The coefficient $C$ is to be determined from EMIC wave and background-plasma characteristics. For typical peak power parameters \cite{Kersten14, Zhang16:jgr} $f_{\rm peak}/f_{cp}\simeq 0.4$, hydrogen band EMIC waves, and 98\% proton $+ 2$ \% helium ions (or $f_{\rm peak}/f_{cp}=0.43$ and $92$\% protons $+ 8$\% helium ions which result in similar values), we get $C\simeq 246$. The corresponding theoretical estimate of the typical peak resonance energy, $E_{\rm peak}$ is shown as a dashed red curve in Panel (d). Next, we used a factor $C/2.8$ instead of $C$, to get our theoretical estimate of the minimum resonance energy $E_{R,\min}$ corresponding to significant wave-driven electron scattering toward the loss-cone. This is shown as a dashed red line in Panel (e) along with the observationally determined energies at half-max precipitating-to-trapped electron flux ratio $E^*_{\min}$ and their average. The factor $C/2.8$ corresponds to electron cyclotron resonance with hydrogen band EMIC waves of frequency $f/f_{cp} \sim 0.56$, well above the peak-power frequency $f_{\rm peak}/f_{cp}\simeq 0.4$, and taken to be a reasonable approximation for $f_{\max}/f_{cp}$ in statistical averages of wave power. At that frequency, wave power is still finite albeit an order of magnitude lower than at peak-power frequency \citep{Zhang16:grl,Zhang21}. This choice is consistent with the smaller precipitating-to-trapped flux ratio at $E_{R,\min}$ (compared to that at $E_{\rm peak}$), despite the theoretically anticipated increase in diffusion rate with decreasing energy for otherwise constant wave power \citep{Mourenas16:grl,Ni15,Summers&Thorne03}. As we see from the panels under discussion, there is a reasonably good match between the theoretical expectation for $E_{\rm peak}$ and $E_{R,\min}$ independently derived from EMIC wave-power statistics alone (the dashed red lines), and the averages of $E^*$ and $E^*_{\min}$ from observations of the precipitating-to-trapped flux ratio at ELFIN (the solid black lines).

In addition, we show in Panel (e) two lower limits of $E_{R,\min}$ estimated based on wave cyclotron damping by cold/cool ions near the proton gyrofrequency (see Section \ref{SecINTER}) at $kc/\Omega_{pi}\sim 1$ (dashed blue line) and at $kc/\Omega_{pi}\sim 2$ (dotted blue line). These two estimates fit well the average $E^*_{\min}$ values (solid black line) and the lower limit of $E^*_{\min}$ data points at $L=4-10$, respectively. This suggests a significant effect of wave cyclotron damping in determining the wave spectrum shape at high frequencies $f>0.6\,f_{cp}$ \citep{Chen13:emic}.

\subsection{Properties of most intense events}\label{sec:intenseevents}

\subsubsection{Dependence on AE, MLT and concurrence with whistlers}

\begin{figure}
  \centering
  \includegraphics[width=0.9\textwidth]{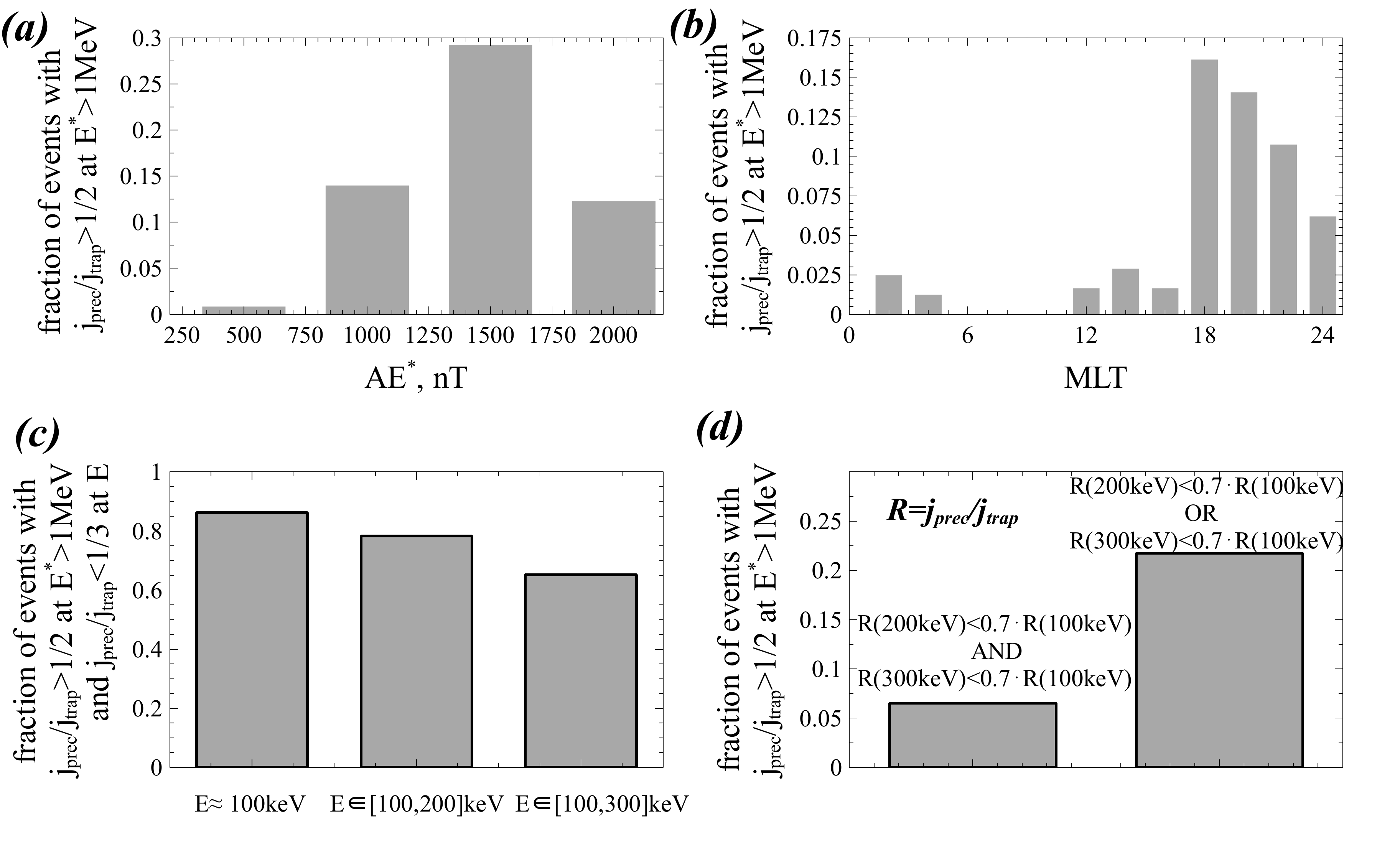}
\caption{Properties of the most efficient EMIC wave-driven precipitation events, those exhibiting highly relativistic (E$^* > $1 MeV), strong (R = jprec/jtrap $>$ 1/2) electron precipitation, observed at L $<$ 7. (a) Fraction of these most efficient events in each $AE^{*}$ bin divided by total number of most efficient events ($AE^{*}$ is the maximum AE in the preceeding 3 hours). (b) Same as (a) but as a function of MLT. (c) Fraction of most efficient events that also have $0\leqq R<1/3$) in the three low-energy range categories listed in the abscissa. The rest (those with $R>1/3$), representing concurrent low-energy moderate or strong precipitation, can still be a significant fraction of the most efficient EMIC wave-driven precipitation events (15 - 35\%, depending on energy range category). (d) Fraction of most efficient EMIC wave-driven precipitation events satisfying two different criteria for $R=j_{prec}/j_{trap}$ at lower energy as defined in annotations. Implications are discussed in the main text.}
\label{fig:statistics2}
\end{figure}

Let us now investigate some properties of the highly relativistic, strong electron precipitation EMIC wave-driven events, those with $R=j_{prec}/j_{trap}>1/2$ at $E^{*}>1$ MeV, of the predominant spatial category of events, the ones with  $L<7$. These represent the most efficient subset of the previously discussed strong precipitation (R$>$0.5) EMIC wave-driven events, those with peak energy $>$1MeV at L-shell $L<7$, but otherwise very similar. Figure \ref{fig:statistics2}(a) shows that when binned as a function of $AE^*$ (the maximum $AE$ in the preceding 3 hours), the fraction of these most efficient events over the total number of events within our database strongly increases from $250$ nT to $\sim1500$ nT and is quite significant for $AE^*>600$ nT. It is noteworthy that the increase of this fraction with $AE^*$ happens despite the rapidly decreasing probability of intervals with large $AE^*$, for values above $\sim200$ nT. Thus the smaller number of events in the highest $AE^*$ bin ($1750-2500$ nT) is likely due to the decrease in occurrence rate of such extremely high $AE^*$ periods. We interpret the result of Figure \ref{fig:statistics2}(a) as a consequence of the fact that EMIC wave power (which determines the efficiency of electron scattering) is itself strongly dependent on geomagnetic activity owing to activity-dependent injections of anisotropic hot ions which provide the free energy for EMIC wave generation \citep{Chen10:emic,Chen11:emic}.

Figure \ref{fig:statistics2}(b) shows the fraction of the most efficient EMIC wave-driven electron precipitation events (normalized to all events within $L<7$) as a function of MLT. It also reflects the relative occurrence probability of such events with MLT, since ELFIN's science zone collections are uniformly distributed in local time. This is much higher in the 18-24 MLT sector than elsewhere. We suggest this is because it is towards this sector that highly anisotropic hot ions drift, after being produced by nightside injections (which are known to peak in occurrence rate around the pre-midnight sector). The probability of efficient precipitation is much weaker at 0-16 MLT, likely because of the reduced anisotropy of the aforementioned hot ion populations (since these ions have had a chance to be modified by EMIC wave scattering at dusk). Thus when the injected ions reach the 12-16 MLT range they can generate only weaker EMIC waves, if any. Conversely, the 0-6 MLT sector can still harbor direct injections from the magnetotail, albeit at a lower probability than at pre-midnight, which could explain the reduced probability of both EMIC wave generation and subsequent efficient electron precipitation in that sector.

Figure \ref{fig:statistics2}(c) shows the fraction of the aforementioned most efficient EMIC wave-driven events that have a low energy $j_{prec}/j_{trap}$ ratio that is clearly low, $0\leqq R <1/3$. This condition is applied in three energy ranges: $E\simeq 100$ keV, E$\in$ [100,200] keV, and  E$\in$ [100,300] keV. The figure reveals that 85\% of events with strong precipitation ratio at $E^{*}> 1$ MeV are accompanied by a significantly weaker precipitation ratio, $<1/3$, at $E\simeq 100$ keV. However, the unity-complement of the third category (E$\in$ [100,300] keV) also shows that $\sim$35\% of the most efficient EMIC wave-driven events still have $j_{prec}/j_{trap}>1/3$ at some energy between $100$ and $300$ keV. Therefore, moderate to strong precipitation is still present at low energies (up to a few hundred keV), for quite a significant fraction of these most efficient, highly relativistic precipitation events.

So, we next investigate whether the precipitation at low ($100-300$ keV) energies we observed here during the most efficient EMIC wave-driven events could be due to electron scattering by whistler-mode chorus or hiss waves that may occur simultaneously with EMIC waves. Since the whistler-mode pitch-angle diffusion rate (and, thus, the precipitation efficiency) decreases with energy, $E$, the precipitating-to-trapped flux ratio $R=j_{prec}/j_{trap}$ should first decrease and reach a minimum somewhere above $100$ keV, before increasing again at relativistic energies due to cyclotron resonance with EMIC waves (e.g., see \citep{Mourenas21:jgr:ELFIN, Mourenas22:jgr:elfin}). We note that such intense sub-MeV precipitation has been reported previously \citep{Hendry17,Hendry19,Capannolo21}, but in the past it was not possible to separate the effects of whistler-mode wave and EMIC wave scattering by studying at sufficient energy resolution and extent this $R-E$ relationship. Here, we explore this relationship next, in Figure \ref{fig:statistics2}(d). The figure shows, on the right, the fraction of all events with $j_{prec}/j_{trap}>1/2$ at $E^{*}> 1$ MeV for which $R(200$ keV)$<0.7\times R(100$ keV) OR $R(300$ keV)$<0.7\times R(100$ keV); in other words, where the spectral slope of the ratio $R$ is decidedly negative ($0.7$ being an arbitrary but reasonable choice). Such a negative slope with energy at the energy range $E<300$ keV likely corresponds to chorus or hiss wave-driven precipitation simultaneous with EMIC wave precipitation at MeV energies. We see that the depicted fraction is $\simeq22\%$, indicating that about one fifth of the most efficient EMIC wave events exhibit signatures of simultaneous whistler-mode chorus precipitation below 300 keV.

The {\it OR} condition has been used above to ensure inclusive accounting of a broad range of low energies and an increased statistical significance of the results. However, it is still possible that sub-MeV precipitation by EMIC waves due to the aforementioned nonresonant \citep{Chen16:nonresonant,An22}, or bounce resonant \citep{Cao17:bounce, Blum19} interactions, might extend down to the 300 keV range of energies. This could provide a positive $R-E$ slope in the 300 keV range but retain a negative slope in the 200 keV range. Therefore, we examine a more stringent criterion, applying the {\it AND} condition to the above quantities. In other words, to examine bona fide concurrent whistler-mode and EMIC wave scattering, we investigate the fraction of putative EMIC wave-precipitation events (those with $j_{prec}/j_{trap}>1/2$ at $E^{*}> 1$ MeV) which exhibit both $R(200$ keV)$<0.7\times R(100$ keV) AND $R(300$ keV)$<0.7\times R(100$ keV). Figure \ref{fig:statistics2}(d), left, depicts this fraction. It shows that whistler-mode wave-driven precipitation may be present at $E\leq 300$ keV for only $\sim6.5$\%, or one sixteenth of all events with $j_{prece}/j_{trap}>1/2$ at $E^{*}>1$ MeV at $L<7$. These results suggest that at least $\sim78$\% and possibly up to $\sim93.5$\% of all intense EMIC wave-driven precipitation events extending down to $200$ keV are likely not due to concurrent whistler-mode and EMIC wave scattering, and deserve further scrutiny.

\subsubsection{Consistency of precipitation with diffusion theory}

The statistically significant and moderately efficient precipitation observed at $\sim$200 keV in the presence of strong precipitation at highly relativistic energies ($E^{*}>1$ MeV) by EMIC waves, does not seem, at first glance, to be similarly due to resonant scattering by EMIC waves: Based on Van Allen Probes statistics \citep{Ross21}, EMIC waves with sufficiently high frequencies (relatively close to the ion, and in particular the proton gyrofrequency) to provide resonant scattering and precipitation at such low energies \citep{Kennel&Petschek66, Mourenas22:jgr:elfin} indeed have very low wave power (at least 10 times lower than that at peak wave power \citep{Zhang16:grl}. In addition, hot plasma corrections to the wave dispersion relation become necessary in the close vicinity of the ion gyrofrequency, where they can suppress the expected resonant scattering \citep{Chen13:emic,Cao17}. Nonresonant scattering by EMIC waves with very sharp edges may be implicated \citep{Chen16:nonresonant,An22}, but the actual presence of such sharp wave-packet edges still has to be verified experimentally and its effects have yet to be compared with theory and simulations. Bounce resonant scattering can, in principle, provide precipitation of sub-MeV energies \citep{Cao17:bounce, Blum19}, but the associated scattering rate is quite small for quasi-parallel EMIC waves and, thus, further statistical investigation of the most effective oblique EMIC waves is needed to evaluate the relative contribution of this mechanism. Nevertheless, some useful insights into the possible origin of the moderate sub-MeV electron precipitation that accompanies strong relativistic electron precipitation by EMIC waves, can be gained from a careful examination of the energy spectrum characteristics of EMIC wave-driven precipitation in our database. We do this next.

\begin{figure}
  \centering
  \includegraphics[width=0.9\textwidth]{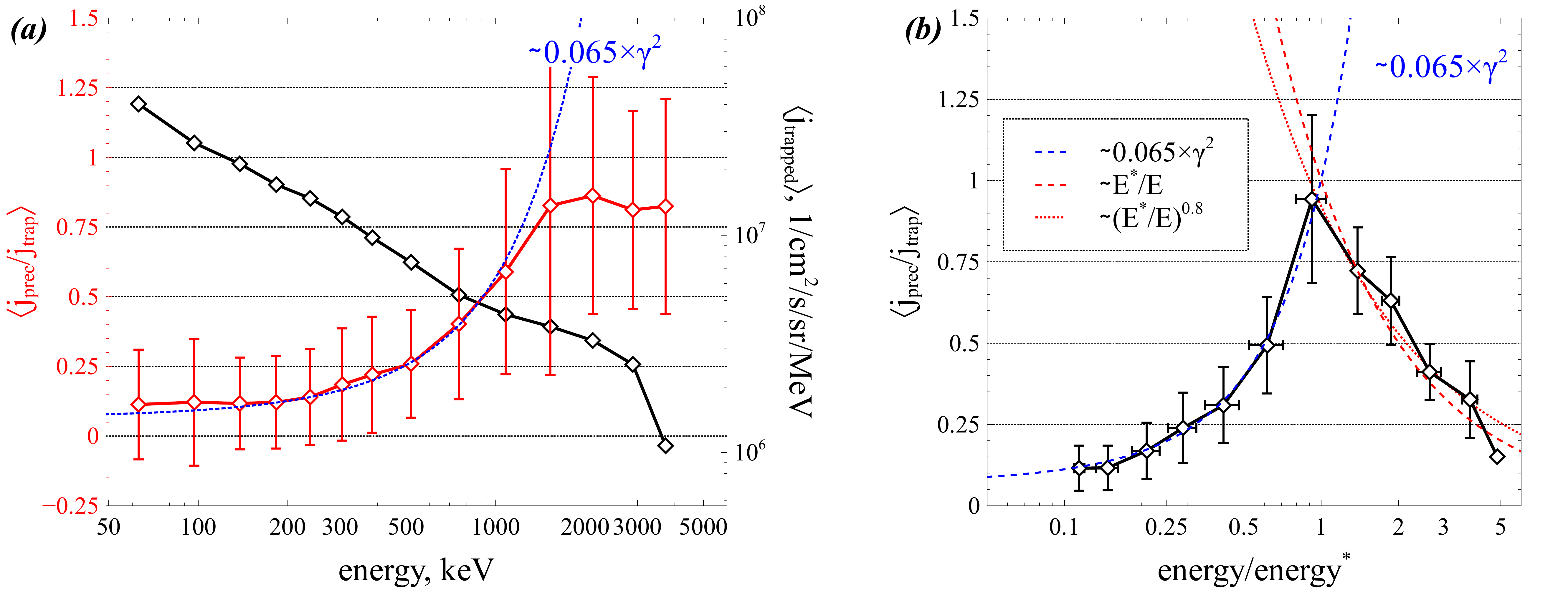}
\caption{Spectral properties of EMIC wave-driven electron precipitation. (a) Average spectrum of trapped fluxes (black curve) and of precipitating-to-trapped flux ratio (red curve) for events with $j_{prec}/j_{trap}>1/2$ at $E^{*}>1$ MeV. The dotted blue curve shows a best least-squares fit to the average flux ratio at low energy. (b) Average precipitating-to-trapped flux ratio for the events of Panel (a), plotted as a function of the normalized energy $E/E^*$, where $E^*$ is the peak precipitating-to-trapped flux ratio energy for each event (a proxy for the minimum resonance energy $E_{R,\min}$ with the most intense waves). Best least-squares fits to the average flux ratio are shown below the peak, $\langle j_{prec}/j_{trap}\rangle\approx \gamma^2(E)/\gamma^2(E^*)\sim0.065\,\gamma^2(E)$ for $E^*=\langle E^*\rangle\sim1.45$ MeV (dashed blue curve), and above the peak, $\langle j_{prec}/j_{trap}\rangle\approx 1.01\cdot (E/E^*)^{-1.01}$ (dashed red curve). Note that if the last point ($E/E^*\approx 4.84$) with a small number of measurements was discarded, the best fit would become $\langle j_{prec}/j_{trap}\rangle\approx 0.92\cdot (E/E^*)^{-0.8}$ (dotted red curve).
}
\label{ratio}
\end{figure}

Figure \ref{ratio}(a) shows the average trapped electron flux (in black) and the average precipitating-to-trapped flux ratio $\langle j_{prec}/j_{trap}\rangle$ (in red) as a function of energy for the dominant category of all highly relativistic electron strong precipitation events (at $L<7$ with $j_{prec}/j_{trap}>1/2$ at $E^*>1$ MeV). The flux ratio, denoting precipitation efficiency, increases approximately as $\gamma^2$ (where $\gamma=1+E/mc^2$ is the Lorentz factor) as kinetic energy $E$ increases (blue dotted line). It starts from a low, approximately constant average value $\langle j_{prec}/j_{trap}\rangle\sim 1/8$ at $E<200$ keV and rises to $\langle j_{prec}/j_{trap}\rangle>1/2$ at $\gtrsim 1$ MeV. It attains a high, approximately constant average value $\langle j_{prec}/j_{trap}\rangle\approx 0.85$ for $E > 1.5$ MeV. The low values of $\langle j_{prec}/j_{trap}\rangle\sim 1/8$ at the low energies could be partly due to the simultaneous presence of chorus wave-driven precipitation during a small fraction $\approx 6.5$\% of these events (as suggested by results in Figure \ref{fig:statistics2}(d)). Since such chorus events should give a higher $j_{prec}/j_{trap}$ ratio at $E<100$ keV \citep{Mourenas22:jgr:elfin}, whereas all the other events without chorus waves should give a slightly smaller $j_{prec}/j_{trap}$ in the same energy range, the resulting average $\langle j_{prec}/j_{trap}\rangle$ can be nearly constant at $50-200$ keV. The flattening of the slope of $\langle j_{prec}/j_{trap}\rangle$ versus energy at $E \geq 1.5$ MeV is consistent with the most common minimum cyclotron resonance energy with intense EMIC waves being $E_{R,\min} > 1$ MeV based on wave statistics from satellites \citep{Summers&Thorne03, Kersten14, Cao17, Ross21}. With some energies below and some above $E_{R,\min}$ in the range $\sim 1.5$ MeV to $4$ MeV in our dataset, the averaging of $\langle j_{prec}/j_{trap}\rangle$ over all events lowers its value (to well below the peak ratio value of $\approx 1$) and flattens its spectrum.

Figure \ref{ratio}(b) shows the average ratio $\langle j_{prec}/j_{trap}\rangle$ as a function of $E/E^*$. For each event, the energy $E$ has been normalized to the event's $E^*$ at maximum $j_{prec}/j_{trap}$, the approximate energy for cyclotron resonance with the most intense EMIC waves. A least-squares fit to the data at $E\geq E^*$ (red dashed line) shows that $\langle j_{prec}/j_{trap}\rangle\sim (E/E^*)^{-1}$. This agrees well with the prediction of quasi-linear diffusion theory, which can be expressed as $j_{prec}/j_{trap} \sim \sqrt{D_{\alpha\alpha}} \sim [\gamma(E)/\gamma(E^*)]^{-5/4} \sim (E/E^*)^{-1}$ for $E > E^*$ and $E^*\in[1,4]$ MeV, where $D_{\alpha\alpha}$ is the electron bounce-averaged quasi-linear pitch-angle diffusion rate at the loss-cone angle $\alpha_{LC}$ and, again, $\gamma$ is the Lorentz factor \citep{Kennel&Petschek66, Ni15, Mourenas16:grl, Mourenas21:jgr:ELFIN, Mourenas22:jgr:elfin}. This suggests that electron precipitation driven by EMIC waves is described well by a quasi-linear diffusion treatment, which was developed under the assumption of low amplitude waves exhibiting low coherence (random phases) leading to slow phase space evolution compared to the wave period.

The precipitation ratio falloff with normalized energy at $E/E^{*}>1$ in Figure \ref{ratio}(b) is consistent with our assertion that the energy $E^*$ likely corresponds to the energy for cyclotron resonance at the frequency of the most intense EMIC waves. This is because, we contend, at lower energies, $E/E^* < 1$, the observed rapid decrease of $\langle j_{prec}/j_{trap}\rangle$ with progressively decreasing energy, down to $E/E^* \sim 0.1$, likely corresponds to cyclotron resonance with less intense EMIC waves at progressively higher frequencies \citep{Zhang16:grl, Ross21}, or to electron nonresonant scattering by the steep edge of some EMIC wave packets \citep{Chen16:nonresonant,An22}, both of which result in a reduced efficiency of scattering compared to that at peak wave power. Thus, viewed as a function of decreasing energy starting from the highest energies, while that ratio increases exponentially at first (above $E^*$) due to the increase in the diffusion rate, it starts to decrease just as quickly later (below $E^*$) due to the fast wave power decrease -- the turning point in that behavior being consistent with $E^*$, the resonance energy at peak wave power.

To check whether the decrease of $\langle j_{prec}/j_{trap}\rangle$ with decreasing energy below $E^*$ is actually consistent with EMIC wave power observations, it is useful to fit that ratio below $E^*$ (blue dashed line in Figure \ref{ratio}(b)). This fit quantifies the ratio's observed energy dependence at $E<E^*$ as $\langle j_{prec}/j_{trap}\rangle\sim \gamma^2(E)/\gamma^2(E^*)\sim 0.065\,\gamma^2(E)$, where we have replaced $E^*$ with its mean value $\langle E^*\rangle \approx1.45$ MeV in our dataset ($E^*$ varies between $\sim1$ MeV and $\sim3$ MeV). With this replacement, the best fit in Figure \ref{ratio}(b) is identical to the best fit in Figure \ref{ratio}(a). The fit value for $\langle j_{prec}/j_{trap}\rangle\sim 1$ actually occurs at an energy $E\sim E^*$. For a peak wave power nearly flat over a significant range of frequencies, $E^*$ represents the lowest energy at which cyclotron resonance with the highest power waves is achieved, because the diffusion rate $D_{\alpha\alpha}(E)$ that characterizes the behavior on the high-energy side decreases toward higher energy even for a flat wave power spectrum $B_w^2(f)$ \citep{Mourenas16:grl}. In that case, $E^*$ corresponds to the highest frequency over the nearly flat region of peak wave power, henceforth simply denoted $f_{peak}$, for which the diffusion rate $D_{\alpha\alpha}(E_{R,\min})\sim B_w^2(f)\cdot f/f_{cp}$ is maximized.

Next, we investigate the simplest explanation for the $\langle j_{prec}/j_{trap}\rangle$ dependence on $E/E^*$, at $E<E^*$: cyclotron resonance with progressively less intense EMIC waves at higher frequencies. This is done in two steps: First, assuming that this precipitation is due to quasi-parallel hydrogen band EMIC waves we use quasi-linear theory to infer from the above fit to the ELFIN measurements the statistical EMIC wave-power ratio $B_w^2(f)/B_w^2(f_{peak})$. And second, we compare this inferred wave power ratio with published statistics of EMIC wave power $B_w^2(f)$ directly observed by the Van Allen Probes near the equator in 2012-2016 \citep{Zhang16:grl}. As a reminder, nonlinear resonant scattering by an ensemble of independent, short-range, and large amplitude wave packets that may partake in the measured average wave spectral shape should produce average $\langle j_{prec}/j_{trap}\rangle$ energy spectra similar to that caused by classical diffusion by waves with the average wave power. In other words, the diffusive formalism, even though borrowed from quasi-linear theory, also applies for such nonlinear interactions if they participate statistically in the process. Yet, since we focus now on the precipitation of sub-MeV electrons, at $E<E^*$, which can reach cyclotron resonance only with high frequency waves, at $f>f_{peak}$, of much lower amplitudes than at peak wave power, the contribution from nonlinear interactions is likely much smaller than in the case of multi-MeV electron precipitation.

Towards the first step, we start from the full quasi-linear expressions for the precipitating to trapped flux ratio at ELFIN $j_{prec}(\alpha)/j_{trap}(\alpha_{trap})$ (\citep{Kennel&Petschek66} and \citep{Li13:POES}). Pitch angles are referenced to the equator and $\alpha_{trap}$ is such that $\ln(\sin\alpha_{trap}/\sin\alpha_{LC}) \sim 1/20$ (where $\alpha_{LC}$ is the equatorial loss-cone angle). Averaging $j_{prec}$ over equatorial pitch-angles $\alpha<\alpha_{LC}$, we get
\[
\frac{j_{prec}}{j_{trap}} \simeq \int_{0}^{1} \frac{I_0(z_0 x)/I_0(z_0)}{1+(z_0/20) I_1(z_0)/I_0(z_0)} dx,
\]
where $z_0 = 2\alpha_{LC}/\sqrt{D_{\alpha\alpha}\tau_B}$ \citep{Li13:POES}, and $\tau_B\sim\gamma/(\gamma^2-1)^{1/2}$ is the bounce period \citep{bookSchulz&anzerotti74}. This integral can be approximated as: $j_{prec}/j_{trap} = 0.9/z_0$ with less than 20\% error for $z_0\in[0.9,8]$, corresponding to $j_{prec}/j_{trap} \in[0.11,1]$. At low energies, $E < E^*$, and for quasi-parallel left-hand-polarized hydrogen band EMIC waves \citep{Kersten14} with a monotonically decreasing power $B_w^2(f)$ toward higher frequencies $f>f_{peak}$ in Van Allen Probe statistics \citep{Zhang16:grl}, the most efficient wave-driven pitch-angle diffusion of electrons near the loss-cone through cyclotron resonance should occur for $E(f)\sim E_{R,\min}(f)$ at similar latitudes $\lambda_R$ close to the equator for all $E$ in that range \citep{Mourenas16:grl}. Bounce-averaging \citep{Mourenas16:grl} the full expression of the local pitch-angle diffusion rate in the cold plasma approximation \citep{Summers&Thorne03, Su12EMIC} then gives
\[
D_{\alpha\alpha} \approx \frac{B_w^2(f_{cp}/f)(1-f/f_{cp})^{3/2}}{(\gamma^2-1)^{1/2}\gamma\,(1-f/2f_{cp})}
\]
at an equatorial pitch-angle $\alpha\simeq\alpha_{LC}$.

Combining the theoretical scaling laws for $j_{prec}/j_{trap}$ and $D_{\alpha\alpha}$ and the fit to ELFIN observations $j_{prec}/j_{trap}\sim\gamma^2$ in Figure \ref{ratio}, gives us the estimate of the wave-power spectrum consistent with these observations:
\[
\frac{B_w^2(E)}{B_w^2(E^*)} = \frac{(1 + 2 E)^4 E(E+1)}{(1 + 2 E^*)^4 E^*(E^*+1)} \cdot \frac{f(1-f_{peak}/f_{cp})^{3/2}(1-f/2f_{cp})}{f_{peak}(1-f/f_{cp})^{3/2}(1-f_{peak}/2f_{cp})}
\]

To obtain the mapping between energy and frequency, we utilize the cyclotron resonance condition coupled to the cold plasma dispersion relation. Its expression for $E<E^*$ and $f>f_{peak}$ is combined with its expression for $E^*$ and $f_{peak}$, the (highest-)frequency of peak EMIC wave-power ($f_{peak}/f_{cp}\sim 0.37-0.41$, as seen in statistical wave observations from the Van Allen Probes \citep{Zhang16:grl}). We also assume an ion composition with $>$94\% protons, as appropriate for when hydrogen band waves are present \citep{Kersten14, Ross22}. This yields a second order equation for $f/f_{cp}$, with solution $f/f_{cp} \simeq 2/(1 +\sqrt{1 + 4 C})$, where $C = (f_{cp}/f_{peak}) (f_{cp}/f_{peak} -1) E(1+E)/\left(E^*(1+E^*)\right)$.

The resonant frequency $f(E)/f_{cp}$, expressed as a function of the resonance energy in the above equation, allows us, by substitution in the previous equation, to obtain the EMIC wave-power ratio $B_w^2(f)/B_w^2(f_{peak})=B_w^2(E)/B_w^2(E^*)$ inferred from ELFIN, with an error smaller than 40\% for $j_{prec}/j_{trap} \in [0.11, 1]$. Note that the above normalizations of $j_{prec}/j_{trap}$ to its level at $E^*$ and of EMIC wave power to its level at $f_{peak}/f_{cp}$ allow us to eliminate the dependencies of $j_{prec}/j_{trap}$ and $D_{\alpha\alpha}$ on $f_{pe}/f_{ce}$ and on the non-normalized EMIC wave power at the equator conjugate to ELFIN, which are unknown. Here we only had to assume that the ensemble of $j_{prec}/j_{trap}$ measurements from ELFIN are statistically compliant to an ensemble of EMIC wave power observations on another, equatorial, platform in order to convert the energy spectrum of the precipitation to a frequency spectrum of wave power that should be consistent with that precipitation.

\begin{figure}
  \centering
  \includegraphics[width=0.9\textwidth]{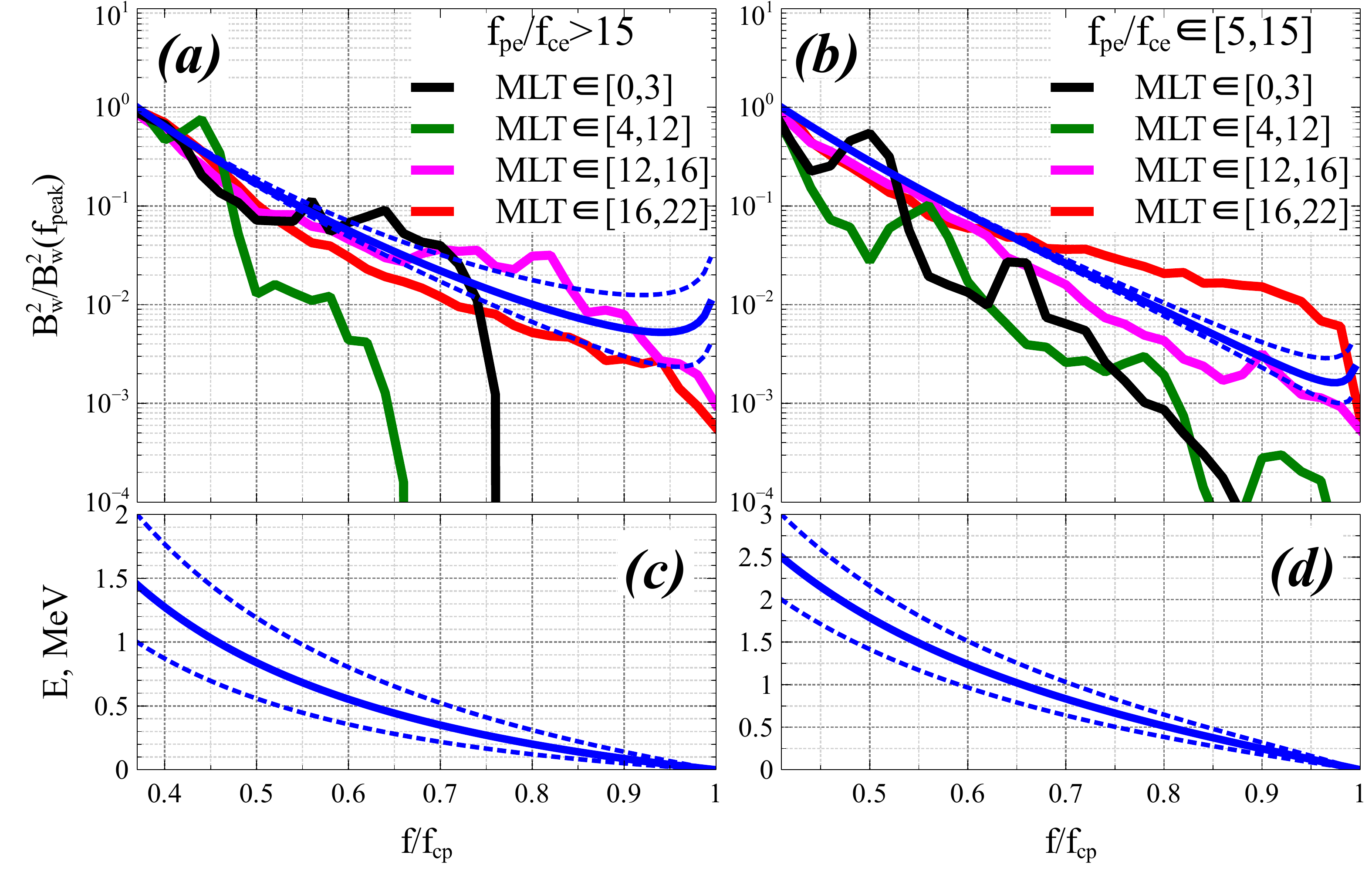}
\caption{(a) EMIC wave power ratio $B_w^2(f)/B_w^2(f_{peak})$ inferred, using quasi-linear theory, from ELFIN statistics of precipitating-to-trapped electron flux ratio $j_{prec}/j_{trap}$ (solid blue curve), and compared with statistical EMIC wave power ratios obtained from Van Allen Probes 2012-2016 observations in four different MLT sectors when $f_{pe}/f_{ce}>15$ (black, green, magenta, and red curves). Specifically, the solid blue curve was derived from the best fit to $\langle j_{prec}/j_{trap}\rangle$ in Figure \ref{ratio}(a) and the dashed blue curves show its uncertainty range, corresponding to the uncertainty in $E^*$ from the measurements, also depicted in Panel (c), below. (b) Same as (a) but for $f_{pe}/f_{ce}\in[5,15]$. (c) Energy $E$ of electrons near the loss cone in cyclotron resonance with EMIC waves at $f/f_{cp}$ in Panel (a), assuming waves at $f_{peak}/f_{cp}\sim0.37$ in resonance with $E^*\sim1.45$ MeV electrons for $f_{pe}/f_{ce}>15$. Dashed curves are based on the measurement uncertainty range in $E^*$. (d) Same as (c) for $f_{peak}/f_{cp}\sim0.41$ and $E^*\sim2.5$ MeV for $f_{pe}/f_{ce}\in[5,15]$, corresponding to wave power ratios in Panel (b).
}
\label{fig:modelspectrum}
\end{figure}

We proceed, now, to the second step in our consistency check between the ELFIN observations of electron precipitation and independently collected EMIC wave-power spectra, in order to assess the validity of quasi-linear diffusion in describing the precipitation down to low energies. Figure \ref{fig:modelspectrum}(a) compares the wave power ratio $B_w^2(f)/B_w^2(f_{peak})$ inferred above from ELFIN statistics of $j_{prec}/j_{trap}$ (solid blue curve) with the measured hydrogen band EMIC wave power ratio from Van Allen Probes statistics when $f_{pe}/f_{ce}>15$ \citep{Zhang16:grl} in different MLT sectors (black, green, magenta, and red curves). In the high density plasmaspheric plume or at the plasmapause, as implied by the $f_{pe}/f_{ce}$ ratio criterion used in this subset of wave data, the cyclotron resonance condition can indeed be satisfied for the optimum parameters $E^*\sim1.45$ MeV in Figures \ref{ratio}(a,b) and $f_{peak}/f_{cp}\sim0.37$ from the above wave statistics. Figure \ref{fig:modelspectrum}(c) indicates that $200$ keV electrons are then in resonance with waves at $f/f_{cp}\sim 0.8$ near the equator. We see in Figure \ref{fig:modelspectrum}(a) that the observed statistical EMIC wave power ratios in the 12-22 MLT sector (magenta and red curves) agree quite well with the power ratios inferred from ELFIN data (blue curve with dashed lines indicating uncertainties) over two decades in power, in the frequency range $f/f_{cp}\sim0.37$ up to $f/f_{cp}\sim0.8-0.95$. Therefore, in high density regions (those with $f_{pe}/f_{ce}>15$), the observed EMIC wave power ratio is consistent with that inferred from ELFIN precipitation of $j_{prec}/j_{trap}$ from $\sim1.45$ MeV down to $200$ keV, sometimes even down to $100$ keV.

In regions of lower plasma density, with $5<f_{pe}/f_{ce}<15$, EMIC wave statistics from the Van Allen Probes show that $f_{peak}/f_{cp}\sim0.41$ \citep{Zhang16:grl}, corresponding to a higher  $E^*\sim2.5$ MeV for $f_{pe}/f_{ce}\sim13-15$. For these parameters, Figure \ref{fig:modelspectrum}(b) shows that in the 12-22 MLT sector the observed EMIC wave power ratios in the range $f/f_{cp}\sim0.41$ to $f/f_{cp}\sim0.90-0.95$ \citep{Zhang16:grl} at Van Allen Probes (magenta and red curves) are consistent with those inferred from ELFIN statistics (blue curves). In the above $f_{pe}/f_{ce}$ range, the frequencies where agreement prevails correspond to $j_{prec}/j_{trap}$ from $\sim2.5$ MeV down to $\sim100-200$ keV.

It is evident from Figures \ref{fig:modelspectrum}(a,b) that the inferred and observed EMIC wave power ratios agree well, and up to high frequencies (low resonance energies), only in the 12-22 MLT sector, i.e., near dusk. This is quite consistent with the fact that most of the EMIC wave-driven events at $L<7$ in our database occurred near dusk (Figure \ref{fig:statistics2}(b)). It is also consistent with the predominance of similar events at dusk in Firebird-II statistics \citep{Capannolo21}. In the 0-3 MLT sector (black lines in Figures \ref{fig:modelspectrum}(a,b)) the observed EMIC wave power is consistent with EMIC wave-driven electron precipitation down to $\sim300$ keV only when $f_{pe}/f_{ce}>15$. Few events exist in this sector in our database. In the 4-12 MLT sector (green lines) the observed EMIC wave power is too weak to drive significant electron precipitation below $\sim1$ MeV, consistent with the absence of precipitation events in that sector in our  database. (Note: Since EMIC wave power spectra depend weakly on $L$ in Van Allen Probes statistics \citep{Zhang16:grl, Ross21}, the present results should hold for $L\in[4,7]$.)

Therefore, we see that the observed, moderately efficient electron precipitation at energies as low as $\sim200-300$ keV could simply be due to quasi-linear scattering by moderate intensity EMIC waves at high frequencies, up to $f/f_{cp}\sim0.8-0.9$, provided that such waves are present during a majority of these events with a similar average intensity as in statistical empirical wave models. Such waves may still be sufficiently below the proton gyrofrequency to evade hot plasma effects, at least to first order \citep{Chen11:emic, Ross21}. Assuming a maximum wavenumber that can be attained in the presence of hot plasma effects given by $kc/\Omega_{pi}\sim2$ (corresponding to a typical cold ion temperature of $\sim 10$ eV in a plasmaspheric plume or just within the plasmasphere) for left-hand-polarized hydrogen band waves in a plasma with more than 94\% protons, the minimum electron energy for cyclotron resonance can indeed be as low as $\sim150$ keV to $350$ keV for $f_{pe}/f_{ce}\sim15$ to $30$, when a sufficient transient hot $H^+$ temperature anisotropy $A>2.3$ generates such high-frequency waves \citep{Chen11:emic, Chen13:emic}. A typical precipitation event of this kind has been analyzed in Section \ref{subsec:event3}.

Another important point is that the measured ratios $j_{prec}/j_{trap}\sim 1/8-1$ from low energy, $E\ll E^*$, to $E^*$ actually correspond to a quasi-linear diffusion close to the {\it strong diffusion regime} with characteristic time scales for partial or full loss-cone filling on the order of a quarter of a bounce period $\tau_B/4\sim 0.25$ s \citep{Kennel&Petschek66}. Such fast diffusive time scales are therefore still consistent with the time scales of the observed electron precipitation at ELFIN. Thus, even in the quasi-linear regime it is possible for temporal or spatial wave power variations encountered by drifting and bouncing electrons to explain the short-lived (sub-spin), bursty nature of the precipitation at ELFIN.

The present results agree with test particle hybrid simulations suggesting that low-amplitude, high-frequency hydrogen band EMIC waves could be the main cause of 0.55 MeV electron precipitation \citep{Denton19}. Several previous studies have also provided hints of a prevalence of hydrogen band waves in driving sub-MeV relativistic electron precipitation \citep{Chen13:emic, Qin18, Zhang21}. Even when such high-frequency waves are not evident at apparently conjugate measurements at the equator, this may be due to mapping uncertainties - the waves could still be present $\gtrsim$ 0.5 Earth radii away where the ionospheric electron precipitation measurements actually map to. Further analysis of EMIC wave data and sub-MeV precipitation with ELFIN's twin spacecraft, providing higher spatio-temporal resolution of these phenomena, will be able to further study these points in greater detail. In addition, one cannot totally rule out a possible role of helium band EMIC waves in driving some sub-MeV electron precipitation, in spite of the expected strong cyclotron damping very close to the helium gyrofrequency \citep{Chen13:emic, Cao17}. A similar study as in Figure \ref{fig:modelspectrum}, but focusing on helium band waves, could be performed to check this point. This would require extensive numerical calculations including hot plasma effects in the dispersion relation \citep{Chen13:emic}.

As a final note, we comment on the global contribution of EMIC waves to sub-MeV electron scattering. Although bursts of EMIC waves can drive relatively intense sub-MeV electron precipitation in a narrow MLT sector, their very weak time- and MLT-averaged power at high frequencies should prevent them from contributing significantly to the global loss rates of sub-MeV electrons up to high equatorial pitch-angles $\alpha\sim 90^\circ$, above and beyond what can already be done by typical chorus waves in the dawnside trough \citep{Mourenas16:grl, Boynton17, Agapitov18:jgr, Drozdov19, Drozdov20, Miyoshi20, Ross21, Zhang22Micro}. This is because at a fixed energy $E\sim E^*$, cyclotron resonance between electrons of equatorial pitch-angle $\alpha$ and hydrogen band EMIC waves at a frequency $f$ indeed corresponds to a scaling $\cos^2\alpha/\cos^2\alpha_{LC} \sim (f_{cp}/f -1)/(f_{cp}/f_{peak} -1) f_{peak}/f $. This implies that electrons of higher $\alpha$ reach resonance with waves of higher frequency $f>f_{peak}$ (e.g. see first standalone equation in \citet{Mourenas16:grl}). Therefore, the steep decrease of EMIC wave power from $f/f_{cp}\sim0.4$ to $\sim0.9$, by two orders of magnitude in Figure \ref{fig:modelspectrum}, implies a rapid decrease of the quasi-linear pitch angle diffusion rate $D_{\alpha\alpha}(\alpha)$ away from the loss-cone \citep{Summers&Thorne03, Mourenas16:grl}. This is consistent with pitch angle bite-out signatures (stronger electron loss near the loss-cone than at higher pitch angles) produced by the most intense EMIC waves \citep{Usanova14, Adair22} at energies of $\sim1-4$ MeV. Such signatures demonstrate the inefficacy of EMIC waves in depleting the relativistic electron flux at equatorial pitch angles far from the loss cone, absent intense chorus waves \citep{Mourenas16:grl, Zhang17}. The same arguments hold at sub-MeV energies, except that cyclotron resonant EMIC waves have higher frequencies than at $\sim1-4$ MeV and, therefore, much lower amplitudes, making them even less efficient than typical intense chorus waves in the dawnside trough in driving the precipitation of electrons near the loss-cone.

\section{Summary and discussion}
Relativistic electrons in the near-Earth environment are an important contributor to space weather and may play a significant role in charged particle energy input to the atmosphere from space. The question of how such electrons, after being accelerated in near-Earth space by waves or transported into it by, say, radial diffusion, are lost remains open. This question is particularly vexing for such electron loss from the outer radiation belt, at (L$<$7), which has the highest fluxes of such electrons and is thus most important for space weather. Aside from magnetopause shadowing (the result of magnetospheric compressions changing electron drift paths from trapped to open), and field-line curvature scattering precipitation at the isotropy boundary, both affecting L-shells beyond the outer radiation belt, and whistler-mode wave-driven precipitation which mostly affects sub-MeV electrons, most attention on relativistic electron loss has been placed on these electrons' interaction with EMIC waves. This is because EMIC waves can resonate with electrons of highly relativistic energies $>1$ MeV under high-density, low magnetic field plasma conditions that can be realized in the outer edge of the plasmasphere, well inside the outer radiation belt at active times. As they can act on trapped, outer radiation belt electrons, such waves can change significantly the outer radiation belt fluxes and therefore are an important process to include in space weather models. The recent launch of the ELFIN mission, with a goal to determine whether EMIC waves are predominantly responsible for this loss or whether other wave processes are implicated, provided a unique dataset of 50 - 5000 keV electrons obtained on a polar, low-altitude orbit, with which we can address this question comprehensively and for the first time. In this paper, we primarily focused on the question of whether EMIC waves can be definitively shown to be responsible for the observed scattering of relativistic electrons, and whether whistler-mode chorus waves (the other candidate for pitch-angle scattering of sub-MeV and up to MeV electrons into the loss cone) simultaneously might be implicated.

After a short review of EMIC wave generation and its interaction with energetic electrons in Section \ref{sec:background}, we presented in Section \ref{sec:casestudies} the first comprehensive examples of typical EMIC wave-driven precipitation of such electrons from ELFIN. The high energy and pitch-angle resolution of ELFIN allowed us to identify and quantitatively investigate the energy range of such precipitation and study its properties using the precipitating-to-trapped flux ratio, $R$, measured by a single detector on a spinning platform, thus avoiding problems with multiple detector inter-calibration. This ratio increasing with energy and peaking in the range $>$0.5 MeV is a tell-tale signature of EMIC wave-driven precipitation. In Section \ref{subsec:event-1} we showed how this spectral signature is differentiated quite well from that of whistler-mode chorus driven precipitation which is a decreasing ratio with energy from as low as 50 keV (the lower energy limit of the ELFIN detector) to $>$0.5 MeV. (Both spectral types can be observed on occasion simultaneously, as expected, since whistler-mode chorus and EMIC waves can co-exist -- in that case, the two spectral shapes can still be well separated.)

Next, in Section \ref{subsec:event0} we used a case study of typical EMIC wave-driven emissions, accompanied by equatorial measurements of EMIC waves on THEMIS to demonstrate that the bursty nature of the precipitation at ELFIN (lasting only 7-14s, or 2.5-5 ELFIN spin periods) is consistent with the spatial extent of the EMIC wave region at the equator (lasting 6min). In this case, the equatorial extent of the EMIC region was $\Delta L \approx$0.23 which is consistent with published statistical averages of $\Delta L \approx$0.5 \citep{Blum16,Blum17}. Thus, the burstiness of EMIC wave-driven precipitation at polar orbiting, low-altitude spacecraft like ELFIN is likely typical. We have also shown that the burstiness can extend to sub-spin time-scales, 1-5 spin sectors, at least in part due to the spatial variability of the wave-power at the equator. This can cause the precipitating-to-trapped flux ratio to be, on occasion, aliased and exceed unity. While there are reported cases when this flux ratio on ELFIN exceeds unity due to nonlinear scattering, an assertion supported by fortuitous measurements of the EMIC wave amplitudes at conjugate observatories \citep{Grach22:elfin}, conclusions on the preponderance and significance of nonlinear effects cannot be drawn from the ratio alone without further statistical or detailed case study analyses, both left for the future. However, we have argued that nonlinear effects (from presumably mainly short wave packets -- e.g., see various examples of such short packets in \citep{Usanova10} and \citep{An22}) are statistically incorporated into the spectral shapes of precipitation at ELFIN, just as these high-amplitude EMIC wave packets are incorporated into the statistical averages of equatorial wave spectra, so that a direct comparison using a diffusive formalism should be possible.

In Section \ref{subsec:event1} we showed that consecutive ELFIN passes over the same MLT region can reveal the EMIC wave-driven precipitation's spatio-temporal evolution in location, extent (in $L$ and $\Delta L$), and intensity. This allows us to study the EMIC wave-generation region in a geophysical context (e.g., during substorms and storms) as well as the potential of the waves for reducing the outer radiation belt flux.

To evaluate the consistency of the observed energy of moderate and strong precipitation at ELFIN (ratios $R\in$[0.3,0.5] and $R>$0.5, respectively) with the resonance energy expected for EMIC wave-driven precipitation, we utilized in Section \ref{sec:conjunctions} conjunctions with ground based or equatorial assets. Using such data-informed estimates of $f_{pe}/f_{ce}$ and of $f_{\max}/f_{ci}$, with only occasional support from models, we examined the resonance energy that would be consistent with these estimates. We found it to be in agreement with the observed energies of moderate and strong precipitation seen at ELFIN. This demonstrates that the observed precipitation as identified by the ratio $R$ is indeed consistent with theoretical expectation from resonant interactions.

We proceeded, in Section \ref{sec:statstudies}, to study statistically an ensemble of EMIC wave-driven events observed by ELFIN. These were identified in two years of ELFIN data (2019-2021) as enhancements in the precipitating-to-trapped flux ratio $R$, peaking at $>0.5$MeV. The average event duration on ELFIN, 17s (6 spins), or $\Delta L \sim$ 0.56, is consistent with published reports of the typical radial extent of EMIC waves in the magnetosphere, $\Delta L \sim$ 0.5 \citep{Blum16,Blum17}, validating our assertion that the bursty nature of the precipitation is due to the spatial localization of the EMIC wave interaction region in the magnetosphere. The most populous category of events is those occurring at pre-midnight and dusk, at $L\sim$5-7. A second class of events was found at midnight, post-midnight and dawn at $L\sim$8-12. These two categories are roughly co-located with the two main populations of EMIC waves in the magnetosphere \citep{Min12}, further solidifying our assertion that our observations of relativistic electron precipitation are caused by EMIC waves.

The peak precipitating-to-trapped flux ratio exceeds one for many of the events studied. Considering that such a ratio, if scrutinized and confirmed, could signify nonlinear interactions, the only means of exceeding the quasi-linear strong diffusion limit (e.g., \citet{Kubota15,Grach&Demekhov20}), we examined further how it may arise. We found that on many occasions extreme burstiness of the precipitation, lasting 1-5 spin phase sectors was evidenced in the data. (Each ELFIN spin period has 16 sectors and the loss cone most often contains 6 such sectors.) Examination of EMIC waves at THEMIS E on one event studied revealed that this burstiness is, in fact, to be expected at ELFIN due to the spatial localization of the EMIC wave coherence and amplitude arising from interference of wave packets in the presence of extreme density gradients. This burstiness in precipitation can result in temporal aliasing of ELFIN's precipitating-to-trapped flux ratio, $R$, resulting in values that can exceed 1 or be well below 1. Careful analysis of individual events together with modeling of measured wave spectra from ancillary datasets can, at times, confirm the nonlinear nature of the scattering (\cite{Grach21:emic}). Either future extensive case-by-case analysis or careful statistical analysis to identify and remove events subject to aliasing is required to establish the preponderance of nonlinear effects. However, we argued that by analogy with similar effects for whistler-mode chorus waves, it is still quite likely that precipitation due to nonlinear EMIC wave scattering can conform to a statistical analysis using a diffusive treatment, because precipitation bursts contribute to the average flux (and flux ratio) at a level commensurate to the contribution of nonlinear wave power bursts in statistical averages of equatorial wave power. We proceeded with such a statistical analysis even though we recognize that in small or large part the precipitation and the associated wave amplitudes considered may incorporate nonlinear effects.

Using this database we have shown (Section \ref{subsec:spatialstats}) that:
\begin{itemize}
\item The typical energy of the peak precipitating-to-trapped flux ratio, $E^*$, a measured proxy for the resonance energy at the frequency of peak wave power, $E_{peak}$, is in the range $\sim 1-3$ MeV. This is consistent with expectation for EMIC wave-electron resonant interactions, based on the cold plasma dispersion relation and prior statistics of wave-power near peak wave-power, $f_{peak}$, at the equator.
\item The above measured energy $E^*$ decreases with $L$-shell as $\sim L^{-0.5\pm0.1}$. This dependence is also in good agreement with theoretical expectations.
\item The typical energy of the half-peak of the precipitating-to-trapped flux ratio, $E^*_{\min}$, a proxy for the minimum resonance energy $E_{R,\min}$ corresponding to significant wave-driven electron scattering toward the loss-cone, is in the range $\sim 0.5-1.5$ MeV and falls off with $L$-shell at the same rate as $E^*$. This too is consistent with theoretical expectations based on equatorial wave-power estimates at the maximum frequency below the ion gyrofrequency where wave-power remains significant. This shows that sub-MeV electron precipitation accompanying multi-MeV electron precipitation, is also likely driven by EMIC waves.
\end{itemize}
We next examined (Section \ref{sec:intenseevents}) the properties of the most intense ($R>$1/2), highly relativistic (having $E^*>1$ MeV), EMIC wave-driven electron precipitation events, of the most populous category in our database (L$<$7). We consider these to be the most efficient EMIC wave-driven events in our database. Such events are highly correlated with geomagnetic activity. And, as expected from the statistics of all events, they are predominantly seen near dusk (MLT $\sim$ 18) with their occurrence rates dropping precipitously (though still finite) at noon and post-midnight. We found that:
\begin{itemize}
\item About 35\% of the time, the most efficient EMIC wave-driven events are still associated with moderate or strong ($R>$1/3) precipitation of electrons with energies as low as 100-300 keV. Only 6.5\% of the time do such events have a definitively negative slope of $R$ versus E, at that low energy range. For most events (at least 78\% and potentially as many as 93.5\%) that slope is positive. This suggests that simultaneous energetic electron scattering by whistler-mode chorus and EMIC waves (as was presented in Figure \ref{fig:intro}, right column, and discussed in Section \ref{subsec:event-1}) are a minority, occurring for only one in every 16 of the highly relativistic, strong EMIC wave-driven events.

\item The average $R$ versus $E$ spectrum of the most efficient EMIC wave-driven events exhibits an exponential increase with energy (more precisely with the square of the Lorentz factor, $R \sim \gamma^2$) up to its peak value $R\sim1$ at $E\sim E^*$ (whose average value is $\langle E^*\rangle\sim 1.45$ MeV). Examined versus energy normalized to the peak-precipitation energy, $E/E^*$, $R$ exhibits an exponential decay with normalized energy away from (on both sides of) unity: ($R\sim (E/E^*)^{-1}$ above and $R\sim(E/E^*)^{+1}$ below $E/E^*=1$). This decay is in agreement with quasi-linear diffusion theory for resonant scattering by a typical equatorial EMIC wave power spectrum that falls off away from its peak wave power, which is consistent with observations.

\item Based on both points above, the majority of the weaker precipitation below $1$ MeV is thus probably due to electron resonant interaction with EMIC waves at frequencies $f>f_{peak}$, i.e., above the peak power frequency of hydrogen band EMIC waves (as was also inferred by the $E^*_{\min}$ decrease as a  function of L-shell in Section \ref{subsec:spatialstats} and summarized above). Even though this wave power is much smaller than at $f_{peak}$, it is still finite \citep{Denton19, Zhang16:grl, Zhang21}. But because of the significantly higher trapped electron flux at lower energies, even waves of such small power may still produce finite precipitating fluxes at $\sim200-500$ keV, as observed.
\item Very low energy ($\sim100$ keV) electron precipitation at times of strong, highly relativistic precipitation, could still be predominantly due to interactions with very low-amplitude high-frequency hydrogen band EMIC waves, excited very near the equatorial gyrofrequency, $f_{cp}$, by $10-100$ eV anisotropic ion populations \citep{Teng19EMIC,Zhang21}. Such waves can more easily experience strong cyclotron damping by low energy ions, likely causing them to be transient and rendering their equatorial observations sparse. Since $\sim 100$ keV electrons cannot reach cyclotron resonance with the most frequent hydrogen band EMIC waves \citep{Zhang16:grl}, other precipitation mechanisms should also be examined and quantified. Those include nonresonant scattering by EMIC wave-packets with sharp edges \citep{Chen16:nonresonant,An22}, which can scatter electrons well below the minimum resonance energy while still exhibiting a rising $R$ versus $E$ spectrum. However, simultaneous whistler-mode wave-driven precipitation which is very efficient at such low energies, is clearly contributing to this precipitation (as discussed above), even though it is likely a minority, and should also be further considered \citep{Artemyev16:ssr,Ma16:diffusion}.
\end{itemize}
Our results confirm the crucial role played by EMIC waves in relativistic electron losses in the Earth’s outer radiation belt (at $L$-shells between 5 and 7, where ELFIN detected most EMIC wave-driven precipitation events). We find that the main predictions of the classical resonant scattering models of EMIC wave-driven precipitation (e.g., typical energy of precipitation peaks, the energy spectra and the $L$-shell dependence of minimum and peak-power resonance energy) are consistent with ELFIN case studies and statistical results presented herein. However, ELFIN's observed fine structure of the precipitation in energy, pitch-angle and time reveal interesting details pertaining to the nature of low-energy precipitation concurrent with strong, highly relativistic electron scattering, and the origin of precipitation exceeding the strong diffusion limit, that deserve further attention.

Towards that end, significant new information on EMIC wave-driven precipitation can be obtained by further combining ELFIN electron measurements and theoretical modelling of wave-particle interactions with several ancillary datasets: First, ELFIN conjunctions with ground-based and equatorial spacecraft such as Van Allen Probes, THEMIS, ERG and MMS exist, can provide important information on the EMIC waves implicated in the scattering. The equatorial spacecraft can also provide local density and magnetic field information, thus constraining the $f_{pe}/f_{ce}$ ratio. The large dataset of ELFIN (spanning four years of operation, from 2018 to 2022) provides ample opportunities for studies using such conjugate observations. These also present good opportunities to investigate the relationship between the fine structure of EMIC waves at the equator (wave packet coherence \citep{Blum16,Blum17}, frequency ranges \citep{Kersten14,Zhang16:grl}, and hot plasma effects \citep{Cao17,Chen19}) and the energy distribution of precipitating electrons at ELFIN. Second, ELFIN's energetic particle detector for ions (EPDI), which measures total ions over the same energy range as electrons (50 - 5000 keV), has been calibrated and has collected data for approximately three months prior to the end of the mission on both satellites. And finally, ELFIN's fluxgate magnetometer (FGM) instrument, which provides field-aligned current information and has the capability of measuring waves up to a Nyquist frequency of 5Hz, has been operating well and has provided data on both satellites for more than a year towards the end of the mission (note that sub-spin resolution calibration to reveal EMIC waves is still on-going). These datasets can provide information on energetic ion scattering and precipitation by the same EMIC waves that scatter energetic electrons in the inner magnetosphere (when the ion resonance energy is sufficiently high at appropriate $f_{pe}/f_{ce}$ ratios), on the location of the EMIC wave excitation relative to large scale magnetospheric current system sources, and on EMIC waves potentially seen at ELFIN at the same time as electron precipitation. Critical open questions are (i) What is the overall contribution of nonlinear interactions and what can we learn about the physics of such interactions from ELFIN's precipitation energy and pitch-angle spectrum? (ii) What effects are responsible for precipitation at sub-MeV energies? (iii) What is the relative contribution of EMIC wave-driven precipitation to the total loss of energetic electrons in the inner magnetosphere and to the total atmospheric energy deposition by energetic particles? There is already good theoretical background for further model development of nonlinear resonant interactions \citep{Albert&Bortnik09,Kubota15,Kubota&Omura17,Grach&Demekhov20,Grach21:emic} and nonresonant interactions \citep{Chen16:nonresonant,An22}. Such development can proceed on solid grounds only if guided by statistically-derived properties of the observed electron precipitation and ancillary measurements afforded by the above datasets.

\begin{acknowledgements}
At UCLA, we acknowledge support by NASA awards NNX14AN68G (5/22/2014 – 7/15/2022) and 80NSSC22K1005 (7/16/2022 - present), NSF award AGS-1242918 (9/24/2014-7/31/2019) and AGS-2019950 (May 2020 – present). We are grateful to NASA’s CubeSat Launch Initiative program for successfully launching the ELFIN satellites in the desired orbits under ELaNa XVIII. We thank the AFOSR for their early support of the ELFIN program under its University Nanosat Program, UNP-8 project, contract FA9453-12-D-0285 (02/01/2013-01/31/2015). We also thank the California Space Grant program for student support during the project’s inception (2010-2014). The work at UCLA was also supported by NASA contract NAS5-02099 for data analysis from the THEMIS mission. We specifically thank: J. W. Bonnell and F. S. Mozer for the use of EFI data in determining spacecraft potential derived density from THEMIS E, J. P. McFadden for use of ESA data in obtaining electron temperature measurements to incorporate into the procedure for electron density determination from the spacecraft potential, and K. H. Glassmeier, U. Auster and W. Baumjohann for the use of FGM data provided under the lead of the Technical University of Braunschweig and with financial support through the German Ministry for Economy and Technology and the German Center for Aviation and Space (DLR) under contract 50 OC 0302. Additionally, we thank the operators of the Magnetic Induction Coil Array, in particular Dr. Marc Lessard of UNH and the operators of the South Pole, PG3, and PG4 sites at NJIT, UNH, and Virginia Tech; and the Sodankyla Geophysical Observatory for providing the Finnish pulsation magnetometer network data. The induction coil magnetometer at South Pole is operated and managed by New Jersey Institute of Technology (NJIT) under NSF grant OPP-1643700.

X.-J.Z. and A.A. acknowledge NASA awards 80NSSC20K1270, 80NSSC22K0517; X.-J.Z. also acknowledges NSF award 2021749; X.A. acknowledges NSF grant NO. 2108582; W. Li acknowledges NSF award AGS-2019950 and NASA award 80NSSC20K1270; M.D.H. was supported by NSF AGS-2027210; X. M. acknowledges NASA contract 80NM0018D0004 to the Jet Propulsion Laboratory, California Institute of Technology; S.R.S. thanks Dasha Gloutak of CU Boulder and Samuel Rickert of UCLA.

We acknowledge the critical contributions and talent of numerous volunteer team members (more than 300 students contributed to this program since its inception) who made this challenging program both successful and fun. ELFIN would not have been possible without the advice and generous time contributions from numerous members of the science and technology community who served as reviewers, became unofficial mentors, or were the bouncing board of ideas during trade studies or tiger team deliberations. Special thanks go to UCLA’s: Steve Joy and Joe Mafi; The Aerospace Corporation’s: David Hinkley, Eddson Alcid, David Arndt, Jim Clemmons (now at UNH), Chris Coffman, Joe Fennel, Michael Forney, Jerry Fuller, Brian Hardy, Petras Karuza, Christopher Konichi, Justin Lee, Pierce Martin, Leslie Peterson, David Ping, Dan Rumsey and Darren Rowen; NASA GSFC’s: Gary Crum, Thomas Johnson, Nick Paschalidis, David L. Pierce, Luis Santos and Rick Schnurr; NASA JPL’s: Matt Bennett, Olivia Dawson, Nick Emis, Justin Foley, Travis Imken, Allen Kummer, Andrew Lamborn, Marc Lane, Neil Murphy, Keith Novac, David R. Pierce, Sara Spangelo, and Scott Trip; AFRL’s: Travis Willett, David Voss and Kyle Kemble; UCB’s: Peter Berg, Manfred Bester, Dan Cosgrove, Greg Dalton, David Glaser, Jim Lewis, Michael Ludlam, Chris Smith, Rick Sterling, and Ellen Taylor; Tyvak’s: Bill McCown, Robert Erwin, Nathan Fite, Ehson Mosleh, Anthony Ortega, Jacob Portucalian, and Austin Williams; Cal State University Northridge’s: James Flynn and Sharlene Katz; JHU/APL’s: Jeff Asher and Edward Russell; Montana State University’s: Adam Gunderson; Space-X’s: Michael Sholl; Planet Labs’: Bryan Klofas; and LoadPath’s: Jerry Footdale.

\end{acknowledgements}
%
\section*{Conflict of Interest}

The authors have no relevant financial or non-financial interests to disclose.

\bibliographystyle{agsm}

%
%

\newpage

\end{document}